\documentclass[rmp,aps,amsmath,amssymb,amsfonts,superscriptaddress,nofootinbib,endfloats,twocolumn,Times]{revtex4-1}

\usepackage{graphics}
\usepackage{epsfig}
\usepackage{enumerate}
\usepackage{color}
\usepackage[colorlinks=true,citecolor=blue,linkcolor=blue,linktoc=none,pdfmenubar=false]{hyperref} 
\usepackage{amsthm} 
 \usepackage{natbib}

\newcommand{\dm}{n}
\newcommand{\hook}{\raisebox{-0.35ex}{\makebox[0.6em][r]
{\scriptsize $-$}}\hspace{-0.15em}\raisebox{0.25ex}{\makebox[0.4em][l]{\tiny
 $|$}}}
\newcommand{\be}{\begin{equation}}
\newcommand{\ee}{\end{equation}}
\newcommand{\ba}{\begin{eqnarray}}
\newcommand{\ea}{\end{eqnarray}}
\newcommand{\cwedge}[1]{\mathop{\wedge}_{{}^{#1}} }
\newcommand{\nn}{\nonumber}
\newcommand{\even}{{\mathrm{e}}}
\newcommand{\odd}{{\mathrm{o}}}
\newcommand{\eps}{\varepsilon}
\newcommand{\A}[1]{A^{\!(#1)}}
\newcommand{\cv}[1]{{\partial}_{#1}}
\newcommand{\KV}[1]{\xi^{(#1)}}
\newcommand{\psc}[1]{\Psi_{#1}}
\newcommand{\lst}[1]{\langle{#1}\rangle}
\newcommand{\psf}[1]{\tilde\Psi_{#1}} 
\newcommand{\mD}{\mathcal{D}} 
\newcommand{\h}[1]{\hat{#1}}
\newcommand{\M}{\mathcal{M}} 
\newcommand{\mP}{\mathcal{P}} 
\def\half{\frac{1}{2}}
\newcommand{\Mb}{\overline{M}} 
\newcommand{\rhob}{\overline{\rho}}

\newtheorem*{Theorem}{Theorem}

\begin{document}
\title{Hidden Symmetries of Dynamics in Classical and Quantum Physics}

\author{Marco Cariglia}
\email{marco@iceb.ufop.br}
\affiliation{DEFIS, Universidade Federal de Ouro Preto,  Campus Morro do Cruzeiro, 35400-000 Ouro Preto, MG - Brasil}

\begin{abstract}  
This article reviews the role of hidden symmetries of dynamics in the study of physical systems, from the basic concepts of symmetries in phase space to the forefront of current research. Such symmetries emerge naturally in the description of physical systems as varied as non-relativistic, relativistic, with or without gravity, classical or quantum, and are related to the existence of conserved quantities of the dynamics and integrability. In recent years their study has grown intensively, due to the discovery of non-trivial examples that apply to different types  of theories and different numbers of dimensions. Applications encompass the study of integrable systems such as spinning tops, the Calogero model, systems described by the Lax equation,  the physics of higher dimensional black holes, the Dirac equation, supergravity with and  without fluxes, providing a tool to probe the dynamics of non-linear systems. 
\end{abstract}                                                                 

\date{September 14, 2014}
\maketitle
\tableofcontents

\section{INTRODUCTION}
\label{sec:introduction}
The role of symmetries in Physics is ubiquitous. They can appear either as a useful tool to describe special systems and configurations, or as a fundamental building block of a theory itself. The Standard Model of particle physics and General Relativity are two examples of the latter type, as well as their supersymmetric extensions. String theory with its considerable amount of symmetries and dualities is another. In particular the usefulness of the idea of symmetry is that it can be fruitfully applied to diverse areas, such as relativistic and non-relativistic theories, classical and quantum, and to different types of systems within each of these areas. 
 
Symmetries have been successfully used in Physics so much that a stage has been reached where a more refined strategy might be needed. Take the example of General Relativity. There the most common meaning of the word symmetry is associated to that of isometry, that is a spacetime diffeomorphism that leaves the metric invariant. A one-parameter continuous isometry is linked to the existence of Killing vectors. Therefore much of the activity in the area related to using symmetries to solve Einstein's equations or the equations of motion of other systems has been directed towards finding metrics admitting Killing vectors. Such activity has probably already reached its maturity and it is likely that few genuinely new insights will emerge from it. However, there are other types of symmetries that might be used. Instead of looking at the symmetries of a spacetime, the isometries, one can consider a physical system evolving in a given spacetime (whose metric for simplicity we assume here to be fixed, in the limit of zero backreaction), and analyse the \textit{symmetries of the dynamics} of such system. By symmetries of the dynamics we mean here, for a classical system, transformations in the whole phase space of the system such that the dynamics is left invariant. For a quantum system instead we mean a set of phase space operators that commute with the Hamiltonian or with the relevant evolution operator, and transform solutions into solutions. In the literature such symmetries are often referred to as \textit{hidden symmetries}. 
 
The present review will focus on this class of symmetries. In earlier literature these have been known and discussed although not systematically, see for example \cite{havas:1975,Woodhouse:1975,Crampin:1984}. However in recent years interest in this subject has been renewed due to the discovery of the non-trivial example of hidden symmetries in higher dimensional rotating black hole metrics \cite{DavidValeri2006}. It has been shown that these metrics admit special tensors such as Killing vectors, Killing tensors and Killing-Yano tensors, which will be discussed in detail in section \ref{sec:finite_dimensional}. Such tensors are responsible for the fact that several physical systems in these spacetimes present hidden symmetries of the dynamics and as a consequence their equations of motion are separable and integrable. Known such systems are: geodesic motion, the Klein-Gordon equation, the Dirac equation, stationary strings, tensor gravitational perturbations, the bosonic sector of the supersymmetric spinning particle \cite{DavidEtAl2007,DavidEtAl2007complete,DavidEtAl2007constants,DavidValeriPavel2006,DavidValeri2007,MarcoDavidPavel2011_2,HariJamesHarveyReall2006,OotaYasui2010, DavidMarco2011}. It is not clear at this stage if the same kind of behaviour applies to spin 1 fields and the remaining gravitational perturbations. Since the appearance of the black hole metrics other examples of systems with hidden symmetries of the dynamics have been discovered, providing metrics with new non-trivial Killing tensors of order $\ge 3$ \cite{DavidGaryClaudeHouri2011,Galajinsky2012,GaryMarco2013}. 
 
When there are enough independent isometries and hidden symmetries a system is integrable. There are known examples of integrable classical  systems in the literature, such as for example special spinning tops, the Calogero model, the inverse square central force motion (Kepler problem and classical hydrogen atom), the problem of geodesic motion on an ellipsoid, first discussed by Jacobi in 1839, the Neumann model, the motion in presence of two Newtonian fixed centres, the quantum dot, the spinning particle, which will be discussed in this review. In the case of higher dimensional black holes the integrability is related to the existence of a non-degenerate principal Killing-Yano tensor. Spacetimes with hidden symmetries generated by degenerate principal Killing-Yano tensors have been discussed in \cite{HouriOotaYasui2009}. They have a richer structure and, on the other hand, in general do not have enough isometries or hidden symmetries to guarantee full integrability of the related dynamical systems. 
 
When considering quantum mechanical systems, the concept of hidden symmetries can still be applied fruitfully. Now one is looking for operators defined on phase space that commute with the appropriate evolution operator, for example the Hamiltonian for the Schr\"{o}dinger equation, the wave operator in the case of the Klein-Gordon equation, the Dirac operator for the Dirac equation. However, in the quantum mechanical case the classical hidden symmetries can be anomalous. In the case of the Klein-Gordon equation for example it is possible to construct operators using Killing tensors, when present, such that in the classical limit they provide hidden symmetries for the theory of the scalar particle. However, an anomalous term arises proportional to a contraction of the Killing tensor and the Ricci tensor of the spacetime. In some cases the anomaly will be zero, for example if the metric is Ricci flat or if the Killing tensor can be written in terms of a Killing-Yano tensor, but in the general case the classical hidden symmetry will be broken by quantum effects. This phenomenon should be contrasted with the remarkable fact that when the symmetry operators can be built using Killing-Yano tensors, for example in the case just mentioned or in the case of the Dirac equation, then there are no quantum anomalies. When considering generalised Killing-Yano tensors in the presence of flux fields the situation is different: already at the classical level it is not always possible to build symmetries that generalise those in the case without flux, and at the quantum mechanical level there can be anomalies \cite{DavidClaudePavel2010}. 
 
There is an interesting connection between hidden symmetries and special geometries. For finite dimensional Hamiltonian systems in the absence of forces other than gravity  hidden symmetries are associated typically to special tensors: Killing and Killing-Yano. In the presence of scalar and vector potentials or other flux fields like torsion for example, but also for more general fields, it is possible to find suitable generalised Killing and Killing-Yano tensors that are associated to hidden symmetries and conserved quantities in the theory of systems evolving in the spacetime in the presence of flux \cite{vanHolten2006,DavidClaudeHouriYasui2010,DavidClaudePavel2010,DavidClaudeHouriYasui2012}. Manifolds equipped with specific G--structures host Killing-Yano tensors and the associated hidden symmetries of the dynamics \cite{Papadopoulos2007,Papadopoulos2011}. Also, Killing-Yano tensors are linked to hidden symmetries of the supersymmetric theory of the spinning particle \cite{GaryetAl1993,DavidMarco2011}. It is worth mentioning the Eisenhart-Duval lift procedure, which allows to start from an $n$ dimensional Riemannian spacetime and associate to it an $n+2$ dimensional Lorentzian one. Correspondingly non-relativistic theories described on the former in the presence of scalar and vector flux can be associated to relativistic theories on the latter, and their hidden symmetries studied. In the case of the Dirac equation for example it is possible to give a geometric interpretation of the fact that some hidden symmetries in the presence of flux are anomalous \cite{Marco2012}.

We should point out that there exists another approach to the subject of hidden symmetries of the dynamics, which originates in the seminal work of Sophus Lie. Lie developed his theory of continuous groups in order to find symmetries of differential equations that transform solutions into solutions. The transformations act both on dependent and independent variables. Lie's great insight was that, even if the full group transformations can in principle be complicated and non-linear, if the symmetry group is continuous then one can look for linear transformations that reflect an appropriate type of infinitesimal transformation. As such, the infinitesimal symmetry conditions are amenable to being solved using standard methods. The theory of Lie groups applied to differential equations can be cast in a form appropriate for Hamiltonian systems that is equivalent to the results we will present. In this review however we have chosen to adopt from the beginning the language of Hamiltonian systems and phase space which we think physicists and applied mathematicians might broadly consider as familiar territory. Examples of books on the subject of Lie theory applied to differential equations are \cite{Olver2000applications,Stephani1989differential}.

The review is organised as follows. In section \ref{sec:finite_dimensional} we discuss hidden symmetries in the context of finite dimensional, classical Hamiltonian systems. We introduce these systems from the point of view of symplectic geometry in phase space, discussing canonical transformations and conserved charges. Conserved charges in phase space that are polynomial in the momenta are associated to Killing tensors, and to canonical transformations that preserve the Hamiltonian. We discuss the formalism of Lax pairs, its relation to integrability and its covariant formulation. Then we present a number of examples of systems of different kind that display hidden symmetries: relativistic, non-relativistic, with or without the presence of gravity, as well as the supersymmetric spinning particle. 
 
Section \ref{sec:Hamilton-Jacobi} is an introduction to the theory of intrinsic characterisation of the separation of variables for the Hamilton-Jacobi equation. The 	Hamilton-Jacobi equation is presented and then the role of Killing vectors, rank-two Killing tensors and their conformal counterparts in the theory of separation of variables is discussed. 	
 
In section \ref{sec:Eisenhart-Duval} we present the Eisenhart-Duval geometric lift, an important geometric construction that is ideal to discuss the full group of symmetries of the dynamics. The lift naturally allows embedding a non-relativistic Hamiltonian system with scalar and vector potential into a relativistic geodesic system. Therefore it can be used to apply known results from the theory of geodesic motion to more complicated systems with interactions. The lift has been originally presented in the 1920s by Eisenhart, then has been rediscovered in the 1980s by Duval and collaborators, and since then has been applied to a variety of different settings, as will be discussed. 
 
In section \ref{sec:special_geometries} we discuss several types of special geometries associated to hidden symmetries. The main example is that of rotating Kerr-NUT-(A)dS black holes in higher dimension. For these the special geometry is generated by a principal Killing-Yano tensor: we describe the tower of special tensors generated from it, and how it is possible to associate to these tensors hidden symmetries for a range of physical systems, such as the geodesic and Hamilton-Jacobi equations, Klein-Gordon and Dirac equations, the bosonic sector of the spinning particle. We then describe other special geometries: geometries with torsion, admitting generalised Killing-Yano tensors and their application in supergravity theories, including a local classification of such metrics when they admit a generalised principal Killing-Yano tensor. Another type of special geometries are those associated to some G--structures, for which Killing-Yano tensors can be found. 
 
In section \ref{sec:quantum_systems} we discuss the extension of the concept of hidden symmetries to quantum systems. Some examples are described in the case of the Schr\"{o}dinger equation, such as the isotropic harmonic oscillator and the hydrogen atom, which is the quantum mechanical version of the classical Kepler problem. Then we deal with the Klein-Gordon equation and the Dirac equation, the main example and application being that of their integrability in the higher dimensional rotating black hole metrics. 
 
Sec.\ref{sec:geodesics_lie_groups} presents recent results in the study of dynamical symmetries for systems that can be described using geodesic motion on some Lie group. We consider the specific physical system represented by the Toda chain. A different type of dynamical symmetries arises when coupling constants in the system are promoted to dynamical variables, and link same energy states of theories in the same family but with different values of the coupling constants. Position dependent couplings are also admitted. There is a natural relation between the Lie group geodesic motion and the Eisenhart lift geodesic motion: the higher dimensional Lie group geodesic motion turns out to be a generalised Eisenhart lift of the Toda chain. 
 
We conclude the review in section \ref{sec:final} with a summary and a final commentary. 
 
There has been much work recently in the area of hidden symmetries and it would be quite hard to give a complete and fully balanced account of all the activity. While we have made efforts at doing so, we are aware that this review is shaped in its form by our own personal trajectory in the field, and that we might have overlooked some important related results, for which we offer our apologies. We have aimed at including a bibliography comprehensive enough to allow the reader to make links with most of the existing results.

\subsection{Some initial words: what hidden symmetries are and what they are not}
Before the main body of the review begins, we would like to spend some words on the role of hidden symmetries. Some of the concepts discussed here will become clearer to the reader after going through the contents of section \ref{sec:finite_dimensional}, however we think it is useful to mention them at this point of the review, so that the reader can keep these issues in mind going along. To some readers it might also be useful to come back to this section and read again parts of it after some  technical concepts have been presented in \ref{sec:finite_dimensional}.

One of the chief objectives of this review is to clarify the notion of hidden symmetries, in particular what distinguishes them from 'ordinary symmetries'. 
As will be seen in the next section, dynamical symmetries are naturally described in terms of a phase space: they are transformations in phase space that map solutions into solutions. As is well known, the phase space description of a dynamical system requires a minimal amount of structure: a symplectic manifold, with a symplectic form and a Hamiltonian function, and this leaves a great deal of freedom in the form of a coordinate free formulation. Therefore, the first natural question the reader might ask is what distinguishes an ordinary symmetry from a hidden one, and whether the distinction is of a fundamental nature or reflects a human point of view or a historical context. In brief, the answer is that in the absence of further structures really all symmetries of dynamics are equivalent, and there is no difference between 'apparent' and ' hidden' symmetries: all of them are described infinitesimally by flows in phase space that leave invariant the symplectic structure and the Hamiltonian. 
 
However, there is a notable case that arises frequently in physical applications, where a distinction arises between ordinary and hidden symmetries: this is when phase space is the cotangent bundle of some well defined configuration space. Typically there will be a metric defined on configuration space. Also, but this is not a crucial element, in these cases a dynamical system that is widely studied is that for which the Hamiltonian function is quadratic in the momenta: such systems are known in the literature as \textit{natural Hamiltonians}, and the metric defines what the quadratic part of the Hamiltonian is. For a cotangent bundle we can define ordinary symmetries as those that admit conserved quantities that are linear in the momenta; they generate transformations that are reducible to configuration space, and for natural Hamiltonians these are associated to Killing vectors and isometries of the metric. Thus ordinary symmetries are \textit{manifest symmetries} in the sense that they are associated to transformations that do not depend on the momenta, that act on some configuration space variables and are not genuine phase space transformations. In other words, for ordinary symmetries the phase space transformations can be lifted from a set of simpler transformations defined on a configuration space, a subspace of the whole phase space. Hidden symmetries will instead be defined as those whose conserved quantities are of higher order in the momenta. Such transformations are genuine phase space transformations and cannot be obtained from the lift of a configuration space transformation. It is in this sense that one can say that ordinary and hidden symmetries are different. While it is possible that for a given symplectic manifold the definition of a configuration space, when it exists, is not unique, this seems rather unlikely. One would need inequivalent global Darboux charts. Even in this case, the statement that in one such global chart there is a conserved quantity that is linear in the momenta is equivalent to having a symmetry transformation on a submanifold with commuting variables, and this is a coordinate independent condition.

It should be remarked that hidden symmetries really are symmetries of the Hamiltonian, in the sense that they are canonical transformations where both positions and momenta change, and that leave the Hamiltonian function unchanged. 
Similar statements, with appropriate modifications, can be made for the quantum mechanical case. We also briefly mention that hidden symmetries are not the same thing as some types of partial symmetries appearing in quantum field theory, for example in the Standard Model, sometimes called accidental symmetries. These are not full symmetries, while hidden symmetries are, and they are broken by quantum effects moving along the renormalization group flow, while hidden symmetries can exist from a purely classical point of view.

\section{FINITE DIMENSIONAL CLASSICAL HAMILTONIAN SYSTEMS\label{sec:finite_dimensional}}

\subsection{Hamiltonian systems\label{sec:Hamiltonian_systems}} 

\subsubsection{Symplectic geometry\label{sec:Symplectic geometry}}  
There are several books that discuss the topic of Hamiltonian systems and symplectic geometry. In the following we refer mainly to  \cite{Arnold:book}, \cite{AbrahamMarsden:book} and \cite{BabelonEtal:book}. 

In Classical Mechanics the possible configurations of a physical system are described by a generalised phase space, given mathematically by a symplectic manifold, and the dynamics is obtained from a Hamiltonian function. 
 
A \textit{symplectic manifold} is given by a pair $\{\mP, \omega \}$, where $\mP$ is a manifold and $\omega : T\mP \times T\mP \rightarrow \mathbb{R}$ is an antisymmetric two-form, the \textit{symplectic form}, that satisfies the following properties: 
\begin{enumerate}[i)] 
\item it is non degenerate, det $\omega \neq 0$; 
\item it is closed, $d\omega = 0$.  
\end{enumerate} 

Local coordinates on $\mP$ are given by $\{ y^a \}$, with $a= 1, \dots, 2\dm$ since property (i) imply that $\mP$ is even dimensional,  and we can write $\omega = \frac{1}{2} \omega_{ab} \, dy^a \wedge dy^b$. Using property (ii), by Darboux's theorem locally we can find coordinates $y^a = (q^\mu, p_\nu)$, $\mu, \nu = 1, \dots, \dm$, such that $\omega = dp_\mu \wedge dq^\mu$. There is a natural volume form on $\mP$: 
\be 
\eta = \frac{(-1)^\dm}{\dm !} \omega^{\wedge \dm} \, , 
\ee 
where $\wedge \dm$ represents the $\dm$-fold wedge product, and therefore $\mP$ must be orientable. We adopt the following convention for the wedge product: when it acts on a $p$--form $\alpha$ and a $q$--form $\beta$ it is defined so that in components 
\be \label{eq:wedge_product_definition} 
{(\alpha\wedge\beta)_{a\dots b\dots} = \frac{(p+q)!}{p!\,q!}\,\alpha_{[a\dots}\,\beta_{b\dots]}} \, . 
\ee 
In fact, any intermediate exterior power $\omega^{\wedge p}$ of $\omega$, $1<p<\dm$, is a closed $2p$-form. Let $\omega^{ab}$ be the inverse components of $\omega$, $\omega^{ab} \omega_{bc} = \delta^a_c$.

A Hamiltonian function is a function $H : \mP \rightarrow \mathbb{R}$. It induces a dynamics on $\mP$ according to the following first order evolution equation: 
\be \label{eq:symplectic_eom}
\frac{d y^a}{d \lambda}  = \omega^{ab} \partial_b H  \, ,  
\ee
or in terms of the local $q,p$ coordinates 
\be \label{eq:eom2} 
\left\{ \begin{array}{rcl} \frac{d q^\mu}{d \lambda} &=& \frac{\partial H}{\partial p_\mu} \, , \\ 
                                         \frac{d p_\nu}{d \lambda} &=& - \frac{\partial H}{\partial q^\nu} \, . 
          \end{array} \right. 
\ee 
In fact, for any phase space function $f: \mP \rightarrow \mathbb{R}$ it is possible to define a \textit{symplectic gradient} which is a vector with components 
\be 
X_f^a = \omega^{ab} \partial_b f \,  . 
\ee 
Then the equations of motion \eqref{eq:symplectic_eom} can be written as 
\be \label{eq:symplectic_eom2}
\frac{d y^a}{d \lambda}  = X_H^a \, . 
\ee 
$X_H$ is the vector representing the \textit{Hamiltonian flow}, and it is tangent to the trajectory in $\mP$ associated to the dynamical evolution of the system. Then for any phase space function $f$ the derivative along the Hamiltonian flow is given by 
\be \label{eq:symplectic_hamiltonian_derivative}
\frac{d f}{d \lambda} = \frac{\partial f}{\partial y^a} \frac{d y^a}{d \lambda} =   X_H \left(f \right) = \{ f, H \} \, . 
\ee 
In particular, $H$ itself is conserved along the evolution \eqref{eq:symplectic_eom2}, $\frac{d H}{d \lambda} = 0$, and can be set to a constant, $H = E$, the energy associated to a particular trajectory. 

One the ring of smooth functions on $\mP$ it is possible to define a bilinear operation, the \textit{Poisson bracket}, that acts according to 
\be \label{eq:symplectic_Poisson_brackets}
(f,g) \mapsto \left\{ f , g \right\} = \omega^{ab}\,  \partial_a f \, \partial_b g = X_g (f) = - X_f(g) \, , 
\ee
and endows it with a Lie algebra structure, the \textit{Poisson Algebra}, as the Poisson brackets are anti-symmetric and satisfy Jacobi's identity. In terms of local coordinates 
\be \label{eq:symplectic_Poisson_brackets_local}
\{f, g \} =  \frac{\partial f}{\partial q^\mu} \frac{\partial g}{\partial p_\mu} - \frac{\partial f}{\partial p_\mu} \frac{\partial g}{\partial q^\mu} \, . 
\ee 

There is an anti-homomorphism from the Poisson Algebra of functions in $\mP$ to the Lie algebra of symplectic gradients: 
\be \label{eq:symplectic_relation_poisson_lie}
\left[ X_f , X_g \right] = -  X_{\left\{ f,g\right\}} \, . 
\ee
This can be seen as, for any smooth function $h$, 
\ba 
&& \left[ X_f , X_g \right] (h) =  (X_f X_g - X_g X_f) (h) \nn \\ 
&& = \left\{ \left\{ h, g \right\}, f \right\} - \left\{ \left\{ h, f \right\}, g \right\} = - \left\{ \left\{ g, f \right\}, h \right\} \nn \\ 
&& - \left\{h, \left\{ f, g \right\} \right\} = - X_{ \{f,g\} } (h) \, , 
\ea 
where the Jacobi identity has been used in the intermediate passage. 
 
A \textit{canonical transformation} is a diffeomorphism $\Phi: \mP \rightarrow \mP$, $y \mapsto y^{\prime} = \Phi(y)$, such that  $\omega$ is preserved under the pullback  action of $\Phi$: $\Phi^* (\omega) = \omega$. This means that under $\Phi$ a Hamiltonian system is mapped into another Hamiltonian system, since the equations of motion \eqref{eq:symplectic_eom} are left invariant in form, as well as the Poisson brackets \eqref{eq:symplectic_Poisson_brackets}. However, the new Hamiltonian system will be in general different from the original one with a Hamiltonian $H^\prime$ such that $H \neq \Phi^* (H^\prime)$. It is worth noticing that under a canonical transformation all exterior powers of $\omega$ are also preserved. 
 
An \textit{infinitesimal canonical transformation} generated by a vector field $X$ is a canonical transformation that maps points of $\mP$ to points along the integral lines of $X$. The attribute 'infinitesimal' arises from the fact that generating integral lines of $X$ from $X$ is a procedure that is well defined at least locally if $X$ is sufficiently regular. Then $\omega$ is preserved under the infinitesimal transformation if $\mathcal{L}_X \omega = 0$, where $\mathcal{L}$ is the Lie derivative. We can rewrite the Lie derivative on forms by introducing the hook operation, or \emph{inner derivative}. This is an action of a vector ${v}$ on any antisymmetric form ${\alpha}$. In components: 
\begin{equation}\label{hook}
    (v\hook\alpha)_{a_1\dots a_{p{-}1}}=v^b\alpha_{b a_1\dots a_{p{-}1}}\;.
\end{equation}
For a scalar ${\varphi}$, we set ${v\hook\varphi=0}$. Then it is possible to show that for any $p$-form $\alpha$ the Lie derivative acts as 
\be 
\mathcal{L}_X \alpha = X \hook d\alpha + d\left( X \hook \alpha \right) \, . 
\ee 
Applying this to $\mathcal{L}_X \omega = 0$ and using the fact that $\omega$ is closed this implies that at least locally $X \hook \omega = df$ for some function $f$ on $\mP$, or in coordinates 
\be 
X^a \omega_{ab} = \partial_b f \, . 
\ee 
Multiplying times $\omega^{bc}$ this gives $X^c = \omega^{bc} \, \partial_b f$ or 
\be 
X = X_{-f} \, . 
\ee 
This shows that symplectic gradients are in one-to-one correspondence with infinitesimal canonical transformations.

\subsubsection{Dynamical symmetries\label{sec:dyn_symm}}
Suppose there is a phase space function $C: \mP \rightarrow \mathbb{R}$ that Poisson commutes with $H$, $\{ H, C \} = 0$. Then according to eq.\eqref{eq:symplectic_hamiltonian_derivative} $C$ is conserved along the Hamiltonian flow and is a constant of motion. On the other hand, $C$ is associated to an infinitesimal canonical transformation whose tangent vector is $X_C$, and the condition $\{H, C \} = 0$ has the meaning that the Hamiltonian is left invariant under the transformation. Such kind of canonical transformation is called a \textit{dynamical symmetry}, since it transforms the original Hamiltonian system into itself, and therefore transforms a trajectory under \eqref{eq:symplectic_eom2} into a different trajectory associated to the same system and with the same value of the energy $E$. 
 
If there are two such conserved quantities, $C_1$ and $C_2$, then using Jacobi's identity also $C_3= \{ C_1, C_2 \}$ will be a conserved quantity. Using eq.\eqref{eq:symplectic_relation_poisson_lie} this means that $\left[ X_{C_1}, X_{C_2} \right] = - X_{C_3}$. Thus the infinitesimal canonical transformations associated to conserved quantities present a Lie algebra structure, and hence define the local action of a group $\mathcal{D}$,  the \textit{dynamical symmetry group}. 
 
A recurring case of symplectic manifold is that of the cotangent space over a base manifold $\M$, $\mP = T^*\M$. $\M$ can be thought of as a configuration space, with local coordinates $q^\mu$. If there is a metric $g = \frac{1}{2} g_{\mu\nu} dq^\mu \otimes dq^\nu$ defined on $\M$, then among all the dynamical symmetries one special class is given by \textit{isometries}. These are generated by constants of motions $C$ such that the metric $g$ is invariant along the $C$ flow, $\mathcal{L}_{X_C} (g) = 0$. Isometries are associated with Killing vectors on the base manifold. Dynamical symmetries that are not isometries are typically called \textit{hidden symmetries}. There exist non-trivial examples of systems exhibiting dynamical symmetries that are not isometries, these are the main object of study of this review. 
In the coming section we present some examples of dynamical symmetries that can be associated to special tensors.

\subsubsection{Special tensors \label{sec:symplectic_special_tensors}} 
In this section we consider the case where $\mP$ is a cotangent bundle, $\mP = T^*(\M)$, where $\M$ is a manifold with a metric $g$. We will see that if $\M$ admits special classes of tensors then conserved quantities can be defined for the free particle or the spinning particle. 

\noindent \textit{Killing vectors}: A Killing vector is a base manifold vector $K^\mu (q) \frac{\partial}{\partial q^\mu}$, satisfying 
\be \label{eq:symplectic_KV}
\nabla_{(\mu} K_{\nu)} = 0 \, , 
\ee
where $\nabla$ is the Levi-Civita connection. In the presence of a Killing vector the system of a free particle with Hamiltonian $H = \frac{1}{2m} g^{\mu\nu} p_\mu p_\nu$ admits the conserved quantity $C = K^\mu p_\mu$. This holds whether the system is relativistic or not, and whether or not the metric the metric $g_{\mu\nu}$ represents a gravitational field. In the presence of a potential $V(q)$ such that $\mathcal{L}_K V =  0$ then $C$ is conserved also for the Hamiltonian $H^\prime = \frac{1}{2m} g^{\mu\nu} p_\mu p_\nu + V$. 
 
The infinitesimal transformation generated by $X_C^a$ is given by 
\be \label{eq:symplectic_transformation_isometry}
\left\{ \begin{array}{rcl} \delta q^\mu &=& \epsilon K^\mu \, , \\ 
                                         \delta p_\nu &=& - \epsilon \frac{\partial K^\mu}{\partial q^\nu} p_\mu \, , 
          \end{array} \right. 
\ee
and the Hamiltonian transforms as 
\ba 
\delta H &=& - \frac{\epsilon}{m} g^{\mu\nu} p_\mu \frac{\partial K^\rho}{\partial q^\nu} p_\rho + \frac{\epsilon}{2m} \partial_\rho (g^{\mu\nu}) K^\rho p_\mu p_\nu \\ 
&=& - \frac{\epsilon}{2m} \nabla^{\mu} (K^\rho) p_\mu p_\rho = 0 \, , 
\ea 
where we used $\nabla g = 0$ and $\nabla^{(\mu} K^{\rho)} = 0$. 
The projection of the transformation \eqref{eq:symplectic_transformation_isometry} on $\M$ is an \textit{isometry}, a transformation well defined on configuration space. This is the most widely known case of dynamical symmetry. \\

\noindent \textit{Killing-St\"{a}ckel tensors}: A Killing-St\"{a}ckel tensor is a rank $r$ symmetric tensor defined on $\M$, $K (q) = \frac{1}{r!} K^{(\mu_1 \dots \mu_r)} (q) \frac{\partial}{\partial q^{\mu_1}} \otimes \dots \frac{\partial}{\partial q^{\mu_r}}$, such that 
\be \label{eq:symplectic_KS}
\nabla^{(\mu} K^{\rho_1 \dots \rho_r)}=0 \, . 
\ee 
This equation is a generalisation of \eqref{eq:symplectic_KV}. If $\M$ admits a Killing-St\"{a}ckel tensor  then the system of a free particle with Hamiltonian $H = \frac{1}{2m} g^{\mu\nu} p_\mu p_\nu$ admits the conserved quantity $C = K^{\mu_1 \dots \mu_r} p_{\mu_1} \dots p_{\mu_r}$ As in the previous example, this holds whether the system is relativistic or not, and whether or not the metric the metric $g_{\mu\nu}$ represents a gravitational field. We will consider here a more general Hamiltonian $\frac{1}{2m} g^{\mu\nu} p_\mu p_\nu + V(q)$. The infinitesimal transformation associated to $C$ is given by 
\be \label{eq:transformation_KS}
\left\{ \begin{array}{rcl} \delta q^\mu &=& \epsilon \, r \, K^{\mu \nu_1 \dots \nu_{r-1}} p_{\nu_1} \dots p_{\nu_{r-1}}  \, , \\ 
                                         \delta p_\nu &=& - \epsilon \frac{\partial K^{\mu_1 \dots \mu_r}}{\partial q^\nu} p_{\mu_1} \dots p_{\mu_r} \, , 
          \end{array} \right. 
\ee
and the Hamiltonian transforms as 
\ba 
\delta H &=&  \frac{\epsilon}{2m} \nabla^{\mu} K^{\rho_1 \dots \rho_p} p_\mu p_{\rho_1} \dots p_{\rho_p} + \nn \\ 
&& + \epsilon \, r \,  K^{\mu \nu_1 \dots \nu_{p-1}} p_{\nu_1} \dots p_{\nu_{p-1}} \partial_\mu V \, . 
\ea 
This will be zero if $K^{\mu \nu_1 \dots \nu_{p-1}} \partial_\mu (V) = 0$. 
 
Since both terms in eq.\eqref{eq:transformation_KS} are proportional to $p$, the projection of the transformation on the base manifold $\M$ is zero, differently from the case of isometries: this transformation is a genuine phase space transformation that cannot be obtained from a configuration space transformation \cite{MarcoDavidPavel2012hidden}. It transforms trajectories into trajectories of the same energy $E$, but differently from an isometry in general it will change their shape, since $\delta g \neq 0$ and the distance between any two points on a trajectory will in general not be preserved.  
 
Killing-St\"{a}ckel tensors form a close algebra with respect to the Schouten-Nijenhuis bracket. For a rank $p$ such tensor $K_p$ and, respectively, a rank $q$ one $K_q$, this is defined as 
\be \label{eq:schouten_nijenhuis} 
\{ C_{K_p} , C_{K_q} \} = [ K_p, K_q ]_{SN}^{\mu_1 \dots \mu_{p+q-1}} p_{\mu_1} \dots p_{\mu_{p+q-1}} \, .  
\ee

A generalisation of eq.\eqref{eq:symplectic_KS} is given by 
\be \label{eq:symplectic_conKS}
\nabla^{(\mu} K_{(c)}^{\rho_1 \dots \rho_r)}= g^{(\mu \rho_1} \Phi^{ \rho_2 \dots \rho_r)} \, ,  
\ee 
where by consistency $\Phi$ is related to the divergence of $K_{(c)}$ and derivatives of its traces. A tensor $K_{(c)}$ satisfying eq.\eqref{eq:symplectic_conKS} is called a \textit{conformal Killing tensor}. In the specific case where $r=2$, we can always ask for $K_{(c)}$ to be traceless since we can add to it the term $a \, g^{\mu\nu}$ for any constant $a$. Then $\Phi$ is given by the simple formula 
\be 
\Phi^\mu = \frac{2}{\dm +2} \nabla^\lambda K_{(c)\lambda} {}^\mu \, . 
\ee 
Repeating the considerations seen above for a Killing tensor $K$, it can be seen that the quantity $\tilde{C} = K_{(c)}^{\mu_1 \dots \mu_r} p_{\mu_1} \dots p_{\mu_r}$ will now be conserved for null geodesics, using \eqref{eq:symplectic_conKS}. 
\\

\noindent \textit{Conformal Killing-Yano tensors}: A Conformal Killing-Yano tensor is an $r$-form defined on $\M$, $h = \frac{1}{r!} h_{[\mu_1 \dots \mu_r]} \, dq^{\mu_1} \dots dq^{\mu_r}$, such that 
\ba \label{eq:symplectic_CKY_definition_0}
\nabla_\lambda \, h_{\mu_1 \dots \mu_r} &=& \nabla_{[\lambda} \,  h_{\mu_1 \dots \mu_r]} \nn \\ 
&& + \frac{r}{\dm-r+1} g_{\lambda [\mu_1} \nabla^\rho h_{|\rho| \mu_2 \dots \mu_r]} \, . 
\ea 
Introducing the degree operator ${\pi}$ that acts on an inhomogeneous form $\alpha = \sum_{p} \alpha^{(p)}$ as $
\pi \alpha = \sum_{p=0} p \, \alpha^{(p)}$, we can write \eqref{eq:symplectic_CKY_definition_0} without using components as 
\be \label{eq:symplectic_CKY_definition}
\nabla_X h = \frac{1}{\pi +1} X \hook d h - \frac{1}{\dm - \pi +1} X^\flat \wedge \delta h \, , 
\ee
for any vector $X$, where $X^\flat$ is the dual form $X^\flat = X_\mu dq^\mu$. It is worth noticing that when $r=1$ this provides a different generalisation of the Killing equation \eqref{eq:symplectic_KV}. When $h$ is co-closed, $\delta h = 0$, $h$ is called a Killing-Yano form \cite{Yano1952}, and when it is closed, $dh =0$, it is called a closed conformal Killing-Yano form \cite{Tachibana1969,Kashiwada1968}. Equation \eqref{eq:symplectic_CKY_definition} is invariant under Hodge duality, interchanging Killing-Yano and closed conformal Killing-Yano tensors. If $\omega = \omega_\mu dq^\mu$ is a one-form, we also define its dual vector $\omega^\sharp = \omega^\mu \partial_\mu$. 
 
Given a Killing-Yano $p$--form $h$ a direct calculation using the defining property eq.\eqref{eq:symplectic_CKY_definition_0} shows that the tensor 
\be \label{eq:KY_squared}
K^{\mu\nu} = h^\mu {}_{\lambda_1 \dots \lambda_{p-1}} h^{\nu \lambda_1 \dots \lambda_{p-1}} \, ,
\ee
is Killing--St\"{a}ckel. Similarly, is $h$ is conformal Killing-Yano then formula \eqref{eq:KY_squared} will give a conformal Killing tensor. 
 
If $\M$ admits a conformal Killing-Yano tensor then the system of a free spinning particle admits a conserved quantity that is also a generator of a supersymmetry transformation that is different from the canonical one, generated by the vielbein. In that case it is also possible to construct a symmetry operator for the Dirac equation on $\M$ . The spinning particle system will be discussed separately in sec.\ref{sec:examples_spinning_particle}, and the Dirac equation in sec.\ref{sec:quantum_dirac_equation}. 
 
In \cite{DavidPavelValeriDonPage2007} it was shown that closed conformal Killing-Yano tensors form an algebra under the wedge product. In particular, closed conformal Killing--Yano tensors of rank 2 that are non-degenerate are called \textit{principal conformal Killing--Yano tensors}. They are crucial for the integrability of various systems in four and higher dimensional black hole spacetimes, see sec.\ref{sec:higher_dimensional_black_holes}.

\subsubsection{The covariant Hamiltonian formalism\label{sec:covariant_formalism}} 
As we will see soon in the examples section \ref{sec:applications_and_examples}, several of the known dynamical systems with non-trivial dynamical symmetries display conserved quantities that are polynomial in the momenta, of the type 
\be 
C = \sum_{i=0}^m \frac{1}{i!} \, T_{(i)}^{\mu_1 \dots \mu_i}(q) p_{\mu_1} \dots p_{\mu_i} \, ,  
\ee 
where $\mP = T^*(\M)$ and the $q^\mu$ are coodinates on the base manifold. In particular, $C$ is a scalar under coordinate changes $q^\prime = q^\prime (q)$, which 
can be absorbed by a canonical transformation  with $p_\mu^\prime (q,p) = p_\nu \frac{\partial q^\nu}{\partial q^{\prime \, \mu}}$. Then the quantities $T_{(i)}$ are tensors defined on $\mathcal{M}$, and the theory is coordinate invariant. However, this is not evident in the standard form of the Poisson brackets \eqref{eq:symplectic_Poisson_brackets_local}. To this extent we define a \textit{natural Hamiltonian} to be a phase space function of the form 
 \be
H = \frac{1}{2}\, g^{\mu\nu}(q)\, \Pi_\mu \Pi_\nu + V(q) \, , 
\ee 
where $V$ is a scalar potential, $\Pi_\mu=p_\mu - e A_\mu$ are the gauge-covariant momenta \cite{JackiwManton1980,DuvalHorvathy2004,HorvathyNgome2009,Visinescu2011,Ngome2009}, $e$ is a charge, and $A_\mu$ a vector potential. For these types of Hamiltonians manifest invariance can be displayed by writing the brackets in a covariant form \cite{GaryetAl1993,vanHolten2006,Visinescu2010}
\be \label{eq:Poisson_covariant} 
\left\{f, g \right\}_{\mathcal{P}} = D_\mu f \frac{\partial g}{\partial \Pi_\mu} -  \frac{\partial f}{\partial \Pi_\mu} D_\mu g + e F_{\mu\nu}(q) \frac{\partial f}{\partial \Pi_\mu} \frac{\partial g}{\partial \Pi_\nu} \, , 
\ee
where $F_{\mu\nu} = \partial_\mu A_\nu - \partial_\nu A_\mu$ is the field-strength and 
the covariant derivatives are defined by
\be
D_\mu f :=  \left. \frac{\partial f}{\partial q^\mu} \right|_\Pi + \Gamma_{\mu\nu}^{\;\;\;\lambda}\, \Pi_\lambda\,\frac{\partial f}{\partial \Pi_\nu} \, . 
\ee 
If $C$ is re-written using the covariant momenta  
\be 
C = \sum_{i=0}^m \frac{1}{i!} \tilde{T}_{(i)}^{\, \mu_1 \dots \mu_i}(q) \Pi_{\mu_1} \dots \Pi_{\mu_i} \, ,  
\ee  
then $D_\mu$  acts   as 
\be 
D_\mu f = \sum_{i=0}^m \frac{1}{i!} \nabla_\mu \tilde{T}^{\nu_1 \dots \nu_i} \Pi_{\nu_1} \dots \Pi_{\nu_i} \, ,  
\ee 
where $\nabla$ is the Levi-Civita covariant derivative acting on tensors. Similar equations for covariant derivatives of phase space functions have been used in \cite{DavidPavelValeriMarco2012} in order to give a geometrical description of Lax pairs associated to covariantly constant phase space tensors. 
 
The condition $\{ C , H \} = 0$ can be re-written as a set of generalised Killing equations for the tensors $\tilde{T}_{(i)}$ \cite{vanHolten2006,Visinescu2010,Cariglia2014killing} 
\ba 
\tilde{T}_{(1)}^{\nu}\, V_{;\nu} &=& 0 \, , \nn \\ 
\tilde{T}_{(0) ;\mu} &=& e\, \tilde{T}_{(1)}^{\nu} F_{\mu\nu} + \tilde{T}_{(2) \mu} {}^{\nu} V_{;\nu} \, , \nn \\ 
\tilde{T}_{(i) (\mu_1 ..\mu_i; \mu_{i+1})} &=& e\, \tilde{T}_{(i+1) (\mu_1...\mu_i} {}^\nu F_{\mu_{i+1})\nu} \nn \\ 
&& \hspace{-0.75cm} + \frac{1}{i+1} \tilde{T}_{(i+2) \mu_1...\mu_{i+1}} {}^\nu V_{;\nu} \, ,  \quad i \ge 1 \, . \label{eq:generalised_Killing_equations} 
\ea	
In particular, the highest rank tensor $\tilde{T}_{(m)}$ will be a Killing tensor. This allows to systematically search for conserved quantities that are polynomial in the momenta.

\subsubsection{Integrable systems and Lax pairs} 
A classical Hamiltonian system is called \textit{Liouville integrable} if there are $\dm$ globally defined, independent functions $F_\mu$, $\mu=1, \dots , \dm$, that mutually Poisson commute, $\{ F_\mu, F_\nu \} = 0$, and the Hamiltonian $H$ is a function of the $F_\mu$. By independent we mean that almost everywhere on $\mP$ the set of one-forms $\{ dF_\mu , \mu = 1, \dots, \dm \}$ spans an $\dm$-dimensional space. 
A typical case is where $F_1 = H$ and there are $\dm -1$ mutually commuting constants of motion $C_\mu$, $\mu = 2, \dots , \dm$. For these systems the solution of the equations of motion can be obtained by a finite number of function inversions and of integrations. Although the maximum number of independent Poisson commuting functions is $\dm$, if the system is integrable and there are some extra independent constants of motion then the system is called \textit{superintegrable}. The maximum number of independent constants of motion is $2\dm -1$. An integrable system with $2\dm -1$ independent constants of motion is called \textit{maximally superintegrable}, an example being the Kepler problem discussed in sec.\ref{sec:examples_Kepler}. 
 
Liouville's theorem shows explicitly how to obtain the solution for an integrable system, by constructing a canonical transformation $(q^\mu, p_\nu) \mapsto (\Psi^\mu, F_\nu)$, where $(q^\mu, p_\nu)$ are the local coordinates of Darboux's theorem, with $\omega = dp_\mu \wedge dq^\mu = dF_\mu \wedge d\Psi^\mu$. Once it is shown that the transformation exists then the new equations of motion become trivial: 
\be \label{eq:symplectic_eom_new_variables}
\left\{ \begin{array}{rcl} \frac{d F_\mu}{d \lambda} &=& \{ F_\mu , H \} = 0 \, , \\ 
                                         \frac{d \Psi^\mu}{d \lambda} &=& \{ \Psi^\mu , H \}  =  \frac{\partial H}{\partial F_\mu} = \Omega^\mu (F) \, ,  
          \end{array} \right. 
\ee 
where $\Omega^\mu (F)$ are constant functions. The solutions are $F_\mu (\lambda) = F_\mu (0)$ and $\Psi^\mu (\lambda) = \Psi^\mu (0) + \Omega^\mu \lambda$. 
 
Given a data set $f = \{ f_\mu , \mu = 1, \dots, \dm \}$, let $\mP_f$ be the set in $\mP$ such that $F_\mu (q,p) = f_\mu$. If the functions $F$ are sufficiently regular and since they are independent this is an $\dm$-dimensional submanifold of $\mP$. Assuming sufficient regularity on $\mP_f$ it is possible to invert the relation $F_\mu (q,p) = f_\mu$ to give $p_\mu = p_\mu (q, f)$, and then extending this to $p_\mu = p_\mu (q, F)$. On $\mP_f$ we can consider a reference point $p_0$, and for any other point $p\in \mP_f$ we can define the function 
\be \label{eq:symplectic_change_of_variables}
S (q, F) = \int_{p_0}^p p_\mu (q, f) dq^\mu \, . 
\ee 
The path used for integration can be deformed as long as $\mP_f$ does not have trivial cycles since $\omega|_{\mP_f} = 0$. In fact, typically there will be non-trivial cycles and $\mP_f$ is topologically an $\dm$-torus. This is related to the possibility of defining action-angle variables, more details can be found for example in \cite{BabelonEtal:book}. To see that $\omega|_{\mP_f} = 0$ first notice that the the tangent space to $\mP_f$ is spanned by the set of vectors $\{ X_{F_\mu} \}$. This is because $X_{F_\mu} (F^\nu) = \{ F_\nu , F_\mu \} = 0$. Then for the same reason $\omega (X_{F_\mu}, X_{F_\nu}) = X_{F_\mu} (F_\nu) = 0$, which proves that $\omega|_{\mP_f} = 0$. Then \eqref{eq:symplectic_change_of_variables} is a well defined, in general multivalued, function defined on $\mP$, whose partial derivatives relative to $q$ gives the function $p(q,f)$. Then we can define a new variable 
\be 
\Psi^\mu = \frac{\partial S}{\partial F_\mu} \, . 
\ee 
Since 
\be 
dS = p_\mu dq^\mu + \Psi^\mu dF_\mu \, 
\ee 
then $d^2 S = 0$ implies that $\omega = dF_\mu \wedge d\Psi^\mu$, and the transformation $(q,p) \rightarrow (\Psi, F)$ is  canonical. Thus the dynamics of the system is known using \eqref{eq:symplectic_eom_new_variables}, which requires the inversion $p_\mu = p_\mu (q, F)$ and the integration \eqref{eq:symplectic_change_of_variables}. This completes the proof of Liouville's theorem. It should be noticed that, since $S$ is in general multivalued, then the variables $\Psi^\mu$ will also be multivalued, the variation over a non-trivial cycle being a function of $F$ variables only. 
 
A recent tool that has been used in the study of integrable systems is that of a Lax pair. A Lax pair consists of two matrices $L$ and $M$ taking values in $\mP$ such that the equations of motion imply the {\em Lax pair equation}
\begin{equation}\label{eq:symplectic_Lax_pair_equation}
\frac{dL}{d \lambda} =[L,M]\;.
\end{equation}
There exist two formulations of the Lax pair method. The stronger formulation requires that eq.\eqref{eq:symplectic_Lax_pair_equation} implies the equations of motion, in which case the Lax pair formulation can be used as a starting point of the description of the dynamical system. A weaker formulation does not require equivalence of \eqref{eq:symplectic_Lax_pair_equation} and the equations of motion. In both cases, the Lax pair matrices satisfying \eqref{eq:symplectic_Lax_pair_equation} play an important role in the study of integrability since they allow a simple construction of constants of motion. Indeed, the solution of \eqref{eq:symplectic_Lax_pair_equation} is of the form ${L(\lambda)=G(\lambda)L(0)G^{-1}(\lambda)}$, where the evolution matrix ${G(\lambda)}$ is determined by the equation
\be \label{eq:Lax_evolution_matrix} 
\frac{dG}{d\lambda} = -MG \, . 
\ee 
Therefore, if $I(L)$ is a function of $L$ invariant under conjugation $L\to G LG^{-1}$, then $I(L(\lambda))$ is a constant of motion. All such invariants can be generated from traces of various matrix-powers of $L$:
\begin{equation}\label{eq:symplectic_traces}
tr(L^j)\;.
\end{equation}
A particular Lax pair may not yield all the constants of motion. However, in such a case it is often possible to upgrade the initial Lax pair so that the upgraded one already yields all the conserved observables of the dynamical system. Since the dimensionality of the Lax matrices is not fixed and the Lax pair equation is linear, two Lax pairs can be easily combined by their direct sum. Another useful method of producing a parametric class of Lax pairs is to introduce so-called spectral parameters, see, e.g., \cite{BabelonEtal:book}.
 
Unfortunately, in general there is no constructive procedure  to find a Lax pair for the given problem or even to determine whether the Lax pair (in its stronger formulation) exists. Moreover, the solution is in no sense unique and even the dimensionality of the matrices may vary. However, when the Lax pair exists, it can be a very powerful tool for dealing with the conserved quantities.
 
When $\mP = T^* \M$ and a metric $g$ is defined on $\M$ then it is possible to define a covariant derivative  that acts on tensors that take value on $T^* \M$ and have configuration space indices. Any such tensor with two indices  that is conserved gives rise to an appropriate Lax pair and is called a \textit{Lax tensor}. A similar construction can be given for antisymmetric tensors with indices on the Clifford bundle of $\M$, for any number of indices, yielding a \textit{Clifford Lax tensor}. closed conformal Killing-Yano and Killing-Yano forms on $\M$, if they exist, give rise to Clifford Lax tensors and, depending on their rank, to Lax tensors \cite{DavidPavelValeriMarco2012}.

\subsection{Applications and examples\label{sec:applications_and_examples}}

\subsubsection{Spinning tops} 
In this section we consider two examples of motions of spinning tops with hidden symmetries of the dynamics: the Goryachev--Chaplygin and the Kovalevskaya's tops. 
 
We begin describing the kinetic energy of a spinning top using Euler angles and $SO(3)$ left invariant metrics. Let $G$ be the centre of mass of the top, and $P$ be the pivot point which we take to be the centre of coordinates. We can consider a body-fixed frame $\bar{S}$, with coordinates $\vec{r}_F$, and an inertial frame $S$, with coordinates $\vec{r}_I$, both with centre in $P$. Then each point of the top moves along a trajectory $t \mapsto \vec{r}_I (t) = O(t) \vec{r}_F$, where $O(t) \in SO(3)$. A change of the inertial frame $\vec{r}_I \rightarrow L \vec{r}_I$ , $L \in SO(3)$, induces a left multiplication $O(t) \mapsto L O(t)$, and similarly a change of the fixed frame $\vec{r}_F \rightarrow R^{-1} \vec{r}_F$ , $R \in SO(3)$, induces a right multiplication $O(t) \mapsto O(t) R$. The kinetic energy is invariant under rotations of the inertial axes and therefore it should be possible write it in terms of the left invariant forms of $SO(3)$ 
\ba 
\sigma_1 &=& \sin \theta \cos \psi d \phi - \sin \psi d\theta \, , \nn \\ 
\sigma_2 &=& \sin \theta \sin \psi d \phi + \cos \psi d\theta \, , \nn \\ 
\sigma_3 &=& d\psi + \cos \theta d\phi \, , 
\ea 
where $\phi \in [0, 2\pi[$, $\theta \in [0, \pi [$, $\psi \in [0, 2\pi [$ are the Euler angles associated to $O$. In fact, the kinetic energy is given by 
\be 
\frac{1}{2} g_{ij} \dot{x}^i \dot{x}^j \, , 
\ee 
where $x^i = (\phi, \theta, \psi )$ and $g_{ij}$ is the left invariant metric on $SO(3)$ 
\be 
g = I_1 \, \sigma_1^2 + I_2 \, \sigma_2^2 + I_3 \, \sigma_3^2 \, , 
\ee 
the $I_j$ being the principal moments of inertia relative to the pivot point $P$. It is also possible to show that the gravitational potential energy  is given by 
\ba
V = mg \Big( && x_F^{(CM)} \sin\theta \cos\psi + y_F^{(CM)} \sin\theta \sin\psi \nn \\ && + z_F^{(CM)}  \cos\theta \Big) \, , 
\ea  
where $m$ is the mass, $g$ the gravitational acceleration, $(x_F^{(CM)}, y_F^{(CM)}, z_F^{(CM)})$ are the coordinates of the centre of mass $G$, relative to $P$, in the body fixed frame. 
 
In the specific case of the Goryachev--Chaplygin top the system can be obtained by constraining the dynamics of a heavy top whose principal moments of inertia are given by $I_1= I_2 = 1 = 4 I_3$, and for which the centre of gravity lies in the plane determined by the two equal moments of inertia, so that we can take $x_F^{(CM)} = 0 = z_F^{(CM)}$, $y_F^{(CM)} = const$. The unconstrained Lagrangian is given by 
\be
L= \half ( \dot \theta ^2 + \sin ^2 \theta \dot \phi^2)   + \frac{1}{8}  (\dot \psi + \cos
\theta \dot \phi  )^2  - \alpha^2 \sin \theta \sin \psi \, , 
\ee    
where $\alpha$ is a constant. The momenta are given by 
\ba
p_\phi &=& \sin ^2 \theta \dot \phi + \frac{1}{4} \cos \theta (\dot \psi
+ \cos \theta \dot \phi) \,, \nn \\ 
p_\theta &=& \dot \theta\,, \nn \\ 
 p_\psi &=& \frac{1}{4} ( \dot \psi + \cos \theta \dot \phi ) \,, 
\ea
from which the Hamiltonian is 
\ba\label{top2}
H&=& \half p_\theta ^2 + 2  p_\psi ^2 + \half
(\frac{p_\phi}{\sin \theta} - \cot \theta p_\psi ) ^2 \nn \\ 
&& +\alpha^2 \sin \theta \sin \psi \nn \\ 
&=&\frac{1}{2}\bigl(M_1^2+M_2^2+4M_3^2\bigr)+\alpha^2 \sin \theta \sin \psi \,, 
\ea
where in the last line we have introduced the moment maps for left actions of $SO(3)$: 
\ba 
M_1 &=& -\sin \psi p_\theta + \frac{\cos \psi}{\sin\theta}p_\phi-\cos\psi\cot \theta p_\psi \,,\nn \\   
M_2 &=&\cos \psi p_\theta + \frac{\sin \psi}{\sin\theta}p_\phi-\sin\psi\cot \theta p_\psi \,, \nn \\ 
M_3&=& p_\psi \,. 
\ea

The coordinate $\phi$ is cyclic and hence $p_\phi$ is a constant of motion. 
The Hamiltonian of the Goryachev--Chaplygin top is obtained by setting $p_\phi=0$\,,  
\be
H_{GC}= \half \bigl( \cot^2 \theta  +4\bigr )p_\psi ^2 + \half p_\theta
^2 + \alpha^2 \sin \theta \sin \psi\,. \label{top}   
\ee

The following remarkable property holds: the function 
\be\label{Ktop2}
{K} =M_3(M_1^2+M_2^2)-\alpha^2M_2x_3
\ee
obeys 
\be\label{HK}
\{H , K\}=\alpha^2 p_\phi M_1\,.
\ee
Hence, for $p_\phi=0$, i.e. for Goryachev--Chaplygin top, (\ref{Ktop2}) is a constant of motion and reads
\be
{K}_{GC}=p_\psi p_\theta^2+\cot^2\!\theta p_\psi^3+\alpha^2\cos\theta\bigl(\sin\psi\cot\theta p_\psi-\cos\psi p_\theta\bigr) \,.  
\ee

The second type of top considered in this section is Kovalevskaya's top. 
In this case $I_1= I_2 = 1 = 2 I_3$, and $y_F^{(CM)} = 0 = z_F^{(CM)}$, $x_F^{(CM)} = const$. The Lagrangian is 
\be
L_{K}= \frac{1}{2}(\dot \theta ^2 + \sin ^2 \theta \dot \phi^2)+\frac{1}{4}(\dot \psi + \cos
\theta \dot \phi)^2   -\alpha^2 \sin \theta \cos \psi \,.
\ee 
Again $\phi$ is ignorable and the Hamiltonian 
\ba
H_{K}&=& \half \left( p_\theta ^2 +(\frac{p_\phi}{\sin\theta}-\cot \theta p_\psi)^2 
 + 2 p_\psi ^2\right ) + \alpha^2 \sin \theta \cos \psi \nonumber\\
 &=& \half \left( M_1^2+M_2^2+2M_3^2\right)+ \alpha^2 \sin \theta \cos \psi
\ea
is constant.
Kovalevskaya found another constant which is quartic in the momenta \cite{Whittaker, Borisov} and reads
\ba
&& K_{K} \left( p_\theta ^2 +(\frac{p_\phi}{\sin\theta}-\cot \theta p_\psi)^2\right)^2
+ 4 \alpha^4 \sin
^2 \theta \nn \\ 
&&  - 2 \alpha^2 \sin \theta \left(e^{i\psi}(\frac{p_\phi}{\sin\theta}- \cot \theta p_\psi + i p_\theta
)^2 + c.c. \right) \, . 
\ea

\subsubsection{Calogero model}  
 The Calogero model describes a set of $\dm$ particles on a line, interacting pairwise with an inverse square potential. It has been first discussed by Calogero in \cite{Calogero1969}. It is super-integrable, both classically \cite{Wojciechowski1983superintegrability}, and quantum mechanically \cite{Kuznetsov1996,Gonera1998}. It has been applied in a wide range of settings, among which black hole physics \cite{Gibbons1999}, gauge theory \cite{Gorsky1994}, fractional statistics \cite{Polychronakos1989}. 
 
The Hamiltonian is given by 
\be 
H = \frac{1}{2} \sum_{i=1}^\dm p_i^2 + g^2 \sum_{i<j} \frac{1}{(x_i - x_j)^2} \, , 
\ee 
where the $x_i$ , and $p_j$ are canonical coordinates and $g$ is the common coupling constant, and the particles' mass has been set to unity. 

The model is conformal: one can define the quantities 
\ba 
K &=& \frac{1}{2} \sum_i q_i^2 \, , \\ 
D &=& - \frac{1}{2} \sum_i p_i q_i \, , 
\ea 
and check that they generate the conformal algebra $sl(2, \mathbb{R})$ 
\ba 
\{ K, H \} &=& - 2 D \, , \\ 
\{ D, H \} &=& - H \, , \\ 
\{D, K \} &=& K \, . 
\ea 
 
Moser \cite{Moser1975}, Regge and Barucchi \cite{BarucchiRegge1977} showed that the system is integrable by displaying the Lax pair: 
\ba
&& L_{jk} =  p_j \delta_{jk} +  (1-\delta_{jk}) \frac{i g}{q_j - q_k} \, ,  \\ 
&& \hspace{-.4cm} M_{jk} = \hspace{-.1cm} g \hspace{-.1cm} \left( \delta_{jk} \sum_{l\neq j} \frac{1}{(q_j - q_l)^2} - (1 - \delta_{jk}) \frac{1}{(q_j - q_k)^2}  \right)  . 
\ea 
Therefore it is possible to build $\dm$ integrals of motion $I_j = \frac{1}{j!} Tr L^j$,  among which $I_1$ is the total momentum and $I_2$ the Hamiltonian. Regge and Barucchi have shown that the $I_j$ are in involution. These integrals of motion are inhomogeneous polynomial in the momenta of order $j$. Galajinsky in \cite{Galajinsky2012} used the fact that for the Eisenhart lift  of the Calogero model these conserved quantities lift to conserved quantities that are homogeneous and of higher order in the momenta, see eqs.\eqref{eq:Eisenhart_conserved_quantity}, \eqref{eq:Eisenhart_lifted_conserved_quantity}. These consequently are in correspondence with Killing Tensors, according to the results of sec.\ref{sec:Eisenhart-Duval_geodesics}. 
 
There also exist $\dm - 1$ extra conserved quantities that make the system super-integrable. These are built as follows: one first defines 
\be 
N_j = \frac{1}{j} \left\{ K , I_j \right\} \, ,  
\ee 
and notices that 
\be 
\{ D, I_j \} = -\frac{1}{2} j I_j \, . 
\ee 
From this and the Jacobi identity it follows that 
\be 
\{ N_j, H \} = I_j \, .  
\ee 
Then the quantities 
\be 
\tilde{I}_j = N_j - t I_j 
\ee 
are conserved. In particular, the $\dm - 1$ quantities $\tilde{I}_i I_j - \tilde{I}_j I_i$ do not depend explicitly on time and are functionally independent \cite{Wojciechowski1983superintegrability}. The $\tilde{I}_j$ also lift and give Killing tensors of the Eisenhart lift metric.

\subsubsection{Kepler problem\label{sec:examples_Kepler}} 
The Kepler problem describes a classical particle of mass $m$ in three-dimensional space, in a central potential that is proportional to the inverse square of the distance from the origin of the coordinates. It is one of the main examples of solvable Hamiltonian systems: it is super-integrable, having the maximum number, 5, of functionally independent constants of motion, its group of dynamical symmetries is well known and has been discussed from several points of view, as well as the quantum version of the problem. For all these reasons it is a typical textbook example, and there are books exclusively dedicated to its study, a good example being the one by Cordani \cite{Cordani2003kepler}. The conserved Runge-Lenz vector of the Kepler problem, that is not linear in the momenta, has been (re)discovered and discussed by various authors in the past and using different points of view, including Hermann, Bernoulli, Laplace, Gibbs, Jacobi. Hulth\'en noticed that for bound orbits the Poisson brackets of the angular momentum and the Runge-Lenz vector form a Lie-algebra isomorphic to that of $O(4)$ \cite{Hulten1933}, and Fock in 1935 showed, analising the quantum problem in the context of the non-relativistic hydrogen atom, that there exist a dynamical symmetry group $O(4)$ that acts on bound states mixing orbitals with the same energy but different angular momentum, and that this can be explained by the fact that the Schr\"{o}dinger equation in momentum space is the stereographic projection of the spherical harmonics on the sphere \cite{Fock1935}. Subsequently Bargmann showed that the constants of motion of the Kepler problem generate Fock's group of transformations \cite{Bargmann1936}. Rogers in 1972 has explicitly constructed the finite dynamical symmetry transformations for negative energy trajectories, using an eight-dimensional enlarged phase space and rewriting the dynamics and the transformations using quaternions \cite{Rogers1973}. Prince and Eliezer have shown that the dynamical symmetries associated to the Runge-Lenz vector can be obtained from Lie's theory of differential equations \cite{PrinceEliezer1979}. 
 
The Hamiltonian of the Kepler problem is given by 
\be 
H (\vec{r}, \vec{p}) = \frac{p^2}{2m} - \frac{k}{r} \, . 
\ee
Trajectories of negative energy are given by ellipses, zero energy one by parabolae and positive energy ones by hyperbolae. 
The angular momentum 
\be \label{eq:L_def}
\vec{L} = \vec{r} \times \vec{p} \, , 
\ee 
and the Runge-Lenz vector 
\be \label{eq:RungeLenz_def}
\vec{A} = \vec{p} \times \vec{L} - mk \, \frac{\vec{r}}{r} \, , 
\ee 
are conserved vectors. Notice that $\vec{A}$ is not linear in the momentum. In fact Duval, Gibbons and Horv\'athy  have shown that the Kepler problem can be lifted to null geodesic motion in $5$ dimensions, and that \eqref{eq:RungeLenz_def} lifts to a $5$-dimensional conserved quantity that is homogeneous and second order in the momenta, generated by a conformal Killing tensor as described in sec.\ref{sec:symplectic_special_tensors} \cite{GaryDuvalHorvathy1991}. This is an example of Eisenhart-Duval lift, that will be described in detail in section \ref{sec:Eisenhart-Duval}. Earlier Crampin noted that the Runge-Lenz vector can be written using a Killing-tensor in $3$ dimensions \cite{Crampin:1984}. The equations he wrote are a special case of those that appear for the Eisenhart non-null lift of section \ref{sec:n_plus_one_Eisenhart_lift}.

$\vec{L}$ and $\vec{A}$ satisfy the following algebra: 
\ba \label{eq:L_A_algebra}
\{ L^i , L^j \} &=& \sum_{k = 1}^3 \epsilon^{ijk} L^k \, , \nn \\ 
\{ L^i , A^j \} &=&  \sum_{k = 1}^3 \epsilon^{ijk} A^k \, ,  \nn \\ 
\{ A^i , A^j \} &=& - 2 m H \sum_{k = 1}^3 \epsilon^{ijk} L^k \, . 
\ea 
If we restrict to solutions with zero energy, $H=0$, the bracket of $A$ with itself is zero and the algebra of $\vec{L}$ and $\vec{A}$ is that of $O(3) \ltimes \mathbb{R}^3$. For solutions with $H = E \neq 0$ one can rescale $\vec{A}$ according to $\vec{B} = \frac{\vec{A}}{\sqrt{2m |E|}}$. Then the Poisson algebra becomes 
\be \label{eq:LB_algebra}
\{ L^i , B^j \} =  \sum_{k = 1}^3 \epsilon^{ijk} B^k \, ,  
\ee 
\be \label{eq:BB_algebra}
\{ B^i , B^j \} = - \text{sgn} (E) \sum_{k = 1}^3 \epsilon^{ijk} L^k \, , 
\ee
 where we defined $\text{sgn}(x) = x / |x|$ for $x\neq 0$, $\text{sgn}(0) = 0$. The algebra is that of $O(4)$ for $E<0$, and $O(1,3)$ for $E>0$. 
According to the results discussed in sec.\ref{sec:Symplectic geometry} $\vec{L}$ and $\vec{A}$ generate infinitesimal canonical transformations of the kind 
\be \label{eq:transformation_generic}
\left\{ \begin{array}{lcl} \delta x^i &=& \epsilon \{  x^i, f  \}  \\ 
			    \delta p_i &=& \epsilon \{  p_i , f \} 
\end{array} \right. \, , 
\ee 
where $f$ is any of the components of $\vec{L}$ or $\vec{A}$ and $\epsilon$ an infinitesimal parameter. Since $\vec{L}$ and $\vec{A}$ are conserved, the transformations will change a trajectory into a trajectory of the same energy. The reader that is not accustomed to dynamical symmetry transformations can find in \cite{MarcoEduardo2013} a pedagogical discussion of the finite form of the transformations of the trajectories for the Kepler problem. Ellipses are transformed into ellipses, parabolae into parabolae, and the same for hyperbolae, but their shape and their eccentricity change, apart from the eccentricity of parabolae which is fixed.

\subsubsection{Motion in two Newtonian fixed centres}  
We use here the same notation used in section \ref{sec:examples_Kepler}. The problem of motion in two Newtonian fixed centres was first studied by Euler in 1760, and then receveid contributions along time by several scientists, including Lagrange, Liouville, Laplace, Jacobi, Le Verrier, Hamilton, Poincar\'e, Birkhoff. It has since being referred to as Euler's three body problem. A detailed review of the Euler problem and related integrable systems is the one by Math\'{u}na \cite{Mathuna2008}.  It has applications both in Newtonian gravity and in the study of a one-electron diatomic molecule in the adiabatic approximation \cite{Dalgarno1956}. As an example, Pauli studied molecular hydrogen in his doctoral dissertation using the classical solutions of the Euler problem.   
 
The Hamiltonian is given by 
\be 
H (\vec{r}, \vec{p}) = \frac{p^2}{2m} - \frac{k_1}{r_1} - - \frac{k_2}{r_2} \, , 
\ee 
where $r_{1,2}^2 = x_1^2 + x_2^2 + (x_3 \mp a)^2$, and $2a>0$ is the distance between the two fixed centres, which we suppose lying on the $x_3$ axis with the origin of coordinates in their midpoint. 
 
Since the potential is axisymmetric $p_3$ is a conserved quantity. One can qualitatively expect that the problem will admit close orbits by considering the simpler case of the Kepler problem. There, we know that the inverse distance potential admits elliptic orbits such that the source of the potential lies in one of the foci. It is also known that during motion the position of the foci does not change, this being associated to the conservation of the Runge-Lenz vector. So placing two separate sources of potential in the two foci one should still have the same type of orbit, according to a theorem due to Legendre and known by the name of Bonnet. Doing this the would still leave the freedom of changing the ellipse's eccentricity. As a matter of fact a third, non-trivial conserved quantity that is quadratic in momenta exists for the problem and is given by \cite{Coulson1967} 
\be 
K = L^2 + a^2 p_3^2 - 2 a m \left( \frac{k_1 \, z}{r_1} - \frac{k_2 \, z}{r_2} \right) \, . 
\ee 
$K$ generates non-trivial dynamical symmetries. The functions $\{ H, p_3, K \}$ are independent and mutually Poisson commute, and therefore the system is integrable. The solutions can be written in terms of elliptic integrals using confocal conic coordinates \cite{Jacobi}.

 The problem has been recently and independently rediscovered by Will \cite{Will2009}. Will studied the motion of a non-relativistic particle in a generic axisymmetric Newtonian potential, asking what type of distribution of mass/charge can give rise to a third conserved quantity in addition to the energy and the component of the angular momentum along the symmetry axis. The answer is that then the potential must be that of the Euler problem. Interestingly, Will finds that the multipole moments of such potential satisfy the same relations of the electric moments associated to the no-hair theorem in the Kerr geometry. Interestingly, the general relativistic analogue of the two-fixed centres, the Bach-Weyl solution \cite{BachWeyl}, does not possess such second order in momenta conserved quantity, thus making even more significant the existing result for the Kerr metric.

\subsubsection{Neumann model}    
The Neumann model \cite{Neumann1859} describes a particle moving on a sphere $S^{\dm -1}$ subject to harmonic forces with different frequencies $\omega_i$, $i=1, \dots, \dm$, in each direction. We assume the frequencies to be ordered as $\omega_1^2 < \omega_2^2 < \dots \omega_\dm^2$. The model is integrable and there are $\dm-1$ independent conserved quantities in involution that are quadratic in the momenta. 
 
The equations of motion are most easily written starting with the Lagrangian version of the theory. One can employ a Lagrange multiplier $\Lambda$ to write the Lagrangian function as 
\be 
\mathcal{L} = \frac{1}{2} \sum_{i=1}^\dm \left( \dot{x}_i^2 - \omega_i^2 x_i^2 \right) + \frac{\Lambda}{2} \left( \sum_{i=1}^\dm x_i^2 - 1 \right) \, . 
\ee 
This gives the equations of motion 
\ba 
&& \sum_{i=1}^\dm x_i^2 - 1 = 0 \, , \nn \\ 
&& \ddot{x}_i = - \omega_i^2 x_i + \Lambda x_i \, . 
\ea 
The second time derivative of the constraint implies that $\sum_i (x_i \ddot{x}_i + \dot{x}_i^2 ) = 0$. Multiplying the second equation of motion times $x_i$ and summing over $i$ we can then solve for $\Lambda = - \sum_i ( \dot{x}_i^2 - \omega_i^2 x_i^2 )$, and the equations of motion are re-expressed as 
\be \label{eq:Neumann_eom}
\ddot{x}_i = - \omega_i^2 x_i - x_i \sum_j ( \dot{x}_j^2 - \omega_j^2 x_j^2 ) \, . 
\ee 
Vice-versa, assuming the equations of motion \eqref{eq:Neumann_eom} and initial conditions $\sum_{i=1}^\dm x_i^2 = 1$, $\sum_i x_i \dot{x}_i = 0$, then the conditions are valid for the whole motion. 
 
To appreciate the presence of hidden symmetries of the dynamics it is easier to work in the Hamiltonian framework. We start with the enlarged phase space $\{ x_i, p_i, i = 1, \dots, \dm \}$ with canonical Poisson brackets $\{ x_i, p_j \} = \delta_{ij}$, and introduce the antisymmetric flat angular momentum quantities 
\be 
J_{lm} = x_l p_m - x_m p_l \, . 
\ee 
From these we build the Hamiltonian 
\be 
H = \frac{1}{2} \sum_{l<m} J_{lm}^2 + \frac{1}{2} \sum_i \omega_i^2 x_i^2 \, . 
\ee 
The associated equations of motion are 
\ba \label{eq:Neumann_hamiltonian_eom_1}
\dot{x}_i &=& - J_{ij} x_j \, , \nn \\ 
\dot{p}_i &=& - J_{ij} p_j - \omega_i^2 x_i \, .  \\  
\ea 
The function $C = \frac{1}{2}\sum_i x_i^2$ is trivially conserved and according to the results of sec. \ref{sec:dyn_symm} it generates the dynamical symmetry $x_i \rightarrow x_i$, $p_i \rightarrow p_i + \lambda \, x_i$.  We use $C$ to perform a Marsden-Weinstein dimensional reduction of the phase space: that is, we work on the level set $C = \frac{1}{2}$ and take the quotient of the level set with respect to the action of $X_C$. This guarantees that we get a new symplectic manifold, the Marsden-Weinstein symplectic quotient. 
 
For each motion it is always possible to choose a representative $\bar{p}_i$ in the quotient such that $\sum_i x_i \bar{p}_i = 0$: setting $\bar{p}_i = p_i + \lambda(t) x_i$, this gives $\lambda(t) = - \sum_i x_i(t) p_i(t)$. Then the equations of motion  \eqref{eq:Neumann_hamiltonian_eom_1} imply $\dot{\lambda} = \sum_i \omega_i^2 x_i^2$. Now we can obtain second order equations for $x$ eliminating $\bar{p}$: first of all we calculate $\dot{\bar{p}}_i = - J_{ij} \bar{p}_j - \omega_i^2 x_i + \left(\sum_k \omega_k^2 x_k^2 \right) x_i$, and then notice that $\dot{x}_i = - J_{ij} x_j = - \bar{J}_{ij} x_j = \bar{p}_i$, so that taking one more derivative one recovers exactly equation \eqref{eq:Neumann_eom}. 
 
Uhlenbeck has shown that the Neumann system admits the following $\dm -1$ independent constants of motion in involution \cite{Uhlenbeck1975}: 
\be 
F_i = x_i^2 +  \sum_{k\neq i} \frac{J_{ik}^2}{\omega_i^2 - \omega_k^2} \, , 
\ee 
with $\sum_i F_i = 1$.

\subsubsection{Geodesics on an ellipsoid}  
The problem of geodesics on an ellipsoid is one of the oldest integrable systems. Jacobi solved it in 1838 for the two-dimensional ellipsoid reducing it to quadratures \cite{Jacobi1839first,Jacobi1839second}, and Moser obtained the solution using the modern theory of Lax pairs and isospectral deformations \cite{Moser1979}. The $\dm$-dimensional case was studied by Kn\"{o}rrer from the algebro-geometric point of view \cite{Knorrer1980}, and by Perelomov using the projection method \cite{Perelomov2000}. 
 
It is useful to write the Hamiltonian for this problem using the notation of the previous section, as the system is contained in the Neumann problem: 
\ba 
H &=&  \sum_{i} \frac{F_i}{\omega_i^2} = (X,X)_\omega - (X,X)_\omega (P,P)_\omega + (X,P)_\omega^2 \nn \\ 
&&
 = (X , X)_\omega \left[ 1 - (\xi ,\xi)_\omega \right]  \, , 
\ea 
where we introduced the $\omega$ scalar product $(V,W)_\omega = \sum_i \frac{1}{\omega_i^2} v_i w_i$, and 
\be \label{eq:xi_def}
\xi = P - \frac{(X,P)_\omega}{(X,X)_\omega} X \, . 
\ee 
One should notice that $\xi$ and $H$ are invariant under the same transformation of the previous section, $x_i \rightarrow x_i$, $p_i \rightarrow p_i + \lambda \, x_i$, and therefore once again we will consider a Marsden-Weinstein dimensional reduction, and assume that $\sum_i x_i^2 = 1$. 
 
We now want to show that choosing the constraint $H=0$ leads to the vector $\xi$  following  geodesics on the ellipsoid $(\xi,\xi)_\omega = 1$.  First of all we remind the reader that the geodesic equation for a geodesic on a submanifold $f(\xi) = 0$ is given by 
\be \label{eq:ellipsoid_xi}
\frac{d}{d\lambda} \frac{\xi^\prime}{\sqrt{\xi^\prime \cdot \xi^\prime}} = \Lambda \nabla f(\xi) \, , 
\ee 
where $\xi^\prime = \frac{d\xi}{d\lambda}$ and $\lambda$ a parameter on the trajectory. The geometrical meaning is that the acceleration of $\xi$ with respect to the arc length parameter is normal to the surface, and can be seen by finding the extrema of the functional 
\be 
\int \left( \sqrt{\xi^\prime \cdot \xi^\prime} + \Lambda f(\xi) \right) d\lambda \, , 
\ee 
where $\Lambda$ is a Lagrange multiplier. 
 
Next, we calculate the time derivative of $\xi$ as follows: 
\ba 
&& \dot{x_i} = \frac{\partial H}{\partial p_i} = - 2 (X,X)_\omega \frac{p_i}{\omega_i^2} + 2 (X,P)_\omega \frac{x_i}{\omega_i^2} \nn \\ 
&& \hspace{0.5cm} = - 2 (X,X)_\omega \frac{\xi_i}{\omega_i^2} \, , \nn \\ 
&& \hspace{-.5cm} \dot{p_i} = - \frac{\partial H}{\partial x_i} = - 2( 1 - (P,P)_\omega ) \frac{x_i}{\omega_i^2} - 2 Q(X,Y)_\omega \frac{p_i}{\omega_i^2} \, , 
\ea 
and then from the definition of $\xi$ \eqref{eq:xi_def} 
\be 
\dot{\xi}_i = - 2 \frac{H}{(X,X)_\omega} \frac{x_i}{\omega_i^2} - \dot{s} x_i \, , 
\ee 
where we defined $s = \frac{(X,Y)_\omega}{(X,X)_\omega}$. $s$ is in fact the arc length parameter since the equation of motion for $\xi$ reduces, when $H=0$, to 
\be 
\frac{d\xi_i}{ds} = - x_i \, 
\ee 
from which $\frac{d\xi_i}{ds} \cdot \frac{d\xi_i}{ds} = 1$. Then, taking one more time derivative 
\be 
\frac{d}{dt} \frac{d\xi_i}{ds} =   2 (X,X)_\omega \frac{\xi_i}{\omega_i^2} \propto \nabla (\xi, \xi)_\omega \, , 
\ee 
which shows that the evolution of $\xi$ is geodesic.

\subsubsection{Quantum Dots} 
The following model of a Quantum Dot has been discussed in \cite{Simonovic2003hidden,Alhassid1987dynamical,Ganesan1989comment,Blumel1989chaos,Zhang2014separability}.  
It describes two charged particles with  Coulomb interaction in a constant magnetic field $B$ and a confining oscillator potential. The Hamiltonian is given by 
\be
H = \sum_{i=1}^2 \left[ \frac{1}{2}\, \vec{\Pi}_i^2 + V(r_i) \right] - \frac{a}{|\vec{r}_1 - \vec{r}_2|} \, , 
\ee
where $a$ is a constant, $\vec{\Pi}$ is the covariant momentum introduced in sec.\ref{sec:covariant_formalism}, the magnetic field points in the $z$-direction, and the confining oscillator potential is axially symmetric: 
\be
V(\vec{r}_a) = \frac{1}{2} \left[ \omega_0^2\, (x_a^2 + y_a^2) + \omega_z^2\, z_a^2 \right] \, . 
\ee 
The problem is reduced to an effective 1-particle problem by realising that the centre of mass coordinates separate. Focussing on relative cylindrical coordinates $(\rho, z, \varphi, \pi_\rho, \pi_z, \pi_\varphi)$ one is left with 
\be 
H_{rel} = \frac{1}{2} g^{\mu\nu} \pi_\mu \pi_\nu + V(\rho, z, \varphi) \, , 
\ee 
where $g_{\mu\nu} = $diag$ (1,1, \rho^2)$ and the potential is given by 
\be 
V(\rho, z, \varphi) = \frac{1}{2} \left[ \omega_0^2\, \rho^2 + \omega_z^2\, z^2 \right] - \frac{a}{\sqrt{\rho^2 + z^2}} \, . 
\ee 
In the formalism of sec.\ref{sec:covariant_formalism} the $z$ component of the angular momentum, a conserved quantity, is written as 
\be 
L_z = L_{(1)}^\mu \pi_\mu + L_{(0)} \, , 
\ee 
where $L_{(1)}^\mu = (0, 0, 1)$ is a Killing vector, $L_{(0)} = \omega_L \rho^2$ its associated scalar, and $\omega_L = \frac{e B}{2}$ is the Larmor frequency. 
 
The Hamiltonian also presents extra constants of motion when the parameter $\tau = \frac{\omega_z}{\sqrt{\omega_0^2 + \omega_L^2}}$ assumes specific values. When $\tau = 2$, as discussed in \cite{Simonovic2003hidden,Alhassid1987dynamical,Ganesan1989comment,Blumel1989chaos,Zhang2014separability}, the system is separable in parabolic coordinates and a new constant of motion that solves the generalised Killing equations \eqref{eq:generalised_Killing_equations} arises, associated to a second rank Killing tensor: 
\ba 
K &=& \frac{1}{2} K_{(2)}^{\mu \nu} \pi_\mu \pi_\nu + K_{(1)}^{\mu} \pi_\mu + K_{(0)} \nn \\ 
&=& z \pi_\rho^2 - \rho \pi_\rho \pi_z + \frac{z}{\rho^2} \pi_\varphi^2 + 2 \omega_L  z \pi_\varphi \nn \\ 
&& - \omega_0^2 \rho^2 z - \frac{a z}{\sqrt{\rho^2 + z^2}} \, . 
\ea 
When $\tau = \frac{1}{2}$ instead it is possible to find a conserved quantity that is quartic in the momenta and originates from a rank four Killing tensor. As we will see in sec.\ref{sec:Hamilton-Jacobi} this means that the system is integrable although not separable. The quartic constant of motion is given by the longer expression \cite{Cariglia2014killing} 
\ba
&& C =  \rho^2 \pi_z^4 - 2 \rho z \pi_{\rho} \pi_z^3 + z^2 \pi_{\rho}^2 \pi_z^2
  + \frac{1}{\rho^2}\, \pi_{\varphi}^4 + \pi_{\rho}^2 \pi_{\varphi}^2 + \nn \\ 
&& \left( 2 + \frac{z^2}{\rho^2} \right) \pi_z^2 \pi_{\varphi}^2 
  +\, 2 \omega_L \pi_{\varphi} \left( \rho^2 \pi^2_{\rho} + (2 \rho^2 + z^2) \pi^2_z \right) \nn \\ 
&& +    \left[ (2 \omega_z^2 - \omega_0^2) z^2 \rho^2 + 2 \omega_L^2\rho^4 - \frac{2a \rho^2}{\sqrt{z^2 + \rho^2}} \right] \pi_{z}^2 
\\ 
 & &  +  \left[ \frac{2}{3} \left( 2 \omega_0^2 - 5 \omega_z^2 + 2 \omega_L^2 \right) z^3 \rho + \frac{2az\rho}{\sqrt{z^2 + \rho^2}} \right] 
  \pi_z \pi_{\rho} \nn \\ 
&& + \left[ \omega_L^2 \rho^4 - \frac{1}{3} \left( \omega_0^2 - 4 \omega_z^2 + \omega_L^2 \right) z^4 \right] \pi_{\rho}^2 
\\ 
 & &  + \left[ 2 \omega_z^2 z^2 + \left( \omega_0^2 - 5 \omega_L^2 \right) \rho^2 - \frac{1}{3} \left( \omega_0^2 - 4 \omega_z^2 + \omega_L^2 \right) 
  \frac{z^4}{\rho^2} \right. \nn \\ 
&& \left. - \frac{2a}{\sqrt{z^2 + \rho^2}} \right] \pi_{\varphi}^2   -\, 2 \omega_L \pi_{\varphi} \left[ \frac{1}{3} \left( \omega_0^2 - 4 \omega_z^2 + \omega_L^2 \right) z^4  \right. \nn \\ 
&& \left.  - 2 \omega_z^2 z^2 \rho^2 + \left( 3 \omega_L^2 - \omega_0^2 \right) \rho^4 + \frac{2a \rho^2}{\sqrt{z^2 + \rho^2}}\right] 
\\
  & &  +\, \omega_z^4 z^4 \rho^2 + 2 \omega_z^2 \omega_L^2 z^2 \rho^4 - \omega_L^2 \left( 3 \omega_L^2 - 4 \omega_z^2 \right) \rho^6 \nn \\ 
&& +\, \frac{2a}{\sqrt{z^2 + \rho^2}} \left[ \omega_z^2 z^2 \rho^2 - \omega_L^2 \rho^4 \right] 
  - \frac{a^2}{2} \frac{z^2 - \rho^2}{z^2 + \rho^2}\, . 
\ea

\subsubsection{Spinning particle\label{sec:examples_spinning_particle}}  
The spinning particle theory describes a particle in a curved spacetime with $\dm$ dimensions using a set of real coordinates $x^\mu$, $\mu = 1, \dots, \dm$, and a set of Grassmannian variables $\theta^a$, $a=1, \dots, \dm$, related to the spin. Its non-relativistic version was first studied by Casalbuoni \cite{Casalbuoni1976ClassicalMechanics}. Casalbuoni also discussed the quantisation of generic systems with fermionic variables, showing that it yields a quantum theory with Fermi operators, and viceversa that the $\hbar \rightarrow 0$ limit of quantum theory is in general given by a pseudo-classical theory with Grassmannian variables \cite{Casalbuoni1976Quantization}. The relativistic version of the theory was introduced in \cite{Brink1976LocalSusy,Berezin1977,Brink1977Lagrangian}, and it can be thought of as a semi-classical description of a Dirac fermion. The Hamiltonian is given by 
\ba
H=\frac{1}{2}\Pi_\mu\Pi_\nu g^{\mu\nu}\,,\ \ 
\Pi_\mu=p_\mu - \frac{i}{2}\theta^a\theta^b\omega_{\mu ab}=g_{\mu\nu}\dot x^\nu\,,\qquad
\ea
where $p_\mu$ is the momentum canonically conjugate to $x^\mu$, $\Pi_\mu$ is the covariant momentum and $\omega_{\mu ab}$ the spin connection. Upon quantisation the $\theta$ variables are lifted to Gamma matrices. Poisson brackets are defined as 
\be\label{brackets}
\{F,G\}=\frac{\partial F}{\partial x^\mu}\frac{\partial G}{\partial p_\mu}-
\frac{\partial F}{\partial p_\mu}\frac{\partial G}{\partial x^\mu}+
i(-1)^{a_F}\frac{\partial F}{\partial \theta^a}\frac{\partial G}{\partial \theta_a}\,,
\ee
where $a_F$ is the Grassmann parity of $F$.
 
There is a fermionic generator of supersymmetry transformations on the worldsheet: 
\be\label{Qdef}
Q=\theta^a e_a{}^\alpha \Pi_\alpha\,,
\ee
which obeys
\be
\{H,Q\}=0\,, \quad \{Q,Q\}=-2iH\,.
\ee

Equations of motion are accompanied by two physical (gauge) conditions
\be\label{gaugecond}
2H=-1\,,\quad Q=0\,,
\ee 
Upon quantisation the second condition maps into the Dirac equation on states.  
 
Hidden symmetries of the dynamics for the spinning particles were discussed for the first time by Gibbons, Rietdijk and van Holten in \cite{GaryetAl1993}. One of the motivations was understanding the classical counterpart of the result by Carter and McLenaghan \cite{CarterMcLenaghan1979}, that showed that in the Kerr-Newman background there exists a linear differential operator which commutes with the Dirac operator. Gibbons and collaborators showed that if a curved background admits a Killing-Yano tensor of rank 2, then this can be used to build a conserved quantity in the theory of the spinning particle, and such conserved quantity is a fermionic generator of extra supersymmetry transformations, the prominent example being that of the Kerr-Newman metric. In fact, Killing-Yano tensors are so special that the result keeps holding for the quantum case, where a linear differential operator that commutes with the Dirac operator can be built every time such a tensor exists, in all dimensions and signatures \cite{BennCharlton1997,BennKress2004} and with no anomalies; this result will be discussed in more detail in sec.\ref{sec:dirac_equation_linear_symmety_operators}. To emphasise how this result is remarkable the reader should notice that in the case of the Klein-Gordon equation candidate symmetry operators  whose classical limit corresponds to conserved quantities are in general anomalous, and commute with the wave operator only if certain conditions hold, see the discussion in sec.\ref{sec:Klein_Gordon}. 
 
Given a rank 2 Killing-Yano tensor $h$ the conserved charge found by Gibbons and collaborators is 
\be 
\mathcal{Q}_h = \Pi^{\lambda} h_{\lambda \mu} \theta^\mu - \frac{i}{3} \nabla_{\mu_1} h_{\mu_2 \mu_3} \theta^{\mu_1} \theta^{\mu_2} \theta^{\mu_3} \, . 
\ee 
This has been generalised to the case of a rank $p$ Killing-Yano tensor by Tanimoto in \cite{Tanimoto1995}: 
\ba \label{eq:spinning_particle_exotic_susy}
\mathcal{Q}_h &=& \Pi^{\mu_1} h_{\mu_1 \dots \mu_p} \theta^{\mu_2} \dots \theta^{\mu_p}  \nn \\ 
&& - \frac{i}{p+1} \nabla_{\mu_1} h_{\mu_2 \dots \mu_{p+1}} \theta^{\mu_1} \dots \theta^{\mu_{p+1}} \, . 
\ea 
  
Recently Cariglia and Kubiz\v{n}\'ak studied the spinning particle theory in the Kerr-NUT-(A)dS spacetimes \cite{DavidMarco2011}. These describe rotating black holes with cosmological constant in arbitrary dimension and are discussed in detail in section \ref{sec:higher_dimensional_black_holes}. In these metrics there exists a complete set of mutually commuting operators, one of which is the Dirac operator, and the Dirac equation is separable, as discussed in sec.\ref{sec:quantum_dirac_equation}. 

In the Kerr-NUT-(A)dS spacetimes with $\dm = 2N + \epsilon$ dimensions there are $(N+\varepsilon)$ Killing vectors $\xi_{(k)}$. Here $\epsilon = 0,1$ parameterises whether the dimension $\dm$ is even or odd. According to eq.\eqref{eq:spinning_particle_exotic_susy} with these one can construct bosonic superinvariants, i.e. invariant quantities that are even in the Grassmannian variables. These invariants are linear in velocities and given by
\be\label{Qk}
Q_{\xi_{(k)}}=\xi_{(k)}^\alpha\Pi_\alpha-\frac{i}{4}\theta^a\theta^b (d\xi_{(k)})_{ab}\,.
\ee 
These can be used to express some components of the velocities $\Pi$ in terms of the conserved quantities and of the $\theta$ variables. In \cite{DavidMarco2011} it was shown that it is possible to find $N$ further bosonic supersymmetric conserved quantities ${\cal K}_{(j)}$, this time quadratic in the velocities. These new quantities will not be conserved nor supersymmetric in a general metric, but they are for Kerr-NUT-(A)dS. The $(N+\eps) + N = \dm$ quantities are all independent and using them it is possible to express all the components of $\Pi$, thus showing that the bosonic sector of the theory is integrable. 
 
The quantities ${\cal K}_{(j)}$ are written as 
\ba\label{quadr}
{\cal K}_{(j)}&=&K_{(j)}^{\mu\nu} \Pi_\mu\Pi_\nu+{\cal L}_{(j)}^\mu\Pi_\mu+{\cal M}_{(j)}\,,\nonumber\\
{\cal L}^\mu_{(j)}&=&\theta^{a}\theta^b L_{(j) ab}{}^\mu\,,\quad  
{\cal M}_{(j)}=\theta^{a}\theta^b\theta^c\theta^d M_{(j)abcd}\,.\quad
\ea
The tensors $K$, $L$ and $M$ are given by 
\ba\label{solution}
K^{\mu\nu}&=& h^{\mu \kappa_1\dots \kappa_{p-1}} h^\nu{}_{\kappa_1\dots \kappa_{p-1}}\,,\label{solK}\nonumber\\
L_{\mu\nu}{}^\rho&=& -\frac{2i}{p+1} h_{[\mu|\kappa_1\dots \kappa_{p-1}|}(dh)_{\nu]}{}^{\rho\kappa_1\dots \kappa_{p-1}}\nonumber\\
&-&\vspace{0.5cm} \frac{2i}{p+1} (dh)_{\mu\nu \kappa_1\dots \kappa_{p-1}} h^{\rho\kappa_1\dots \kappa_{p-1}}\,,\qquad \label{solL}\\
M_{\mu\nu\rho\sigma}&=&-\frac{i}{4} \nabla_{[\mu} L_{\nu\rho\sigma]}\, , \nonumber\label{solM}
\ea
where $h_{\mu_1\dots \mu_p}$ is the rank-$p$ Killing--Yano tensor present in the spacetime, for $p=2, 4, \dots, 2N$ in even dimensions and $p= 1, 3, \dots, 2N - 1$ in odd dimensions, see section \ref{sec:higher_dimensional_black_holes}, and in particular eqs.\eqref{eq:h_canonical_basis}, \eqref{eq:h_tower}.

\section{HAMILTON-JACOBI EQUATION AND HIDDEN SYMMETRIES\label{sec:Hamilton-Jacobi}} 
In this section we describe the theory of separation of variables for the Hamilton-Jacobi equation and its links with hidden symmetries of the dynamics and with the Eisenhart-Duval lift, which will be discussed in sec.\ref{sec:Eisenhart-Duval}.

\subsection{Hamilton-Jacobi equation} 
The Hamilton-Jacobi equation can be regarded as one of the deepest formulations of classical Hamiltonian dynamics. Let $\mathcal{P}$ be a symplectic manifold  with $2\dm$ dimensions and consider a Hamiltonian system with Hamiltonian function $H: \mathcal{P} \rightarrow \mathbb{R}$, in local coordinates $H = H(q^\mu, p_\nu)$. The Hamilton-Jacobi equation is the following first order partial differential equation in $\dm +1$ variables, in general non-linear, 
\be \label{eq:HJ_definition}
\frac{\partial S}{\partial t} + H \left(q^\mu, \frac{\partial S}{\partial q^\nu}, t \right) = 0 \, , 
\ee 
for a function $S= S(t, q^\mu)$. A \textit{complete} integral is defined as a function $S = S(t, q^\mu, P_\nu) + C$, where the $P_\nu$ and $C$ are constants.  
Once we have a complete integral, we can interpret it as the generating function of a canonical transformation between the original Hamiltonian system and a new system with variables $Q^\mu, P_\nu$ and $H^\prime$. The link between old and new variables is given by 
\ba  \label{eq:HJ_canonical_transformation}
p_\nu (t, q, P)  &=& \frac{\partial S}{\partial q^\nu}  \, \nn \\ 
Q^\mu (t, q, P)  &=& \frac{\partial S}{\partial P_\mu}  \, \nn \\ 
H^\prime &=& H + \frac{\partial S}{\partial t} \, . 
\ea 
Then, eq.\eqref{eq:HJ_definition} implies $H^\prime = 0$, which means that the function $S$ generates a canonical transformation that trivialises the system. Then $Q$ and $P$ are constants of motion and, assuming the second equation in \eqref{eq:HJ_canonical_transformation} is invertible, one can obtain $q = q(Q,P, t)$, and then from the previous equation $p = p(Q,P,t)$.

Eq.\eqref{eq:HJ_definition} is formally similar to the Schr\"{o}dinger equation, and in fact there are links to it and to quantum mechanics. For a time independent Hamiltonian we can always separate variables by setting $S(q,P,t) = W(q, P) - E t$, where we chose $P_1 = E$. The Hamilton-Jacobi equation becomes 
\be 
H \left(q^\mu, \frac{\partial W}{\partial q^\nu} \right) = E \, . 
\ee 
Then for every function $f(P)$ we can consider the wavefronts that are given by the hypersurfaces in position space 
\be 
S(q,P,t) = W(q,P) - E t = f( P) \, . 
\ee 
The normal to this surfaces is given by $\partial_q W = p$ for any choice of $f$. The connection with quantum mechanics arises as follows. For simplicity we consider a single particle interacting with a potential, with Hamiltonian of the kind $H= \frac{p^2}{2m} + V(q)$, however the reasoning can be generalised to multiple interacting particles. We can  consider the associated Schr\"{o}dinger equation 
\be 
i \hbar \frac{\partial \psi}{\partial t} = - \frac{\hbar^2}{2m} \nabla^2 \psi + V(q) \psi \, , 
\ee 
and parameterise the wave function as 
\be 
\psi(q,t) = A(q,t) \exp\left( \frac{i}{\hbar} \varphi(q,t) \right) \, . 
\ee 
Then, Schr\"{o}dinger's equation splits into the following two equations: 
\ba 
\frac{\partial \rho}{\partial t} + \vec{\nabla}_q \cdot \vec{J} &=& 0 \, , \nn \\ 
\frac{1}{2m} \left(\nabla_q \varphi \right)^2 + V(q) + \frac{\partial \varphi}{\partial t} &=& \frac{\hbar^2}{2m} \frac{\nabla_q^2 A}{A} \, , 
\ea 
where 
\be 
\rho = A^2, \quad \vec{J} = Re \left(\frac{\hbar}{im} \psi^* \nabla_q \psi \right) = \rho \frac{\nabla_q \varphi}{m} \, . 
\ee 
In the small $\hbar$ limit the second equation decouples and becomes the Hamilton-Jacobi equation for $\varphi$. Once $\varphi$ is known, the first equation can be solved for $A$ giving a semi-classical approximation for $\psi$. It is interesting noticing that $\vec{J} = \rho \vec{v}$, where in the semi-classical approximation $\vec{v} = \vec{p}/m$, the classical velocity. Also it is important noticing that imaginary solutions of the Hamilton-Jacobi are relevant in this context, being associated to quantum-mechanical tunneling.

\subsection{Separability of the Hamilton-Jacobi equation} 
Hidden symmetries of the dynamics play a natural role in the separation of variables of the Hamilton-Jacobi equation. Loosely speaking, this is associated to the presence of a complete set of conserved quantities of order one or two in the momenta, with appropriate compatibility conditions. 
 
There is a fair amount of literature on the separability of the Hamilton-Jacobi equation, in particular for \textit{natural Hamiltonians} of the kind $H = \frac{1}{2} g^{\mu\nu} p_\mu p_\nu + V(q)$ there exists a well understood theory. When $V=0$ we will call $H$ a \textit{geodesic Hamiltonian}. We will not present here the complete theory of separation of variables for natural Hamiltonians, rather we will discuss in a certain detail the case when $g$ is a positive definite, time independent Riemannian metric, and provide a description and enough references to the literature for what concerns more general cases. 
 
\subsubsection{Riemannian metric and orthogonal coordinates\label{sec:Hamilton-Jacobi_riemannian_orthogonal}} 
 
We begin with separation of variables in orthogonal coordinates, that is coordinates such that $g^{\mu\nu} = 0$ for $\mu\neq \nu$. Following Benenti \cite{Benenti_RoyalSociety} we will say that the coordinate system is \textit{separable} if the geodesic Hamilton-Jacobi equation 
\be 
\frac{1}{2} g^{\mu\nu} \frac{\partial S}{\partial q^\mu} \frac{\partial S}{\partial q^\nu} = E 
\ee 
admits a complete solution of the form 
\be 
S(q, P) = \sum_{\mu = 1}^\dm S_\mu (q^\mu, P) \, , 
\ee 
with the invertibility condition $\det \left[ \frac{\partial^2 S}{\partial P \partial q} \right] \neq 0$. In our discussion of the separability of the Hamilton-Jacobi equation we will assume  $E \neq 0$ until section \ref{sec:Hamilton-Jacobi_null} where we will discuss the null case. We will make the same assumption for natural Hamilton-Jacobi equations as well.

St\"{a}ckel \cite{Stackel1893} showed that in the case of orthogonal coordinates the geodesic Hamilton-Jacobi is separable if and only if 
\be \label{eq:Stackel_condition_1}
S_{\mu\nu} ( g^{\rho \rho}) = 0 \, , \quad \mu \neq \nu \, , 
\ee 
with no sum over $\rho$. The $S_{\mu\nu}$ are St\"{a}ckel operators that act on a smooth function $f= f(q)$ as 
\be 
S_{\mu\nu} (f) = \frac{\partial^2 f}{\partial q^\mu \partial q^\nu} - \frac{\partial \ln g^{\nu\nu}}{\partial q^\mu} \frac{\partial f}{\partial q^\nu} - \frac{\partial \ln g^{\mu\mu}}{\partial q^\nu} \frac{\partial f}{\partial q^\mu} \, , \quad \mu \neq \nu \, . 
\ee 
This is equivalent to say that the metric elements can be written as $g_{\mu\mu} =  h_\mu^2$ (no sum over $\mu$) where 
\be 
h_\mu^2 = \frac{\Theta}{\Theta_{\mu}^{1}} \, .  
\ee 
In the equation above $\Theta_{\mu}^{1}$ is the $(\mu,1)$ cofactor of a matrix $\theta_{\mu}^{(\nu )}$ that is called a \textit{St\"{a}ckel matrix}. This by definition is a square $\dm \times \dm$ matrix whose $\mu$-th row only depends on $q^\mu$,  $\theta_{\mu}^{(\nu )} = \theta_{\mu}^{(\nu )} (q^\mu)$. Lastly, $\Theta = \det \left[ \theta_{\mu}^{(\nu)} \right]$ is called a \textit{St\"{a}ckel determinant} \cite{Stackel1891}. Equivalently, the conditions above are the same as saying that 
\be 
\sum_\mu g^{\mu\mu} \theta_\mu^{(\nu)} = \delta_1^\nu \, . 
\ee  

When considering a natural Hamiltonian with scalar potential then condition \eqref{eq:Stackel_condition_1} has to be supplemented by 
\be 
S_{\mu\nu} (V) = 0 \, , \quad \mu \neq \nu \, , 
\ee 
as a necessary and sufficent condition for separability. This is equivalent to the existence of a \textit{St\"ackel vector} $w$ whose $\mu$-th component depends only on $q^\mu$, $w^\mu = w^\mu (q^\mu)$, such that 
\be 
\sum_\mu g^{\mu\mu} w^\mu = V \, . 
\ee 
Later on several other authors examined the orthogonal case, we mention \cite{LeviCivita1904,Eisenhart1934,Eisenhart1949} as a non-exhaustive list. In particular Levi-Civita showed a set of necessary and sufficient condition for the separability of the Hamilton-Jacobi equation for generic coordinates: 
\ba \label{eq:Levi-Civita_conditions}
L_{\mu\nu }(H) :&=& \frac{\partial H}{\partial q^\mu} \frac{\partial H}{\partial q^\nu} \frac{\partial^2 H}{\partial p_\mu p_\nu} + \frac{\partial H}{\partial p^\mu} \frac{\partial H}{\partial p^\nu} \frac{\partial^2 H}{\partial q_\mu q_\nu}  \nn \\ 
&& \hspace{-1.5cm} -  \frac{\partial H}{\partial q^\mu} \frac{\partial H}{\partial p_\nu} \frac{\partial^2 H}{\partial p_\mu q^\nu} -  \frac{\partial H}{\partial p_\mu} \frac{\partial H}{\partial \partial q^\nu} \frac{\partial^2 H}{\partial q^\mu \partial p_\nu} = 0 \, , 
\ea 
with no sum over $\mu$ or $\nu$. These are known as \textit{separability equations of Levi-Civita}. For natural Hamiltonians  the separability conditions above are in general polynomial expressions of degree four in the momenta, in particular the fourth degree homogeneous part is given by $L_{\mu \nu} \left( \frac{1}{2} g^{\rho \sigma} p_\rho p_\sigma \right)$. Therefore separability of the geodesic Hamilton-Jacobi equation is a necessary condition for separability of the natural Hamilton-Jacobi equation with a potential. This explains the importance given in the literature to the geodesic case. For us, the geodesic case is also important in view of the Eisenhart lift procedure, discussed in sec.\ref{sec:Eisenhart-Duval}, which describes a natural Hamiltonian system in terms of geodesics in a higher dimensional space. A magnetic field can be added naturally in the description. 

Separability of the Hamilton-Jacobi equation is in correspondence with a specific type of hidden symmetries, those generated by conserved quantities of order two or one in the momenta. In the geodesic case this was showed by Eisenhart for the orthogonal case \cite{Eisenhart1934,Eisenhart1949}, and by Kalnins and Miller for the generic case \cite{KalninsMiller1980,KalninsMiller1981}. If a symmetric tensor $K$ is diagonalised in orthogonal coordinates then its components will satisfy $K^{\mu\mu} = \rho^\mu g^{\mu\mu}$, no sum over $\mu$, for appropriate functions $\rho$. If now $K$ is a Killing tensor then one can show that the Killing equation reduces to 
\be \label{eq:KE_equations}
\partial_\mu \rho^\nu = \left(\rho^\mu - \rho^\nu \right) \partial_\mu \ln g^{\nu\nu} \, , 
\ee 
with no sum over repeated indices. These have been called \textit{Killing-Eisenhart equations} in \cite{Benenti2002a}. The integrability conditions for this linear system of first order partial differential equations are, remarkably, 
\be 
\left(\rho^\mu - \rho^\nu \right) S_{\mu\nu} \left( g^{\rho\rho} \right) = 0 \, . 
\ee 
Therefore, if all the $\rho$ functions are different then the geodesic Hamilton-Jacobi is integrable and, vice-versa, if the geodesic Hamilton-Jacobi is integrable then it is always possible to build a solution of \eqref{eq:KE_equations} with different values for the $\rho$ variables, at least locally. Therefore we have the following result: \textit{the geodesic Hamilton-Jacobi is separable  in a system of orthogonal coordinates if and only if there exists a Killing tensor which is diagonal in these coordinates and with point-wise simple eigenvalues}. 
 
Since the Killing-Eisenhart equations are linear, a Killing tensor as above is equivalent to an $\dm$-dimensional linear space $\mathcal{K}$ of Killing tensors which are all diagonalised in the same orthogonal coordinates, a \textit{Killing-St\"{a}ckel space}. Also, all the Killing tensors in a Killing-St\"{a}ckel space are all in involution, and such a space is closed under the Schouten-Nijenhuis bracket \eqref{eq:schouten_nijenhuis} between symmetric tensors. The notion of separability of the geodesic Hamilton-Jacobi equation and of Killing-St\"{a}ckel space is preserved under a separated change of coordinates $q^\mu = q^\mu (\tilde{q}^{\mu})$ (no sum), and therefore it is appropriate talking about equivalence classes of separable orthogonal coordinates. 
 
It is possible to give an intrinsic, coordinate independent characterisation of the separability noticing that geometrically an equivalence class of orthogonal coordinates is associated to a set of mutually orthogonal vectors that are orthogonal to hypersurfaces, or \textit{normal} mutually orthogonal vectors. Then follows the result of \cite{KalninsMiller1980,Benenti1992}: \textit{the geodesic Hamilton-Jacobi equation is separable in orthogonal
coordinates if and only if there exists a Killing 2-tensor with simple eigenvalues and
normal eigenvectors}. Such a tensor is also called a \textit{characteristic Killing tensor}. The $\dm$ orthogonal foliations it generates are called a \textit{St\"{a}ckel web}. If instead we prefer to  work with a Killing-St\"{a}ckel space of $\dm$ Killing tensors $K_i$ then an intrinsic characterisation of the fact that they have common eigenvalues is that any two $K_i$, $K_j$ must commute as linear operators. Kalnins and Miller have shown that $\dm$ independent Killing tensors in involution and with the same eigenvectors are necessarily normal \cite{KalninsMiller1980}, and Benenti and collaborators showed that if the eigenvectors are orthogonal then one can even relax the requisite of the symmetric tensors being Killing  \cite{Benenti2002a}. Therefore a second, alternative intrinsic characterisation of orthogonal separability of the geodesic Hamilton-Jacobi equation follows: \textit{the geodesic Hamilton-Jacobi equation is separable in orthogonal coordinates if and only if the Riemannian manifold admits a Killing-St\"{a}ckel-space i.e., an $\dm$-dimensional linear space $\mathcal{K}$ of Killing tensors commuting as linear operators and in involution}. 
 
This concludes our overview of orthogonal separability of the geodesic Hamilton-Jacobi equation. Next we can study a natural Hamiltonian of the type $H = \frac{1}{2} g^{\mu\nu} p_\mu p_\nu + V(q)$, still in orthogonal coordinates for the moment. We consider a second order candidate conserved quantity of the kind $C = \frac{1}{2} K^{\mu\nu}(q) p_\mu p_\nu + C_0(q)$. Then a direct calculation shows that this is conserved if and only if the following two separate conditions hold: 
\be \label{eq:orthogonal_natural_hamiltonian_2nd_order}
\left\{ \frac{1}{2} K^{\mu\nu}(q) p_\mu p_\nu , \frac{1}{2} g^{\rho\sigma} p_\rho p_\sigma \right\} = 0 \, , \quad d C_0 = K \cdot dV \, . 
\ee 
The first equation means that $K$ must be a Killing tensor. Since we know that a necessary condition for the separability of the natural Hamilton-Jacobi equation is that the associated geodesic Hamilton-Jacobi equation also separates, we can assume $K$ is a characteristic Killing tensor. Then the second equation in \eqref{eq:orthogonal_natural_hamiltonian_2nd_order} written in separable orthogonal coordinates becomes 
\be 
\frac{\partial C_0}{\partial q^\mu} = \rho^\mu \frac{\partial V}{\partial q^\mu} \, , \quad \forall \mu \; \text{(no sum)} 
\ee 
whose integrability conditions are $\partial_\nu \partial_\mu C_0 - \partial_\mu \partial_\nu C_0 = (\rho^\mu - \rho^\nu ) S_{\mu\nu} (V)$, no sum. Therefore we can state that: \textit{the natural Hamilton-Jacobi equation is separable in orthogonal coordinates if and only if there exists a Killing 2-tensor $K$ with simple eigenvalues and normal eigenvectors such that $d (K \cdot dV) = 0$} \cite{Benenti_RoyalSociety}. 
  
\subsubsection{Riemannian metric and generic coordinates} 
The results presented so far can be generalised to the case of non-orthogonal coordinates and a Riemannian metric. The techniques used are similar and the only main difference is that in the general case ignorable coordinates may arise, associated to Killing vectors. Kalnins and Miller proved general results on the relationship between the geodesic Hamilton-Jacobi separation in generic coordinates and the presence of Killing vectors and Killing tensors for Riemannian metrics \cite{KalninsMiller1981}. Benenti re-phrased the intrinsic characterisation of the separability for natural Hamiltonians with scalar potential in terms of separable Killing webs and characteristic Killing tensors \cite{Benenti1997}. 
 
In general for non-orthogonal variables one distinguishes between first and second class coordinates. A coordinate $q^\mu$ is of \textit{first class} if 
\be 
 \frac{\partial H}{\partial q^\mu} \left(\frac{\partial H}{\partial p_\mu}\right)^{-1}  
\ee 
is a linear function of the momenta $p$ (no sum over $\mu$). In particular \textit{ignorable} coordinates, such that $\frac{\partial H}{\partial q^\mu} = 0$, are first class. For a geodesic Hamiltonian ignorable coordinates are associated to Killing vectors. Coordinates that are not of first class are of \textit{second class}. We will use the following notation: $q^a$, $a = 1, \dots, m$ are second class coordinates, and $q^\alpha$, $\alpha = m+1, \dots, \dm$, with $r = \dm - m$, are first class coordinates. Two different, overlapping systems of coordinates such that the Hamilton-Jacobi equation is separable in both are called \textit{equivalent}  if they yield the same complete solution in the intersection of their domains. 
 
It can be shown that the Levi-Civita separability conditions \eqref{eq:Levi-Civita_conditions} for a generic Hamiltonian implies that the numbers $(m,r)$ are the same in two equivalent systems of separable coordinates, and that there always exists a coordinate transformation that preserves separability such that all the first class coordinates become ignorable \cite{Benenti1980,Benenti1991}. It can also be shown that for a geodesic Hamiltonian the second class coordinates are orthogonal, $g^{ab} = 0$ for $a\neq b$, and that in two equivalent separable systems the second class coordinates are related by separated transformations, $q^a = q^a (\tilde{q}^a)$, no sum. There exist \textit{normal separable coordinates} such that the $q^\alpha$ are ignorable and $g^{a \alpha} = 0$, so that the inverse metric tensor assumes a standard block diagonal form with the $g^{ab}$ part in orthogonal form. 
  
Using these concepts one can prove the following result on the intrinsic characterisation of the separability of a geodesic Hamiltonian \cite{BenentiFrancaviglia1979}. 
\begin{Theorem} 
The geodesic Hamilton-Jacobi equation is separable on a Riemannian manifold if and only if the following conditions hold: 
\begin{enumerate} 
\item 
There exist $r$ independent commuting Killing vectors $V_\alpha$, $\left[V_\alpha, V_\beta \right] = 0$. 

\item There exist $m = \dm - r$ independent rank 2 Killing tensors $K_a$, satisfying 
\be 
\left[ K_a, K_b \right]_{SN} = 0 \,, \quad \left[ K_a, V_\alpha \right]_{SN} = 0 \, , 
\ee 
where $\left[ \cdot, \cdot \right]_{SN}$ is the Schouten-Nijenhuis bracket, eq.\eqref{eq:schouten_nijenhuis}. 
 
\item The Killing tensors have $m$ common  eigenvectors $W_a$ such that 
\be 
\left[ W_a, V_\alpha \right] = 0 \, , \quad \left[ W_a, W_b \right] = 0 \, , \quad g (W_a, V_\alpha ) = 0 \, . 
\ee 
\end{enumerate} 
\end{Theorem} 

The result above is a specific case of the more general structure of a \textit{non-degenerate Killing web}, which generalises that of a St\"{a}ckel web to non-orthogonal coordinates. A Killing web is given by two sets of objects. First, a set $\mathcal{S}_m = \{ \mathcal{S}^1, \dots, \mathcal{S}^m \}$ of $m$ pairwise orthogonal foliations of connected submanifolds of co-dimension $1$. Orthogonal means that their normal vectors are orthogonal, and therefore since the metric is Riemannian they are pairwise transversal, i.e. the span of their tangent spaces is the whole ambient tangent space. Second, a Killing web has an $r$-dimensional abelian algebra $\Delta_r$ of Killing vectors tangent to the leaves of $\mathcal{S}_m$. The Killing vectors span a distribution $\Delta$ of constant rank $r$. A distribution $d$ of dimension $s$ is called \textit{orthogonally integrable} if the orthogonal distribution $d^{\perp}$ is completely integrable, i.e. there exists a foliation of $\dm - s$-dimensional manifolds orthogonal to $d$. In particular for a non-degenerate Killing web $\Delta_r$ is orthogonally integrable. $r$ is called the \textit{degree of symmetry} of the web. A Killing web determines a set of coordinates $(q^a, q^\alpha)$ \textit{adapted} to it. First one chooses $m$ independent functions $q^a$ that are constant to the leaves of the foliations $\mathcal{S}^a$. Then, given a basis $V_\alpha$ for $\Delta_r$ one can consider a local section $X$ of the orbits of $\Delta$, that is an $m$-dimensional submanifold that is transversal to the orbits. Using $X$ as a starting set, one can consider the integral curves of $\Delta_r$ and take the affine parameters on these as ignorable coordinates $q^\alpha$. The web is a \textit{separable Killing web} if the geodesic Hamilton-Jacobi is separable in adapted coordinates. As in the case of orthogonal coordinates, a separable web can be characterised by a characteristic Killing tensor \cite{Benenti1997}: 
\begin{Theorem} 
A non-degenerate Killing web $(\mathcal{S}, \Delta)$ is separable if and only if there exists a $\Delta$ invariant characteristic Killing tensor $K$ with pairwise and pointwise distinct real eigenvalues, corresponding to eigenvectors $W_a$ orthogonal to the leaves of $\mathcal{S}$. 
\end{Theorem} 
 
For a natural Hamiltonian with scalar potential one has the extra condition that the potential $V$ is $\Delta_r$ invariant and that $d(K \cdot V) = 0$.

\subsubsection{Generic signature metric and generic coordinates} 
This case has been treated for example in \cite{Benenti1997}. One still distinguishes between first and second class coordinates, but now refines the classification into \textit{null} second class coordinates $q^{\overline{a}}$ such that $g^{\overline{a} \overline{a}} = 0$ (no sum),  and the remaining non-null second class coordinates $q^{\hat{a}}$, with $\hat{a} = 1, \dots, m_1$, $\overline{a} = m_1 +1, \dots, m_1 + m_0 = m$. While it is always the case that $m_0$ cannot be higher then absolute value of the signature of the metric, it can also be shown that it must be $m_0 \le r$. This follows from the fact that normal separable coordinates still exist, with the $q^\alpha$ ignorable and $g^{\hat{a} \alpha} = 0$, so that the standard form of the inverse metric in these coordinates is 
\be 
g^{\mu\nu} = \left( 
  \begin{array}{ccc} g^{\hat{a}\hat{a}} & 0 & 0 \\ 
                              0 & 0 & g^{\overline{b} \beta} \\ 
                              0 & g^{\alpha \overline{a}} & g^{\alpha \beta} 
  \end{array} \right) \, , 
\ee 
and $m_0 > r$ would imply $\det (g^{\mu\nu}) = 0$. 
 
Now a \textit{Killing web} is defined similarly to the non-degenerate Killing web of the previous section. There are $m$ foliations $\mathcal{S}_m = \{ \mathcal{S}^1, \dots, \mathcal{S}^m \}$ of $m$ pairwise orthogonal foliations of connected submanifolds of co-dimension $1$ that are pairwise transversal and orthogonal. Notice that we need to ask transversality since this is not implied by orthogonality once the metric has generic signature. Next, there is an $r$-dimensional abelian algebra $\Delta_r$ of Killing vectors tangent to the leaves of $\mathcal{S}_m$. The Killing vectors span a distribution $\Delta$ of constant rank $r$. For a generic signature metric we now consider the distribution $I = \Delta \cap \Delta^\perp$, which is in general non-zero and generated by null vectors. By definition we ask that $I$ has constant rank $m_0$.  This time $\Delta_r$ is orthogonally integrable if the orthogonal distribution $\Delta^{\perp}$ is completely integrable, which includes the $I$ distribution. 
 
The intrinsic characterisation of separability of the natural Hamilton-Jacobi equation given in \cite{Benenti1997} is the following: 
\begin{Theorem} 
The natural Hamilton-Jacobi equation is separable in a coordinate system adapted to a Killing web if and only if: 
\begin{enumerate} 
\item 
there exists a $\Delta$ invariant characteristic Killing tensor $K$ with pairwise and pointwise distinct real eigenvalues, corresponding to eigenvectors $W_a$ orthogonal to the leaves of $\mathcal{S}$ and, for $m_0 > 1$, $d (K \cdot g^{\alpha \beta}) = 0$ for any basis $V_\alpha$ of $\Delta_r$; 
 
\item the potential $V$ is $\Delta_r$ invariant and $d(K \cdot V) = 0$ 
\end{enumerate} 
 
\end{Theorem}

\subsubsection{Null Hamilton-Jacobi equation in orthogonal coordinates\label{sec:Hamilton-Jacobi_null}} 
The null case $E=0$ of the Hamilton-Jacobi equation is best discussed separately. The theory of the null Hamilton-Jacobi equation in orthogonal coordinates has been discussed in \cite{Benenti2005,ChanuRastelli2007}. 
 
One can prove that the null geodesic Hamilton-Jacobi equation is separable in orthogonal coordinates $q^\mu$ if and only if these coordinates are \textit{conformally separable}, i.e. the metric is conformal to another orthogonal metric such that the new coordinates are standard separable orthogonal coordinates as seen in the previous sections. A similar result applies for natural Hamiltonians of the type $H= \frac{1}{2} g^{\mu\nu} p_\mu p_\nu + V$ with $V-E \neq 0$: the natural Hamilton-Jacobi equation is separable in orthogonal coordinates $q^\mu$ if and only if the conformal metric $\overline{g} = \frac{g}{V - E}$ (also called Jacobi metric) defines a separable geodesic Hamilton-Jacobi equation. 
 
An intrinsic characterisation of these results can be done in terms of conformal tensors. For null geodesics one needs a \textit{characteristic conformal Killing tensor}, which is a conformal Killing tensor with simple eigenvalues and normal eigenvectors. Similarly, one can invoke $\dm$ pointwise independent conformal Killing tensors $K_{(\mu )}$ that have common eigenvectors and are in conformal involution, meaning there exist vector fields $\{ V_{(\mu\nu)} \}$ such that 
\be 
\left[  K_{(\mu )} , K_{(\nu )} \right]_{SN}^{\rho \sigma \tau} = V_{(\mu\nu)}^{(\rho} g^{\sigma \tau )} \, . 
\ee 
In presence of a potential $V$ such that $V-E \neq 0$ we need, in addition to the above, that the characteristic conformal Killing tensor that also satisfies 
\be 
\left[ g^{-1}, K \right]_{SN}^{\lambda \mu\nu} = \frac{2}{E-V} \left( K \cdot dV \right)^{(\lambda} g^{\mu \nu )} \, . 
\ee 
To conclude, we mention that the important case of a Hamiltonian with both a scalar and vector potential has been discussed in \cite{Benenti2001variable} for a positive definite metric, using the Eisenhart-Duval lift metric that will be presented in sec.\ref{sec:Eisenhart-Duval}. One obtains the same type of generalised Killing equations with flux of sec.\ref{sec:covariant_formalism}, for the special case of a conserved quantity that is of second order in the momenta.

\subsubsection{The role of higher order conserved quantities} 
At this point the reader may wonder what is the role played by conserved quantities that are of higher order than two in the momenta. After all, as we have seen all conserved quantities, independent of their order, generate infinitesimal canonical transformations in phase space that preserve the energy. On the other hand, in this section we have learned that it is only quantities that are at most second order in the momenta that are involved in the variable separation of the Hamilton-Jacobi equation. The answer in brief is that variable separation depends on a certain choice of position variables that is made after having chosen a set of position-momenta coordinates in phase space. In fact, canonical transformations are so general that, if we allow arbitrary transformations, a local theory of variable separation becomes trivial, as remarked by Benenti in \cite{Benenti2002outline}: for any Hamiltonian $H(q,p)$ we can locally find $\dm$ integrals of motion and a canonical transformation $(q,p) \mapsto (a,b)$ such that the $a$ variables are ignorable, meaning that $H = h(b)$, and with respect to such variables the Levi-Civita conditions \eqref{eq:Levi-Civita_conditions} are satisfied. However, finding such a canonical transformation is equivalent to solving the Hamilton-Jacobi equation itself, and so the result is not constructive.  
 
Another way to rephrase the concept is that existence of dynamical symmetries is an intrinsic property of a dynamical system in phase space, while existence of dynamical symmetries of a certain order in the momenta is not intrinsic, but depends on a certain choice of  canonical variables. We illustrate this with a known example, that of the H\'enon-Heiles system. 
 
The H\'enon-Heiles system is a two degrees of freedom system that was introduced in 1964 \cite{HenonHeiles1964}, with a potential that is cubic in the position variables $q_1$, $q_2$. The original system described the motion of a star in the axisymmetric potential of the galaxy, where $q_1$ represented the radius and $q_2$ the altitude. H\'einon and Heiles found that for low energies the system appeared integrable, as numerically integrated trajectories stayed on well defined two-dimensional surfaces in phase-space. They however found that for large energies many of the integral regions were destroyed, giving rise to ergodic behaviour. A certain amount of effort has been devoted to finding those deformations of the potential that give rise to exactly integrable systems. What is now known as the generalised H\'enon-Heiles system is defined by the Hamiltonian \cite{VerhoevenMusetteConte2002} 
\be \label{eq:generalised_HH}
H = \frac{1}{2} \left( p_1^2 + p_2^2 + c_1 q_1^2 + c_2 q_2^2 \right) + a q_1 q_2^2 - \frac{b}{3} q_1^2 + \frac{1}{32 a^2} \frac{\mu}{q_2^2} \, , 
\ee 
The original H\'enon-Heiles Hamiltonian corresponds to $c_1 = 1 = c_2$, $a=1$, $b=1$, $\mu = 0$. The Hamiltonian \eqref{eq:generalised_HH} is Liouville integrable in three cases \cite{BountisSegurVivaldi1982,ChangTaborWeiss1982,GrammaticosDorizziPadjen1982}:  
\ba 
(SK) && \; \; b = - a \, , \quad c_1 = c_2 \, , \nn \\ 
(KdV_5) && \;\; b = - 6 a \, , \quad c_1 , c_2 \; \text{arbitrary} \, , \nn \\ 
(KK) && \; \; b = - 16 a \, , \quad c_1 = 16 c_2 \, . \nn \\ 
\ea 
Fordy  showed that these three cases arise from the reduction $x - ct = \xi$ of three fifth order soliton equations, in order the Sawada-Kotera, $KdV_5$ and Kaup-Kupershmidt \cite{Fordy1991}. 
 
As well as the Hamiltonian itself, these three systems possess a second, independent conserved quantity $C$. The conserved quantity for the $KdV_5$ system is quadratic in the momenta, as well as the one for the $SK$ system when $\mu = 0$. This means that integrability can also be associated to separation of variables for the Hamilton-Jacobi equation according to the theory presented so far. However, for the $SK$ system with $\mu \neq 0$ and for the $KK$ system the conserved quantity is quartic, for example for the $KK$ system it is given by 
\ba 
C &=& \left( 3 p_2^2 + 3 c_2 q_2^2 + \frac{3\mu}{16a^2 q_2^2} \right)^2 + 12 a p_2 q_2^2 \left(3 q_1 p_2 - q_2 p_1 \right) \nn \\ 
&& - 2 a^2 q_2^4 \left( 6 q_1^2 + q_2^2 \right) + 12 a q_1 \left( - c_2 q_2^4 + \frac{\mu}{16a^2} \right) \nn \\ 
&& - \frac{3}{4} \frac{c_2 \mu}{ a^2} \, . 
\ea 
 A non-trivial canonical transformation that is of higher order in the momenta transforms the system in a new Hamiltonian system where the two conserved quantities are of second order in the new momenta \cite{VerhoevenMusetteConte2002}. With respect to the new variables now the system is also separable. A similar procedure works in the case of the $SK$ system for $\mu \neq 0$. Integrable models with higher order conserved quantities are still actively studied, as an example of recent work see \cite{Galajinsky2013}.

\section{GEOMETRY OF DYNAMICS\label{sec:Eisenhart-Duval}} 
\subsection{Eisenhart-Duval Lift} 
The Eisenhart-Duval lift is a geometric construction that on one side realises a geometrisation of interactions, similar in spirit to that of Kaluza-Klein geometries but employing a non-compact null-coordinate for the dimensional reduction, and on the other allows embedding non-relativistic systems into relativistic ones. For this reason it has been applied to the non-relativistic AdS/CFT correspondence, as a non-exhaustive list of examples and related works the reader can refer to \cite{HerzogRangamaniRoss2008,Goldberger2009,Duval2009non,Duval2012schrodinger} and references therein. 	
 
The geometrical idea has been historically introduced in a seminal paper by Eisenhart in \cite{Eisenhart1928}. His work remained largely unnoticed, and it was a number of years after that the same idea was independently re-discovered in \cite{DuvalBurdetKunzlePerrin1985,GaryDuvalHorvathy1991}, under the framework of a \textit{Bargmann structure}, where it has been used to discuss Newton-Cartan structures. From there it prompted further work, among which \cite{GaryPeterHorvathyZhang2012,Duval1994conformal,Duval2009non,Duval1993galilean}. In particular, it is important mentioning that the Eisenhart lift allows easily to discuss the full group of dynamical symmetries, which for example in the case of the free particle in flat spacetime is the Schr\"{o}dinger group \cite{Jackiw2008introducing,Niederer1972,Hagen1972scale}, and also allows incorporating a scalar and vector potential in the discussion of the separability of the Hamilton-Jacobi and Schr\"{o}dinger equation. 
 
During the years several different types of applications have been found for the lift in addition to the non-relativistic AdS/CFT correspondence, among which one can mention the following, non-exhaustive, examples: simplifying the study of symmetries of a Hamiltonian system by looking at geodesic Hamiltonians \cite{Benenti1997,Benn2006};  new Lorentzian pp-wave metrics solutions of the Einstein-Maxwell equations \cite{GaryChrisPope2010,GaryPeterHorvathyZhang2012}; a geometrical approach to protein folding \cite{MazzoniCasetti2008}, rare gas crystals \cite{CasettiMacchi1996} and chaotic gravitational $N$-body systems \cite{Cerruti-SolaPettini1995}. Moreover, using the lift it is possible to buid new examples of spacetimes with non-trivial hidden symmetries of higher order have been found: among the lifted systems when can mention the Goryachev-Chaplygin and Kovalevskaya's tops \cite{DavidGaryClaudeHouri2011}. Benenti, Chanu and Rastelli used the Eisenhart-Duval lift metric to reduce the study of separability of the Hamilton-Jacobi equation for a natural Hamiltonian with scalar and vector potential to that of a purely geodesic Hamiltonian, that follows the methods described in sec.\ref{sec:Hamilton-Jacobi} \cite{Benenti2001variable}. 
  
\subsubsection{The geometry} 
We begin with notation. The base manifold $\M$ is $\dm$--dimensional and it is endowed with a metric $g$. We indicate with $\h{\M}$ its $(\dm + 2)$--dimensional  Eisenhart-Duval lift, with metric $\h{g}$ is defined. We will denote $\dm +2$--dimensional quantities with a hat symbol. Indices $\mu, \nu, \dots$, from the lowercase Greek alphabet represent spacetime indices on $\M$, while $M, N, \dots$, from the uppercase Latin alphabet spacetime indices on $\h{\M}$. Local coordinated used for $\M$ are $\{ x^\mu \}$, and for the lift we introduce new variables $v, t$ so that $\{ \h{x}^M \} = \{ v, t, \{x^\mu \}\}$ are local coordinates on $\h{\M}$. In order to work with the Dirac equation and Gamma matrices we also denote locally flat indices: $a = 1, \dots , \dm$ for $\M$ and $A = +, -, 1, \dots, \dm)$ for $\h{\M}$. 

On $\M$ we can consider a natural Hamiltonian 
\be \label{eq:EisenhartDuval_Hamiltonian_base_manifold}
H = \frac{1}{2m} g^{\mu\nu} \left(p_\mu - e A_\mu \right) \left(p_\nu - e A_\nu \right) + V(x) \, , 
\ee
where $p_\mu$ is the canonical momentum. This describes a particle of mass $m$, electric charge $e$, with a potential $V(x)$ and vector potential $A_\mu(x)$. We could introduce gauge covariant momenta $\Pi_\mu = p_\mu - e A_\mu$ as discussed in sec.\ref{sec:covariant_formalism}, however this is not needed to the extent of the calculations done here, and we will work with the canonical $p_\mu$ variables. 
 
Eisenhart showed that solutions of the natural Hamiltonian  \eqref{eq:EisenhartDuval_Hamiltonian_base_manifold} are in correspondence with geodesics of a $\dm +2$-dimensional pseudo-Riemannian metric \cite{Eisenhart1928} 
\ba  \label{eq:Eisenhart_metric}
\h{g} &=& \h{g}_{MN} d \h{x}^M d \h{x}^N \nn \\ 
&&  \hspace{-0.75cm} =  g_{\mu\nu} dx^\mu dx^{\nu} + \frac{2e}{m} A_\mu dx^\mu dt + 2 dt dv - \frac{2}{m}V dt^2 \, . 
\ea  
This is a metric first discussed by Brinkmann who studied the problem of finding Einstein spaces conformally related \cite{Brinkmann1925einstein}. 
 
The associated geodesic Hamiltonian is 
\ba
\hat{H} &=& \frac{1}{2m} \h{g}^{MN} \h{p}_M \h{p}_N \nn \\ 
&=& \frac{1}{2m}  g^{\mu\nu} \left(p_\mu - \frac{e}{m} A_\mu p_v \right) \left(p_\nu - \frac{e}{m} A_\nu p_v \right) \nn \\ 
&& + \frac{1}{m} p_v p_t + \frac{1}{m^2}V p_v^2 \, , 
\ea 
where $\h{p}_M = (p_v, p_t, p_{\mu_1}, \dots, p_{\mu_\dm})$. 
 
In order to see the relation between the two systems we first point out that $\xi = \partial/\partial v$ is a covariantly constant Killing vector. The momentum $p_v$ is therefore a constant of motion. We then consider null geodesics of $\h{H}$ for which $p_v = m$, and re-write the condition $\h{H} = 0$ as 
\be 
\frac{1}{2m} g^{\mu\nu} \left(p_\mu - e A_\mu  \right) \left(p_\nu - e A_\nu \right) +  p_t + V = 0 \, , 
\ee 
or equivalently 
\be 
 p_t = - H \, . 
\ee 
So we can identify the $-t$ variable in the higher dimensional system with the time variable in the lower dimensional one, for which $H$ is a generator of time translations. The minus sign is mainly due to historical convention. 

From the geometrical point of view $\h{\M}$ is a bundle over $\M$, with projection $P: (t,v, x^\mu) \mapsto x^\mu$. Then for $f$ a $p$--form defined on $\M$ we will indicate with $\h{f}$ its pull-back on $\h{\M}$ under the map $P^*$. 
 
$dt$ is the 1-form associated to the Killing vector $\partial_v$: $(\partial_v)^\flat = dt$. We make the following choice for the vielbeins: 
\be \label{eq:Eisenhart_vielbeins}
\begin{split}
& \hat{e}^+ =  dt \, , \\ 
& \hat{e}^- = dv - \frac{V}{m} dt + \frac{e}{m} A_\mu dx^\mu \, , \\ 
& \hat{e}^a = e^a \, , 
\end{split}
\ee
where $\left\{e^a, a=1, \dots, n\right\}$ is a set of vielbeins for $\mathcal{M}$, and the $n+2$--dimensional Minkowski metric $\hat{\eta}_{AB}$ has the following non-zero entries: $\hat{\eta}_{+-}=\hat{\eta}_{-+}=1$, $\hat{\eta}_{ab} = \eta_{ab}$. Associated to the vielbeins are dual basis vectors: 
\be
\begin{split}
& \left( \hat{e}^+  \right)^\sharp = \h{X}^+ =  \partial_v \, , \\ 
& \left( \hat{e}^- \right)^\sharp = \h{X}^- = \frac{V}{m} \partial_v + \partial_t \, , \\ 
& \left(\hat{e}^a \right)^\sharp = \h{X}^a =  - \frac{e}{m} A^a \partial_v + \left(e^a\right)^{\sharp g}   \, ,  
\end{split}
\ee 
where the $\sharp$ operation has been discussed in section \ref{sec:symplectic_special_tensors}, below equation \eqref{eq:symplectic_CKY_definition}. 
These are related to the inverse vielbein $\h{E}^M_A$ by 
\be 
\h{E}_A = \hat{\eta}_{AB} \h{X}^B \, . 
\ee 

Taking the exterior derivative of eq.\eqref{eq:Eisenhart_vielbeins} we can find the non-zero coefficients of the spin--connection: 
\be \label{eq:EisenhartDuval_spin_connection}
\begin{split}
& \h{\omega}_{+a} = - \frac{1}{m} \partial_a V \, \hat{e}^+ + \frac{e}{2m} F_{ab} e^b \, , \\ 
& \h{\omega}_{ab} = \omega_{ab} - \frac{e}{2m} F_{ab} \hat{e}^+ \, . 
\end{split} 
\ee

As seen above null geodesics on $\h{\M}$ relative to $\h{g}$ generate massive geodesics on $\M$ relative to $g$. As we are going to show in sec.\ref{sec:Eisenhart-Duval_geodesics} it is in fact possible to do more: given a generic conserved quantity for the motion on $\M$ that is a non-homogeneous polynomial in momenta, in other words a hidden symmetry for geodesic motion, this can be lifted to an appropriate hidden symmetry on $\h{\M}$ that is homogeneous in momenta and that therefore is associated to a Killing tensor. The Poisson algebra on $\M$ of conserved charges for the original motion then is the same as the Schouten-Nijenhuis algebra of the Killing tensors associated to lifted conserved charges  \cite{DavidGaryClaudeHouri2011}. This means that the dynamical evolution on $\M$ as described in full phase space can be embedded in the higher dimensional phase space. This does not happen in the case of the Dirac equation as will be shown in sec.\ref{sec:Eisenhart-Duval_CKY},  as in general it will not be possible to lift all hidden symmetries from $\M$ to $\h{\M}$. Also, when performing the dimensional reduction of the Dirac equation on $\h{\M}$ a non-trivial projection is required in phase space in order to recover the Dirac equation with $V$ and $A$ flux on $\M$, see eq.\eqref{eq:spinor_projection}. This projection is not compatible with all hidden symmetry transformations, but only with those that generate on $\M$ genuine hidden symmetries of the Dirac equation with flux. For a discussion of the Dirac equation with flux the reader is referred to sec.\ref{sec:Dirac_equation_with_flux}.

\subsubsection{Conserved quantities for geodesic motion\label{sec:Eisenhart-Duval_geodesics}} 
We can rewrite the Hamiltonian \eqref{eq:EisenhartDuval_Hamiltonian_base_manifold} on $\M$ as a polynomial of degree two in momenta:
\begin{equation}
H = H^{(2)} +H^{(1)}+H^{(0)}, \label{eq:EisenhartDuval_Hdef}
\end{equation}
where $H^{(i)}$ has degree $i$ in momenta, and the lifted Hamiltonian as 
\begin{equation}
\h{H}= \h{H}^{(2)} +\frac{p_v}{m} \h{H}^{(1)}+ \frac{p_v^2}{m^2} \h{H}^{(0)} + \frac{p_v p_t}{m}.
\end{equation}
If the symplectic form on $\M$ is $d\omega = dq^\mu \wedge dp_\mu$
 the new symplectic form is $\omega' =
\omega + dt \wedge dp_t + dv \wedge dp_v$, with associated Poisson
bracket $\{, \}'$. 

We now show the link between conserved quantities for the two systems that are polynomial in the momenta. Consider such a conserved quantity for the system $(H, T^*\M)$: 
\begin{equation}  \label{eq:Eisenhart_conserved_quantity}
K = \sum_{i=0}^{k} K^{(i)}.
\end{equation} 
We can calculate the time derivative of $K$ and organise the terms according to their degree in
momenta, obtaining 
\ba
0&=&\frac{dK}{dt} =\{K, H\}+\frac{\partial K}{\partial t} = \sum_{i=0}^k\left [ \{K^{(i-1)}, H^{(2)}\} \right. \nn \\ 
&& \left. + \{K^{(i)},
  H^{(1)}\}+ \{K^{(i+1)}, H^{(0)}\}  +\frac{\partial
  K^{(i)}}{\partial t}\right ] . 
\ea 
Terms of different order in the momenta will vanish independently since $K$ is conserved for all trajectories. $K$ can be lifted to the extended phase space by the formula 
\begin{equation} \label{eq:Eisenhart_lifted_conserved_quantity}
\h{K} = \sum_{i=0}^{k} \left(\frac{p_v}{m}\right)^{k-i} K^{(i)}.
\end{equation}
Taking the derivative of $\h{K}$ along the higher dimensional null geodesics yields 
\ba
 \frac{d\h{K}}{d\lambda} &=&\{\h{K}, \h{H}\}' =  \sum_{i=0}^k p_s^{k-i+1} \left [ \{K^{(i-1)}, H^{(2)}\} \right. \nn \\ 
&& \hspace{-0.25cm} \left. +  \{K^{(i)},
  H^{(1)}\}+ \{K^{(i+1)}, H^{(0)}\} +\frac{\partial
  K^{(i)}}{\partial t}\right ] . 
\ea 
Therefore it is clear that $\h{K}$ is constant if and only if $K$ is. Since by construction $\h{K}$ is a homogeneous polynomial in the momenta, then it corresponds to a Killing tensor of the metric $\h{g}$. 
 
 A similar
calculation shows that for constants of the motion for the original
system $K_1, K_2, K_3$ which lift to $\h{K}_1, \h{K}_2,
\h{K}_3$ we have
\begin{equation}
\{ K_1, K_2 \} = K_3, \quad \Leftrightarrow \quad \{ \h{K}_1, \h{K}_2 \}' = \h{K}_3.
\end{equation} 
 
The degree of integrability of the system is unchanged by the lift, since the dimension of the configuration space has increased by two, but at the same time two new constants of motion have been introduced: $p_s$ and $p_t$. In particular if the
original system is Liouville integrable, with  $\dm$ functionally independent constants of motion in involution, or super-integrable, then so will be the higher dimensional system, and vice-versa.

\subsubsection{Conformal Killing-Yano tensors\label{sec:Eisenhart-Duval_CKY}} 
In this section we will relate the conformal Killing-Yano equation on $\h{\M}$ to appropriate equations on $\M$. We start with the defining equation  
\be \label{eq:EisenhartDuval_CKY}
\h{\nabla}_{\h{X}} \h{f} = \frac{1}{\pi +1} \h{X} \hook \hat{d} \h{f} - \frac{1}{(\dm +2) - \pi +1} \h{X}^\flat \wedge \hat{\delta} \h{f} \, , 
\ee 
$\forall \, \h{X}$ vector. It is not restrictive to consider a homogeneous form $\h{f}$, since  the equation is linear in $\h{f}$. 
 
The general case has been discussed in \cite{Marco2012}, showing how eq.\eqref{eq:EisenhartDuval_CKY} splits into a set of equations for forms defined on $\M$. Here we consider four simplified ans\"{a}tze for the higher dimensional conformal Killing-Yano form, as they exemplify important features of the lift procedure. Given a  $p$--form $f = f(x)$ defined on the base manifold $\M$ we can build the following higher dimensional forms: 
\ba 
\h{f}_1 &=& f \, ,  \label{eq:EisenhartDuval_ansatz1} \\ 
\h{f}_2 &=& \h{e}^+ \wedge f \, , \label{eq:EisenhartDuval_ansatz2} \\ 
\h{f}_3 &=& \h{e}^- \wedge f \, , \label{eq:EisenhartDuval_ansatz3} \\ 
\h{f}_4 &=& \h{e}^+ \wedge \h{e}^- \wedge f \, , \label{eq:EisenhartDuval_ansatz4} 
\ea 
where the symbol $f$ on the right hand side is used to represent the canonical embedding in $\h{\M}$ of a form originally defined on $\M$. A form of type $\h{f}_1$ is mapped into one of type $\h{f}_4$ by Hodge duality, with $f \rightarrow *_{\M} f$. Also, forms of the type $\h{f}_2$ and $\h{f}_3$ are mapped by Hodge duality to themselves. 
 
From eqs.\eqref{eq:EisenhartDuval_ansatz1} and \eqref{eq:EisenhartDuval_ansatz4} it is possible to generate Killing-Yano and, respectively, closed conformal Killing-Yano tensors on $\h{\M}$ when $f$ is Killing-Yano and, respectively closed conformal Killing-Yano on $\M$. In particular this gives new examples of Lorentzian metrics with conformal Killing-Yano tensors by lifting known conformal Killing-Yano tensors in Riemannian signature, for example when $\M$ is the Kerr-NUT-(A)dS metric or the Taub-NUT metric \cite{GaryRuback1987,Cordani1988,VanHolten1995,VamanVisinescu1997,BaleanuCodoban1998,Visinescu2000}. 
 
However, a generic conformal Killing-Yano tensor on $\M$ cannot be lifted to a conformal Killing-Yano tensor on $\h{\M}$: as will be seen in section \ref{sec:Dirac_EisenhatDuval}, when considering the Dirac equation the process of lift and its inverse, reduction, involves a non-trivial projection in phase space, eq.\eqref{eq:spinor_projection}. This is not compatible with all lower and higher dimensional hidden symmetries, see \cite{Marco2012} for a detailed discussion.

We start considering the first ansatz, with a $p$--form $\hat{f}_1$ on $\h{M}$ given by \eqref{eq:EisenhartDuval_ansatz1}. The conformal Killing-Yano equation \eqref{eq:EisenhartDuval_CKY} splits into three types of equation, one for each of $\h{X} = \h{X}^+$, $\h{X} = \h{X}^-$, and $\h{X} = \h{X}^a$. To analyse these we re-writing the covariant derivatives of $f$ on $\h{\M}$ in terms of covariant derivatives on $\M$, $\partial_v$ and $\partial_t$ derivatives, and flux terms. The explicit formulas can be found in \cite{Marco2012}. 
The $\h{X} = \h{X}^+$ component gives 
\be \label{eq:EisenhartDuval_ansatz1_useful0}
\partial_v f = 0 \, , \quad 
\delta f = 0 \, ,  
\ee
the $\h{X} = \h{X}^-$ component 
\be \label{eq:EisenhartDuval_ansatz1_useful_dV}
 dV^\sharp \hook f  = 0 \, ,  \quad 
F \cwedge{2} f = 0  \, , 
\ee 
\be \label{eq:simple_ansatz1_t_derivative}
\partial_t  f +  \frac{p+1}{p}  \frac{e}{2m} F \cwedge{1} f = 0 \, , 
\ee 
where 
\be 
F \cwedge{1} f= \frac{1}{(p-1)!} F^\lambda {}_{\mu_1} f_{\lambda \, \mu_2 \dots \mu_p} dx^{\mu_1} \wedge dx^{\mu_p} \, , 
\ee 
and 
\be 
F \cwedge{2} f= \frac{1}{(p-2)!} F^{\lambda_1 \lambda_2}  f_{\lambda_1 \lambda_2 \, \mu_1 \dots \mu_{p-2}} dx^{\mu_1} \wedge dx^{\mu_{p-2}} \, . 
\ee
In general one can introduce a $\cwedge{m}$ contraction for differential forms, this is defined in a later section in  \eqref{cwedgedef}. Lastly, the $\h{X} = \h{X}^a$ component gives  
\be 
\frac{e}{2m} (X_a \hook F) \cwedge{1} f =  \frac{1}{p+1} X_a \hook \partial_t f \, ,
\ee 
\be \label{eq:EisenhartDuval_ansatz1_KY}
\nabla_a f = \frac{1}{p+1} X_a \hook df   \, .  
\ee 
This last is the Killing-Yano equation on the base. The former instead implies \eqref{eq:simple_ansatz1_t_derivative}. 

Eq.\eqref{eq:simple_ansatz1_t_derivative} is compatible with the fact that $f$ is Killing-Yano only if $F \cwedge{1} f$ is Killing-Yano as well, which in general will not be the case. Then it must be that separately 
\be 
\partial_t f = 0 \, , \quad 
F \cwedge{1} f = 0 \, .  
\ee  
Then there is no $v$, $t$ dependency and we can lift a Killing-Yano form on the base manifold $\M$ to a Killing-Yano form on $\h{M}$, since the conditions found imply $\h{\delta} \h{f}_1 = 0$. This form can be used to build a symmetry operator for the Dirac equation on $\h{\M}$, and when $p$ is odd such operator strictly commutes with the Dirac operator $\h{D}$, as will be seen in section  \ref{sec:dirac_equation_linear_symmety_operators}. Also for $p$ odd it is guaranteed we can build a symmetry operator for the Dirac equation with $V$ and $A$ flux that is not anomalous, thanks to the conditions found above. This is in correspondence with the results on the Dirac equation with flux of sec.\ref{sec:Dirac_equation_with_flux}. These two facts are related, since in the case of $p$ odd it is possible to dimensionally reduce such hidden symmetry operator on $\h{\M}$ to get a hidden symmetry operator with flux on $M$, \cite{Marco2012}. This is not possible if $p$ is even, which goes in agreement with the fact that the conditions required on an even conformal Killing-Yano tensor with flux in sec.\ref{sec:Dirac_equation_with_flux} are different. Such other conditions are related to the fourth ansatz \eqref{eq:EisenhartDuval_ansatz4}. The conditions found in this section are more restrictive than those in for the generic Dirac equation with flux of sec.\ref{sec:Dirac_equation_with_flux}: in principle one could have tensors that satisfy the less restrictive conditions and that generate symmetries of the Dirac equation with flux on $\M$, but that at the same time do not satisfy the conditions of this section and therefore cannot be lifted to $\h{\M}$.

Ans\"{a}tze 2 and 3 can be examined with the same techniques and give the trivial result that $f$, $\h{f}_2$ and $\h{f}_3$ are covariantly constant forms. 

Lastly, in the case of Ansatz 4 one recovers the Hodge dual of the conditions found for Ansatz 1, with the understanding that the form $f$ of Ansatz 1 is related to the $f$ form of Ansatz 4 by Hodge duality on the base manifold $\M$, namely one gets: 
\be 
\partial_{v,t} f = 0 \, , \quad 
d f = 0 \, , \quad 
dV \wedge f = 0 \, , 
\ee 
\be 
(X_a \hook F ) \wedge g = 0 \, , \quad 
F \wedge f = 0 \, \quad 
F \cwedge{1} f = 0 \, 
\ee  
\be \label{eq:EisenhartDuval_ansatz4_CKY}
\nabla_a f = - \frac{1}{\dm - p +1} e_a \wedge \delta f \, . 
\ee  
 Eq.\eqref{eq:EisenhartDuval_ansatz4_CKY} is the closed conformal Killing-Yano equation on the base manifold, this is linked to the fact that $\dm - p + 1 = (\dm +2) - (p+2) + 1$. The other conditions, according to the results of sec.\ref{sec:Dirac_equation_with_flux}, imply that for $p$ even it is possible to build a symmetry operator of the Dirac equation with $V$ and $A$ flux on $\M$. Similarly to the case of Ansatz 1, these conditions are related to  the dimensional reduction to $\M$ of a symmetry operator on $\h{\M}$. In this case too the conditions are stronger than those found in sec.\ref{sec:Dirac_equation_with_flux}.

\subsection{Another type of Eisenhart lift \label{sec:n_plus_one_Eisenhart_lift}}  
Eisenhart in his seminal paper \cite{Eisenhart1928} introduced a different type of geometric lift for the case of a natural Hamiltonian with no magnetic field and which is independent of time.  In his own words this is "treated not as a special case of the more general theory, but on an independent basis". For a natural Hamiltonian $H = \frac{1}{2} g^{\mu\nu} p_\mu p_\nu + V(q)$ one considers the $\dm +1$ dimensional metric 
\be \label{eq:Eisenhart_metric_2nd_type}
\hat{g} = g_{\mu\nu} \, dq^\mu dq^\nu +  \frac{dy^2}{2V} \, , 
\ee 
 where a new coordinate $y$ has been introduced, an we write $\hat{q} = (q^\mu, y)$. The metric will be globally well defined is $V > 0$  everywhere, or if it is bounded from below using the fact that $V$ is defined modulo a constant. In case this does not happen, the metric will only be defined in some region, differently from the previous, Lorentzian type of Eisenhart lift, which is always well defined. 
 
We can consider the associated higher dimensional free Lagrangian 
\be 
\hat{\mathcal{L}} = \frac{1}{2} \hat{g}_{MN} \dot{\hat{q}}^M \dot{\hat{q}}^N = \frac{1}{2} g_{\mu\nu} \dot{q}^\mu \dot{q}^\nu + \frac{\dot{y}^2}{4V} \, . 
\ee 
From this we calculate the momentum $p_y = \frac{\dot{y}}{2V}$ and the geodesic Hamiltonian 
\be \label{eq:Eisenhart_Hamiltonian_2nd_type}
\hat{H} = \frac{1}{2} \hat{g}^{MN} \hat{p}_M \hat{p}_N = \frac{1}{2} g^{\mu\nu} p_\mu p_\nu + p_y^2 \, V \, . 
\ee
 Since the coordinate $y$ is ignorable in \eqref{eq:Eisenhart_metric_2nd_type} then $p_y$ is constant. Then, if we set $p_y = 1$ in \eqref{eq:Eisenhart_Hamiltonian_2nd_type} and follow the evolution of the $(q, p_q)$ variables then we recover the original system. Notice how the metric \eqref{eq:Eisenhart_metric_2nd_type} is not necessarily Lorentzian as in the previous Eisenhart lift of section \ref{sec:Eisenhart-Duval}, nor its geodesics necessarily null. For example, for a potential $V$ that is always positive the metric will be Riemannian.  Changing $p_y$ is equivalent to an overall rescaling of the potential in the original system. 
 
Now suppose there is a  quantity $K$ for the original system that is polynomial in the momenta and of degree $p$: 
\be \label{eq:Eisenhart_2nd_type_conserved_quantity}
K = \sum_{i=0}^p K_{(i)} \, , 
\ee 
where 
\be 
K_{(i)} = \frac{1}{i!} K_{(i)}^{\lambda_1 \dots \lambda_i} p_{\lambda_1} \dots p_{\lambda_i} \, . 
\ee 
We lift it to the following homogeneous quantity 
\be 
\hat{K} = \sum_{i=0}^p K_{(i)} p_y^{p-i} \, . 
\ee 
This will be conserved in higher dimension if and only if 
\ba 
0 = \{ \hat{K} , \hat{H} \} = \frac{\partial \hat{K}}{\partial q^\mu} \frac{\partial \hat{H}}{\partial p_\mu} - \frac{\partial \hat{K}}{\partial p_\mu} \frac{\partial \hat{H}}{\partial q^\mu} \, . 
\ea 
Rearranging the terms in powers of $p_y$, and writing $H_{(f)} = \frac{1}{2} g^{\mu\nu} p_\mu p_\nu$ we get 
\ba 
&& \{ \hat{K} , \hat{H} \} = - \frac{\partial K_{(1)}}{\partial p_\mu} \frac{\partial V}{\partial q^\mu} p_y^{p+1} \nn \\ 
&& + \sum_{i=0}^{p-2} \left( \frac{\partial K_{(i)}}{\partial q^\mu} \frac{\partial H_{(f)}}{\partial p_\mu} - \frac{\partial K_{(i)}}{\partial p_\mu} \frac{\partial H_{(f)}}{\partial q^\mu} - \frac{\partial K_{(i+2)}}{\partial p_\mu} \frac{\partial V}{\partial q^\mu} \right) p_y^{p-i} \nn \\ 
&& + \sum_{i=p-1}^{p} \left( \frac{\partial K_{(i)}}{\partial q^\mu} \frac{\partial H_{(f)}}{\partial p_\mu} - \frac{\partial K_{(i)}}{\partial p_\mu} \frac{\partial H_{(f)}}{\partial q^\mu}  \right) p_y^{p-i} \, . 
\ea 
Since the metric $g$ is covariantly constant the $\partial H_{(f)} /\partial q$ terms can be exchanged with Levi-Civita connection terms, and the equation can be re-written in terms of covariant quantities as 
\ba 
&& \{ \hat{K} , \hat{H} \} = - K_{(1)}^\mu  \nabla_\mu V  p_y^{p+1} \nn \\ 
&& + \sum_{i=0}^{p-2} \left( \frac{1}{i!} \nabla_{(\lambda_1} K_{(i) \lambda_2 \dots \lambda_{i+1} )} \right. \nn \\ 
&& \hspace{0.5cm} \left. - \frac{1}{(i+1)!} K_{(i+2) \lambda_1 \dots \lambda_{i+1}}{}^\rho \nabla_\rho V \right) p^{\lambda_1} \dots p^{\lambda_{i+1}} \, p_y^{p-i} \nn \\ 
&& +\sum_{i=p-1}^{p}  \frac{1}{i!} \nabla_{(\lambda_1} K_{(i) \lambda_2 \dots \lambda_{i+1} )}  p^{\lambda_1} \dots p^{\lambda_{i+1}} \, p_y^{p-i}  \, . 
\ea 
This will be zero for a generic geodesic if and only if each term with a different power in $p_y$ is zero separately. The conditions obtained are also found to be the conditions that make \eqref{eq:Eisenhart_2nd_type_conserved_quantity} conserved in the original system. These equations are exactly the generalised Killing equations with flux \eqref{eq:generalised_Killing_equations} in the limit of zero field strength, $F = 0$.  Crampin in 1984 had discovered a special case of these equations when the only non-vanishing terms in \eqref{eq:Eisenhart_2nd_type_conserved_quantity} are $K_{(p)}$ and $K_{(0)}$, one application of his results being the Runge-Lenz quadratic conserved quantity for the Kepler problem \cite{Crampin:1984}.  
  
This second kind of Eisenhart lift is particularly interesting for the following reason. Suppose the potential $V$ in the Hamiltonian can be split into $m$ separate potentials, $V (q) = \sum_{i=1}^m g_i V_i (q)$, where the $g_i$ are coupling constants. Then one can separately lift each coupling constant and consider an enlarged $\dm + m$-dimensional space with a \textit{generalised Eisenhart lift metric} 
\be \label{eq:Eisenhart_2nd_type_generalised_metric}
\hat{g} = g_{\mu\nu} \, dq^\mu dq^\nu + \sum_{i=1}^m \frac{dy_i^2}{2 g_i V_i} \, . 
\ee 
This will yield the higher dimensional Hamiltonian 
\be \label{eq:Eisenhart_2nd_type_generalised_Hamiltonian}
\hat{H} = \frac{1}{2} g^{\mu\nu} p_\mu p_\nu + \sum_{i=1}^m p_{y_i}^2 \, g_i V_i \, . 
\ee
Geodesics in the higher dimensional space for different values of the conserved momenta $p_{y_i}$ correspond to solutions of the original equations of motion for different values of the coupling constant. In particular, if there are non-trivial isometries of \eqref{eq:Eisenhart_2nd_type_generalised_metric} that are not just isometries of $g$, then these give rise to non-trivial transformations between the $q$ and $y$ variables that leave the energy of the original system constant. Since $\dot{y_i} = 2 g_i V_i p_{y_i}$, a change in the $y$ variables in general will also amount to a change in the $p_{y_i}$, and therefore to a transformation that changes the coupling constants of the original system. This has been applied to the generalised Eisenhart lift of the Toda chain in \cite{GaryMarco2013}. The Toda chain lift will be described in more detail in section \ref{sec:geodesics_lie_groups}.

\section{HIDDEN SYMMETRIES, GRAVITY AND SPECIAL GEOMETRIES\label{sec:special_geometries}}  
 
\subsection{Taub-NUT and generalised Runge-Lenz vector}

The self-dual Taub-NUT 4-dimensional metric has found several applications, among which one is that it can be identified with a gravitational instanton, since it satisfies the Euclidean Einstein equations with self-dual Riemann tensor when the mass parameter is positive,  \cite{Hawking1977gravitational}. Another is its application to the study of the interaction between two well separated monopoles BPS monopoles in Euclidean flat space when the mass parameter is negative. The BPS monopoles are solitonic solutions of $SU(2)$ Yang-Mills theory that for large separations behave like particles. Manton has observed in 1982 that the low energy dynamics of these monopoles could be described by an appropriate evolution on the parameter space of static multi-monopole solutions, the evolution being geodesic with respect to the restriction of the kinetic energy to the moduli space \cite{Manton1982}. For the scattering of two monopoles the full moduli space metric is the Atiyah-Hitchin metric \cite{AtiyahHitchin1985}, that in the limit of well separated monopoles reduces to the Taub-NUT metric with negative mass parameter. Gibbons and Manton showed that this metric admits a conserved vector of the Runge-Lenz type \cite{Gibbons:1986df}, Feher and Horvathy studied its dynamical symmetries, showing that together with angular momentum it generates an $O(4)$ or $O(1,3)$ algebra analogous to that of the Kepler problem, in particular allowing to calculate the bound-state spectrum and the scattering cross section \cite{Feher1987dynamical,Feher1987erratum}. More generic Kepler-type dynamical symmetries applied to the study of monopole interactions have been considered in \cite{Feher1988non,Cordani1988,Cordani1990kepler}. The study of the Dirac equation in Taub-NUT space has been considered for example in  \cite{Comtet1995dirac,vanHolten1995supersymmetry}, as well as supersymmetry of monopoles and vortices \cite{Horvathy2006dynamical}. Dynamical symmetries for the analogous problem of well separated monopoles in hyperbolic space have been studied in \cite{Gibbons:2006mi}. 
 
The Taub-NUT with negative mass is a four-dimensional hyper-K\"{a}hler manifold with metric 
\be 
g = V \left( dr^2 + r^2 \left( \sigma_1^2 + \sigma_2^2 \right) \right) + 4 \, V^{-1} \sigma_3^2 \, , 
\ee 
where $V = 1 - \frac{2}{r}$ and the $\sigma_i$ are $SU(2)$ left invariant forms given by 
\ba 
\sigma_1 &=& \sin\theta \sin\psi d\phi + \cos\psi d\theta \, , \nn \\ 
\sigma_2 &=& \sin\theta \cos\psi d\phi - \sin\psi d\theta \, , \nn \\ 
\sigma_3 &=& d\psi + \cos\theta d\phi \, , 
\ea 
with $0 \le \phi < 2 \pi$, $0 \le \theta < \pi$, $0 \le \phi < 2 \psi$. 
  
The geodesic Lagrangian is 
\be 
L = \frac{1}{2} \left[ V \dot{\vec{r}}^{\, 2} + 4 V^{-1} \left( \dot{\psi} + \cos\theta \dot{\phi} \right)^2 \right] \, . 
\ee 
There is a number of expected conserved quantities: one is related to the fact the coordinate $\psi$ is ignorable, 
\be 
Q =  4 V^{-1} \left( \dot{\psi} + \cos\theta \dot{\phi} \right) \, , 
\ee 
then there is the energy 
\be 
E = \frac{1}{2}  V \left( \dot{\vec{r}}^{\, 2} + \frac{Q^2}{4} \right) \, , 
\ee 
and the angular momentum 
\be 
\vec{J} = V \vec{r} \times \dot{\vec{r}} + q \hat{r} \, . 
\ee 
There are however extra conserved quantities that are of higher order in the velocities 
\be 
\vec{A} = V \dot{\vec{r}} \times \vec{J} + \left( 2 E - \frac{Q^2}{2} \right) \hat{r} \, . 
\ee 
These are associated to a triplet of rank-2 Killing tensors $K^{\mu\nu}_{(i)}$. Moreover, the Killing tensors can be obtained from a rank-2 Killing-Yano tensor $Y_{\mu\nu}$ and three complex structures $F_{(i) \mu\nu}$, $i=1,2,3$, according to 
\be 
K_{(i)}^{\mu\nu} = Y^{\lambda (\mu} F_{(i) \lambda} {}^{\nu)} \, . 
\ee 
$\vec{J}$ and $\vec{A}$ generate an $\mathfrak{su}(2) \otimes \mathfrak{su}(2)$ Lie algebra. Explicitly 
\be 
Y = r (r-1) (r-2) \sin\theta d\theta \wedge d\phi + 2 dr \wedge (d\psi + \cos\theta d\phi ) \, , 
\ee 
and 
\be 
F_{(i)}= 2 (d\psi + \cos\theta d\phi ) \wedge dx_i + \frac{V}{2} \epsilon_{ijk} dx_j \wedge dx_k \, . 
\ee

\subsection{Higher dimensional rotating black holes} 
In this section we present some recent results in gravitational physics that have attracted much attention to the subject of hidden symmetries of the dynamics. Solutions of Einstein's equations in higher dimensions that generalise the Kerr rotating black hole, with in addition a cosmological constant and NUT parameters, have been found to possess a tower of Killing-Yano and Killing tensors, associated to multiple hidden symmetries of the dynamics that make several equations of physical interest, among which Hamilton-Jacobi, Klein-Gordon, Dirac, integrable and actually separable when written in the so called \textit{canonical coordinates}. We begin with a brief historical survey of the special properties of the Kerr solution, and then of higher dimensional black hole geometries. After that  we will present the construction of hidden symmetries for Kerr-NUT-(A)dS black holes and the more general off-shell \textit{canonical metrics}.

\subsubsection{The Kerr metric} 
In 1963 Kerr presented a solution of Einstein's equations for a spinning mass, the celebrated Kerr black hole \cite{Kerr1963}. This metric is astrophysically relevant since it can describe observed rotating black holes. Although the metric is non-trivial, it is endowed with properties that Chandrasekhar called 'miraculous' \cite{Chandrasekhar1984,Chandrasekhar1984mathematical}. 
 
It describes a rotating black holes that is stationary and axisymmetric, with two associated Killing vectors $\partial_t$, $\partial_\phi$. Carter in 1971 showed that, provided a black hole had an axis of symmetry, then it had to be described by the Kerr solution which depends only on two parameters, the mass and angular momentum \cite{Carter1971}, and Robinson in 1975 that it was the only pseudo-stationary black hole solution with a non-degenerate horizon \cite{Robinson1975}. The Weyl tensor of the metric is algebraically special and of type D in Petrov's classification \cite{Petrov2000,Petrov1969}. The metric can be written in Kerr-Schild form 
\be 
g_{\mu\nu} = \eta_{\mu\nu} + 2 H l_\mu l_\nu \, , 
\ee 
where $\eta$ is a flat metric and the vector $l$ is null with respect to both $g$ and $\eta$ \cite{KerrSchild1965,DebneyKerrSchild1969}. 
 
Going back to hidden symmetries, Carter in 1968 that showed the Hamilton-Jacobi and Klein-Gordon equations are separable in the Kerr metric \cite{Carter1968global,Carter1968hamilton}, thus reducing the problem of calculating geodesics to one of quadratures. In 1970 Walker and Penrose gave an explanation for that, showing that the Kerr metric admits a rank two Killing tensor, which can be used to build an extra conserved quantity in addition to the energy, angular momentum and the particle's rest mass \cite{WalkerPenrose1970}; this quantity is commonly referred to as Carter's constant. Subsequently, Penrose and Floyd showed in 1973 that the Killing tensor can be obtained from a Killing-Yano tensor \cite{Floyd1974,Penrose1973}. It was also demonstrated that the Killing-Yano tensor is responsible for several of the special properties of the Kerr metric: Hughston and Sommers showed that both isometries in the metric can be generated from the Killing-Yano tensor \cite{HughstonSommers1973} (see sec.\ref{sec:higher_dimensional_black_holes} below for a discussion in the $\dm$-dimensional case). This implies that all the conserved quantities responsible for integrability of geodesic motion are derivable from the Killing-Yano tensor. Colinson showed in 1974 that the integrability conditions for a non-degenerate Killing-Yano tensor imply that the spacetime is of type D in Petrov's classification \cite{Colinson1974}. Other types of important equations of physics simplify considerably in the Kerr metric:  Teukolsky in 1972 showed that the electromagnetic and gravitational perturbation equations decouple and their master equations separate \cite{Teukolsky1972rotating}, Teukolsky and Unruh showed separation for the massless neutrino equations \cite{Teukolsky1973perturbations,Unruh1973separability}, Chandrasekar and Page discussed the massive Dirac equation \cite{Chandrasekhar1976solution,Page1976dirac}. As will be shown later in this section, the Kerr metric satisfies the geometrical conditions described in sec.\ref {sec:Hamilton-Jacobi} for the separability of the Hamilton-Jacobi equations, and being a vacuum metric it also automatically allows for separation of variables of the Klein-Gordon equation, as will be discussed in sec.\ref{sec:Klein_Gordon}. The $\dm$-dimensional generalisations described in sec.\ref{sec:higher_dimensional_black_holes} also allow for separability of the Klein-Gordon equation, even when the Ricci tensor is non-zero, since the Robertson condition discussed in sec.\ref{sec:Klein_Gordon} holds, as will be shown below. The separability of the Dirac equation in these metrics is discussed in detail in sec.\ref{Dirac:KerrNUTAdS}, where it is shown it can be associated to appropriate symmetry operators built using the Killing-Yano (principal) tensor, the semi-classical counterpart of this result being the symmetry operators for the spinning particle theory described in sec.\ref{sec:examples_spinning_particle}. 
Similar symmetry operators for other non-scalar equations have been discussed in the literature, see \cite{Kamran1984symmetry,Kamran1985separation,Castillo1988separability,Kalnins1989killing,Kalnins1996intrinsic}. Jezierski and Lukasik discussed conserved quantities in the Kerr geometry generated by the Killing-Yano tensor.

\subsubsection{Black holes in higher dimensions\label{sec:higher_dimensional_black_holes}} 
In this section we describe Kerr-NUT-(A)dS black holes, which are higher dimensional black holes with rotation parameters, cosmological constant and NUT charges. Such solutions of the Einstein equations became ever more relevant in recent times with the advent of String Theory and of of braneworld cosmological models. An example of application is that of the AdS/CFT correspondence, where one considers their BPS limit. In the odd dimensional case this limit yields Sasaki-Einstein spaces \cite{Hashimoto2004sasaki,Cvetic2009new,Cvetivc2005new}, while in the even dimensional one Calabi-Yau spaces \cite{Oota2006explicit,Lu2007resolutions}. Attempts have been made to find similar solutions to the Einstein-Maxwell equations analytically  \cite{Aliev2004five,Aliev2006slowly,Aliev2006rotating,Aliev2007electromagnetic,Kunz2006rotating, Chen2008kerr} or numerically \cite{Kunz2005five,Kunz2006charged,Kunz2007higher,Brihaye2007charged,Kleihaus2007rotating}.  See also \cite{Charmousis2004axisymmetric,Podolsky2006robinson,Pravda2007type,Ortaggio2008robinson,Chen2006general,Lu2009new}   . 
 
Historically however, it was already in 1963 that Tangherlini described a higher dimensional black hole solution that generalised the Schwarzschild metric \cite{Tangherlini1963schwarzschild}. Myers and Perry in a 1986 paper \cite{Myers1986black} added an electric charge to the Tangherlini solution and also obtained another solution, nowadays referred to as the Myers-Perry metric, which is higher dimensional and has $[(\dm - 1)/2]$ rotation parameters, corresponding to the independent planes of rotation, thus generalising Kerr's solution. Frolov and Stojkovi\'{c} later showed that the 5 dimensional Myers-Perry metric admits a Killing tensor, anticipating the more general result on Kerr-NUT-(A)dS metrics that we will discuss in this section \cite{FrolovStojkovic:2003b}. Galajinsky, Nersessian and Saghatelian discussed superintegrable spherical mechanics models associated with near horizon extremal Myers-Perry black holes in arbitrary dimensions \cite{Galajinsky2013_near_horizon}. 
 
The first example of black hole solution in higher dimension with cosmological constant was provided by Hawking, Hunter and Taylor-Robinson \cite{Hawking1999rotation}, who obtained a 5-dimensional metric  with rotation parameters and cosmological constant, that generalises the Kerr-(A)dS metric of \cite{Carter1968hamilton}. The generalisation of this metric to arbitrary number of dimensions was obtained by Gibbons, L\"{u}, Page and Pope \cite{Gibbons2004rotating,Gibbons2005general} in 2004, and finally NUT charges where added by Chen, L\"{u} and Pope in 2006 \cite{Chen2006general,Chen2007kerr,Chong2005separability}. Good reviews on the subject of black holes in higher dimensions and hidden symmetries of dynamics are \cite{EmparanReall:2008,Yasui2011hidden}; another review is currently in preparation \cite{DavidEtAlInPreparation2014}. 
 
We describe here directly the full Kerr-NUT-(A)dS solution. While it is known that in higher dimensional gravity there exist various kinds of rotating black objects, the Kerr-NUT-(A)dS metric describes the most general rotating black object with spherical horizon topology and cosmological constant that solve Einstein's vacuum equations and that are known analytically, to date. In fact, we present the metric of what are known as canonical spacetimes: these belong to a family of spacetimes that are in general off shell, in that they do not solve the vacuum Einstein equations, and when they do they reduce to the Kerr-NUT-(A)dS spacetimes. Canonical spacetimes are the natural generalisation of Kerr-NUT-(A)dS spacetimes with respect to their geometrical properties and to the hidden symmetries of the dynamics of non-backreacting fields propagating on them. We will work in Riemannian signature as this makes the metric more symmetric. 
 
It is useful to parameterise the spacetime dimension $\dm$ as $\dm = 2N + \epsilon$, where $\epsilon = 0,1$. Then the metric is given by 
\be  \label{metric}
ds^2
  = \sum_{\mu=1}^N\biggl[ \frac{d x_{\mu}^{\;\,2}}{Q_\mu}
  +Q_\mu\Bigl(\,\sum_{j=0}^{N-1} \A{j}_{\mu}d\psi_j \Bigr)^{\!2}  \biggr]  + \eps S \Bigl(\,\sum_{j=0}^N \A{j}d\psi_j \Bigr)^{\!2}.
\ee
Here, coordinates $x_\mu\, (\mu=1,\dots,N)$ stand for the radial coordinate and longitudinal angles, and Killing
coordinates $\psi_k\; (k=0,\dots,N-1 +\eps)$ denote time and azimuthal angles associated with Killing vectors
${\KV{k}} =\cv{\psi_k}$\footnote{In this section and in other sections where the Kerr-NUT-(A)dS metric is used, we indicate generic indices with $a,b, c, \dots$ and adopt the split of spacetime coordinates indices into those of type $\mu$ and type $i$. This is different from the convention used in the rest of the review, where greek letters $\mu, \nu, \dots$ refer to a generic spacetime index. However, the split adopted for Kerr-NUT-(A)dS metrics occurs frequently in the literature and we feel that the benefit of using a notation that can be directly compared with that of existing works outweighs the local loss of consistency in notation. We also suspend the Einstein sum convention when applied to indices of the type $\mu$ or $i$}. The metric functions are defined by 
\ba
Q_\mu&=&\frac{X_\mu}{U_\mu}\,,\quad U_{\mu}=\prod\limits_{\nu\ne\mu} (x_{\nu}^2-x_{\mu}^2)  \;,\quad S = \frac{-c}{\A{N}} \, ,\label{eq:UandS_def}\\
\A{k}_{\mu}&=&\hspace{-5mm}\!\!
    \sum\limits_{\substack{\nu_1,\dots,\nu_k\\\nu_1<\dots<\nu_k,\;\nu_i\ne\mu}}\!\!\!\!\!\!\!\!\!\!
    x^2_{\nu_1}\cdots\, x^2_{\nu_k}\;,\ \
\A{k} = \hspace{-5mm} \sum\limits_{\substack{\nu_1,\dots,\nu_k\\\nu_1<\dots<\nu_k}}\!\!\!\!\!\!
    x^2_{\nu_1}\cdots\, x^2_{\nu_k}\; .\label{eq:A_def}\quad
\ea
The quantities ${X_\mu}$ are generic functions of a single variable ${x_\mu}$, and $c$ is an arbitrary constant.
The vacuum black hole geometry with a cosmological constant is recovered by setting
\begin{equation}\label{BHXs}
  X_\mu = \sum_{k=\eps}^{N}\, c_{k}\, x_\mu^{2k} - 2 b_\mu\, x_\mu^{1-\eps} + \frac{\eps c}{x_\mu^2} \; .
\end{equation}
This choice of $X_\mu$ describes the most general known Kerr-NUT-(A)dS spacetimes in all dimensions \cite{Chen2006general}. The constant $c_N$ is proportional to the cosmological constant and the remaining constants are related to angular momenta,
mass and NUT parameters. The coordinates used here generalise the \textit{canonical coordinates} introduced by Carter in 4D  \cite{Carter1968,Debever1971,Plebanski1975}.

These metrics admit a special type of closed conformal Killing-Yano tensor: the tensor 
\be
h=db\,,\quad b=\frac{1}{2}\sum_{j=0}^{N-1}A^{(j+1)}d\psi_j\,.
\ee  
is closed, of rank two and non-degenerate, meaning that it has rank $2N$. This defines a \textit{principal conformal Killing-Yano tensor}. Frolov and Kubiz\v{n}\'ak showed that the principal conformal Killing-Yano tensor exists for Myers-Perry metrics \cite{DavidValeri2006} and Kerr-NUT-(A)dS ones in \cite{DavidValeri2006KerrNUTAdS}.  Kubiz\v{n}\'ak and Krtou\v{s} studied the four dimensional Pleba\'nski-Demia\'nski family of metrics, which includes the Kerr metric, and found a conformal generalisation of the Killing-Yano tensor that reduces to the Killing-Yano tensor in absence of acceleration \cite{DavidPavel2007conformal}. Jezierski \cite{Jezierski1997} showed that for a rank-2 conformal Killing-Yano tensor $h$ the vector 
\be 
\xi^a = \frac{1}{\dm -1} \nabla_b h^{ba} \, , 
\ee 
where $a=1, \dots, \dm$, satisfies 
\be 
\nabla_{(a} \xi_{b)} =  \frac{1}{\dm-2} R_{c(a} h_{b)} {}^c \, , 
\ee 
this being a consequence of the integrability condition of the conformal Killing-Yano equation for $h$. 
Therefore for Einstein spaces like Kerr-NUT-(A)dS $\xi$ is a Killing vector. In fact, it can be proved that $\xi$ is a Killing vector also for the canonical metrics that are not on-shell \cite{DavidValeriPavel2008}. $\xi$ is called \textit{primary Killing vector} for these metrics.

Riemannian metrics that admit a principal conformal Killing-Yano tensor with independent, non-constant eigenvalues have been classified \cite{DavidValeriPavel2008,HouriOotaYasui2007}, and are exactly given by the canonical spacetimes above. 
 
Schematically, the classification works as follows \cite{DavidValeri2008,DavidValeriPavel2008}. Since $h$ is a closed conformal Killing-Yano tensor, using eq.\eqref{eq:symplectic_CKY_definition_0} it can be seen that the tensor $H^{\mu} {}_\nu = h^{\mu} {}_{\rho} h_\nu{}^\rho$ is a conformal Killing tensor, see the definition in eq.\eqref{eq:symplectic_conKS}. Its eigenvalues are real and non-negative, we indicate them by $\{ x_i^2 \}$. Then there exists an orthonormal basis such that $h$ can be written as $diag(\Lambda_1, \dots, \Lambda_p)$ for $\epsilon = 0$, and $diag(0,\Lambda_1, \dots, \Lambda_p)$ for $\epsilon = 1$, where 
\be 
\Lambda_i = \left( \begin{array}{cc} 0 & - x_i I_i \\ x_i I_i & 0 \end{array} \right) \, , 
\ee 
and the $I_i$ are unit matrices. Such a basis is called a Darboux basis, see \cite{Prasolov1994}. When the eigenvalues are all different we denote them by $x_\mu$, $\mu = 1, \dots, N$, and the associated orthonormal eigenvectors pairs by $E_\mu$, $E_{\hat{\mu}}$. When $\epsilon = 1$ there is a further eigenvector $E_0$ associated to the zero eigenvalue. Let  $E^\mu$, $E^{\hat{\mu}}$ and $E^0$ be the associated 1-forms. Then in this basis the metric and principal conformal Killing-Yano tensor take the form 
\ba 
g &=& \sum_{\mu=1}^N \left( E^\mu \otimes E^\mu + E^{\hat{\mu}} \otimes E^{\hat{\mu}} \right) + \epsilon E^0 \otimes E^0 \, , \nn \\ 
\label{eq:h_canonical_basis}
h &=& \sum_{\mu=1}^N x_\mu E^\mu \wedge E^{\hat{\mu}} \, . 
\ea 
This is a local description of the metric and the principal conformal Killing-Yano tensor. In order to move from a local description to a description that holds in a finite spacetime domain, Frolov, Krtou\v{s} and Kubiz\v{n}\'ak  in \cite{DavidValeriPavel2008} additionally include in the definition of the tensor $h$ that its eigenvalues $x^\mu$ are functionaly independent and non-constant, so that they can be used as coordinates. Then they are able to show that they can be complemented by another set of coordinates $\psi_k$, such that the $\partial_{\psi_k}$ are Killing vectors. In particular, $\partial_{\psi_0} = \xi$.  
 
In \cite{DavidValeri2006,DavidValeri2006KerrNUTAdS} Frolov, Krou\v{s}, Kubiz\v{n}\'ak and Page showed that any spacetime with a principal Killing-Yano tensor $h$ admits a tower of Killing-Yano and Killing tensors. The simplest way to see this is to use the fact mentioned in sec.\ref{sec:symplectic_special_tensors} that the wedge product of closed conformal Killing-Yano tensor is still a closed conformal Killing-Yano tensor. Therefore it is possible to build the following tower of closed conformal Killing-Yano tensors: 
\be\label{eq:h_tower}
{h}^{(j)}\equiv {h}^{\wedge j}=\underbrace{{h}\wedge \ldots \wedge
{h}}_{\mbox{\tiny{total of $j$ factors}}}\, , 
\ee
where ${h}^{(j)}$ is a $(2j)$-form, and ${h^{(0)}=1}$, ${h}^{(1)}={h}$. Crucially, since ${h}$ is non-degenerate it is possible to build a set of $N$ non-vanishing forms. When $\epsilon = 0$  then ${h}^{(N)}$ is proportional to the totally antisymmetric
tensor, while for $\epsilon = 1$ it is dual to a Killing vector in odd dimensions. In both cases
${h}^{(N)}$ is trivial and can be excluded from the tower, and we take $j=1,\dots, N-1$.
 
According to the results mentioned in sec.\ref{sec:symplectic_special_tensors} these tensors give rise to the Killing tensors
${K}^{(j)}$
\ba\label{Kj}
K^{(j)}_{ab} &\equiv & \frac{1}{(n-2j-1)!(j!)^2} \nn \\ 
&& \hspace{0cm} (*h^{(j)})_{a c_1\ldots c_{n-2j-1}}
(*h^{(j)})_b {}^{c_1\ldots c_{n-2j-1}}\, , 
\ea 
where $(*h^{(j)})$ is the Hodge dual of $h$, a Killing-Yano tensor. 
The coefficient in this definition is chosen so that 
\ba
{K}^{(j)}&=&\sum_{\mu=1}^N A^{(j)}_\mu\bigl({E}^{\mu}\otimes{E}^{\mu} + 
{\tilde E}^{\mu}\otimes{\tilde E}^{\mu}\bigr) \nn \\ 
&& + \eps A^{(j)}E^0\otimes E^0\,.
\ea
We also define ${K}^{(0)}=g$, so that we can take the range of $j$ as $j=0,\dots N-1$. The Killing tensors are functionally independent and in involution, and together with the $N+\epsilon$ Killing vectors $\partial_{\psi_i}$ form a complete set of integrals of motion for the geodesic motion, thus explaining the complete integrability of geodesic motion and the separability of the Hamilton-Jacobi equation in these metrics \cite{DavidEtAl2007,DavidEtAl2007complete,DavidEtAl2007constants,DavidEtAl2007separability}. Moreover, since they are constructed by taking the square of Killing-Yano tensors, they satisfy the Robertson condition that will be defined in sec.\ref{sec:Schrodinger_general_theory}, and as discussed there this implies that the Klein-Gordon equation is separable as well \cite{DavidEtAl2007separability}. The Killing vectors can all be obtained from the principal Killing vector and the		 Killing tensors according to the formula \cite{DavidEtAl2007,DavidValeriPavel2008} 
\be 
\left( \partial_{\psi_j} \right)^a = K^{(j) a} {}_b \xi^b \, , \quad j = 1, \dots, N-1 \, . 
\ee

Connell, Frolov and Kubiz\v{n}\'ak and Krtou\v{s} showed that in spacetimes with an arbitrary number of dimensions that admit a non-degenerate principal conformal Killing-Yano tensor one can also solve the parallel transport equations by reducing them to a set of first order ordinary differential equations \cite{Connell2008parallel,Connell2009parallel}. Kubiz\v{n}\'ak \cite{Kubizvnak2009supersymmetric} studied a scaling limit of the Kerr-NUT-(A)dS metrics that yields Einstein-Sasaki spaces originally constructed by L\"{u}, Chen and Pope in \cite{Chen2006general} that admit Killing spinors and a tower of Killing-Yano tensors of increasing rank.

Generalisations of the canonical metrics can be obtained for example relaxing the condition that $h$ is non-degenerate and that all of its eigenvalues are non-constant \cite{HouriOotaYasui2008,HouriOotaYasui2009}: then one gets bundles which have K\"{a}hler fibers over Kerr-NUT-(A)dS spaces. Another possible choice is that of considering covariant derivatives different from the Levi-Civita one, this appears naturally in supergravity theories and will be discussed in sec.\ref{sec:torsion_and_supergravity}. There is a number of very good reviews on the subject with much information and of different size: see \cite{DavidThesis2008,Frolov2012,DavidValeri2008higher,Yasui2011hidden,Frolov2008hidden}. See also the short review \cite{MarcoDavidPavel2012hidden}.

\subsection{Spacetimes with torsion and supergravity\label{sec:torsion_and_supergravity}} 
In sec.\ref{sec:higher_dimensional_black_holes} we discussed work classifying all Riemannian metrics admitting a principal  conformal Killing-Yano tensor. While this is a very satisfactory result, it does mean that to study Killing-Yano type symmetries beyond this class of metrics we have to loosen our definition to some extent. In this section we shall consider extending the notion of conformal Killing-Yano tensor by relaxing the condition that the connection in \eqref{eq:symplectic_CKY_definition} be the Levi-Civita connection. We shall see that this generalisation is a natural one in the context of certain supergravity theories. Certain of the properties of conformal Killing-Yano tensors transfer to this generalised setting, but there are some interesting differences.

\subsubsection{The connection with torsion and generalised conformal Killing-Yano forms\label{connection_with_torsion}}

Examining the definition of a conformal Killing-Yano tensor \eqref{eq:symplectic_CKY_definition}, there are two ingredients from the geometry of the underlying manifold. Clearly we require a connection (in this case the Levi-Civita connection) in order to make the left hand side meaningful. Less obviously, we also require a metric in order to make sense of the right hand side: in particular we need the metric to calculate $X^\flat$ and $\delta h$. These two objects are compatible, in the sense that the Levi-Civita connection preserves the metric. Let us now consider generalising the Levi-Civita connection $\nabla$ appearing in \eqref{eq:symplectic_CKY_definition}. Any two connections differ by a tensorial quantity, so we may define our new connection by $\nabla^T_X Y = \nabla_X Y + T(X, Y)$ where $T\in T^{1}_{2}(\mathcal{M})$. We will make the further assumption that $T$ is totally anti-symmetric. That is
\be
g(T(X, Y), Z) = -g(T(Y, X), Z) = -g(T(X, Z), Y)
\ee
for all vector fields $X, Y, Z$. A consequence of this assumption is that $\nabla^T$ is metric, i.e.\ $\nabla^T g = 0$, and furthermore the autoparallel curves of $\nabla^T$ are geodesics of $g$. Notice that we can identify $T$ with a three-form in an obvious way.

We're obviously able to generalise the left hand side of \eqref{eq:symplectic_CKY_definition} by simply changing $\nabla \to \nabla^T$. In order to generalise the right hand side, we need to briefly consider the representation theoretic origin of the conformal Killing-Yano equation. In general for a Riemannian manifold, 
one can decompose $T^*\mathcal{M} \otimes \Lambda^p T^*\mathcal{M}$ as an $O(n)$ representation as follows
\begin{equation}
T^*\mathcal{M} \otimes \Lambda^p T^*\mathcal{M} \cong \Lambda^{p+1}T^*\mathcal{M}\oplus \Lambda^{p-1} T^*\mathcal{M} \oplus \Lambda^{p,1} T^*\mathcal{M} 
\end{equation}
where  $\Lambda^{p,1} T^*\mathcal{M}$ consists of those elements $\alpha \otimes \psi$ 
of $T^*\mathcal{M} \otimes \Lambda^p T^*\mathcal{M} $ which satisfy $\alpha \wedge \psi = 0$, 
$\alpha^\sharp \hook \psi=0$. Applying this to $\nabla {h}$, 
one identifies the projection into $\Lambda^{p+1} T^*\mathcal{M}$ as the exterior differential $d {h}$ 
and the projection into $\Lambda^{p-1} T^*\mathcal{M}$ as the co-differential $\delta {h}$, 
up to multiples. The conformal Killing-Yano equation expresses 
the requirement that the component of $\nabla{h}$ transforming 
in the $\Lambda^{p,1}T^*\mathcal{M}$ representation vanishes. Motivated by this, we define
\be
d^T h = e^i \wedge \nabla^T_{X_i} h, \qquad \delta^T h = -X^i \hook \nabla_{X^T_i} h
\ee
to be the projections of $\nabla^T h$ into $\Lambda^{p+1} T^*\mathcal{M}$, $\Lambda^{p-1} T^*\mathcal{M}$ respectively. Here $X_i$ is a orthonormal basis for $T\mathcal{M}$ with a dual basis $e^i$. These definitions reduce to the standard definitions for $d, \delta$ when $T=0$.

Motivated by the above, we thus define a generalised conformal Killing-Yano tensor to be a $p-$form satisfying the generalised conformal Killing-Yano equation:
\be \label{eq:generalised conformal Killing-Yano_definition}
\nabla^T_X h = \frac{1}{\pi +1} X \hook d^T h - \frac{1}{\dm - \pi +1} X^\flat \wedge \delta^T h \, , 
\ee
for any vector $X$. If additionally $d^Th=0$, $h$ is a generalised Killing Yano tensor, while if  $\delta^Th=0$, $h$ is a generalised closed conformal Killing Yano tensor. 

Generalised conformal Killing-Yano forms were first introduced in this fashion in \cite{KubiznakEtal:2009b} to better understand the symmetries of the Chong-Cveti\'c-L\"u-Pope black holes \cite{ChongEtal:2005b}. Previous work which considered related objects includes \cite{BochnerYano, Wu:2009a}. 

\subsubsection{Properties of generalised conformal Killing-Yano tensors}
Similarly to the torsionless case, the existence of a generalised conformal Killing-Yano tensor has several nice consequences for the underlying spacetime. In particular \cite{Houri:2010fr,DavidClaudeHouriYasui2012}:
\begin{enumerate}[i)]
\item The generalised conformal Killing-Yano equation is invariant under Hodge duality: if $h$ is a generalised conformal Killing-Yano tensor, so is $\star h$.
\item The generalised conformal Killing-Yano equation is conformallly invariant: if a $k-$form $h$ is a generalised conformal Killing-Yano tensor of the metric $g$, with torsion $T$, then $\Omega^{k+1} h$ is a generalised conformal Killing-Yano tensor of the metric $\Omega^2 g$ with torsion $\Omega^2 T$.
\item Generalised conformal Killing-Yano tensors form a (graded) algebra under the wedge product.
\item If $h_1$, $h_2$ are generalised conformal Killing-Yano $k-$forms, then the symmetric tensor $K_{\mu\nu} = h_1{}_{(\mu | \rho_1, \ldots \rho_{k-1}} h_2{}_{\nu)}{}^{\rho_1, \ldots \rho_{k-1}}$ is a conformal Killing tensor.
\item If $h_1$, $h_2$ are generalised Killing-Yano $k-$forms, then the symmetric tensor $K_{\mu\nu} = h_1{}_{(\mu | \rho_1, \ldots \rho_{k-1}} h_2{}_{\nu)}{}^{\rho_1, \ldots \rho_{k-1}}$ is a Killing tensor.
\item generalised conformal Killing-Yano forms can be constructed from spinors $\psi$ which solve the twistor equation with torsion 
\be 
\nabla_X^T \psi - \frac{1}{N} X^\flat D^T \psi = 0 \, ,  
\ee 
where $D^T$ is the Dirac operator with torsion that will be defined shortly below. 
\end{enumerate}
In contrast to the situation for Killing-Yano tensors, the Killing tensor generated by squaring a generalised Killing-Yano need not give rise to a symmetry of the Klein-Gordon equation.

One key defining feature of conformal Killing-Yano tensors is their importance in constructing operators which commute with the Dirac equation. The situation when torsion is present is a little more subtle \cite{HouriEtal:2010a, DavidClaudePavel2010}. Since the torsion field naturally couples to particles' spin, it is perhaps 
not very surprising that the Dirac operator should pick up a torsion correction. 
It was argued in \cite{HouriEtal:2010a} that in the presence of torsion the natural Dirac operator to consider is 
\be \label{eq:natural_Dirac_torsion}
D^T=\gamma^\mu \nabla_\mu-\frac{1}{24}T_{\lambda\mu\nu}\gamma^{\lambda\mu\nu}\,.
\ee  
In the context of certain supergravities,  this is also a natural Dirac operator
to consider as it gives the equation of motion for the linearized
gaugino field in the string frame. See for example \cite{Benmachiche:2008ma, Strominger:1986uh}. This operator also occurs in the context of complex geometries with torsion \cite{ApostolovEtal:2006,Agricola2006srni}. 

It was shown in \cite{HouriEtal:2010a} that given a rank-$p$ generalised conformal Killing-Yano tensor ${h}$ and provided that the corresponding anomaly terms\footnote{%
It is only the first condition, ${A}_{(cl)}=0$, which emerges from the classical spinning particle approximation.  Correspondingly, we call ${A}_{(cl)}$ a  
{\em classical anomaly} and  ${A}_{(q)}$ (which appears only at the operator level) a `{\em quantum anomaly}'.
}
\ba
{A}_{(cl)}({h})\!\!&=&\!\!\frac{{d}(d^T {h})}{p+1}-\frac{{T}\wedge \delta^T {h}}{\dm-p+1}-\frac{1}{2}
{dT}\cwedge{1}{h}\,,\label{Acl}\\
{A}_{(q)}({h})\!\!&=&\!\!\frac{{\delta}(\delta^T {h})}{\dm-p+1}-\frac{1}{6(p+1)}{T}\cwedge{3}d^T {h}\nonumber \\&&+\frac{1}{12}{dT}\cwedge{3}{h}\,,\label{Aq}
\ea
vanish, one can construct an operator $L_{h}$ which (on-shell)
commutes with $D^T$, $[D^T,L_k]=0$. The $\cwedge{m}$ contraction is defined in eq.\eqref{cwedgedef}. Such an operator provides an
on-shell symmetry operator for a massless Dirac equation. When
${h}$ is in addition $d^T$-closed or $\delta^T$-coclosed the
operator $L_h$ may be modified to produce off-shell (anti)-commuting
operators $M_k$ or $K_k$. 

From this construction it is possible by a formal semi-classical limiting process to show that generalised conformal Killing-Yano tensors give rise to symmetries of the supersymmetric spinning particle mechanics. This can also be shown directly, as in \cite{KubiznakEtal:2009b}.

Finally, we note that a partial classification of metrics admitting non-degenerate generalised conformal Killing-Yano two forms, that are analogues of the principal  conformal Killing-Yano form, has been undertaken in \cite{DavidClaudeHouriYasui2012}. Matters are complicated here compared to the case without torsion as there is considerably more freedom inherent in the ability to specify a torsion.

\subsubsection{Examples of spacetimes admitting generalised conformal Killing-Yano tensors}

The Chong-Cveti\'c-L\"u-Pope black holes \cite{ChongEtal:2005b} are solutions of $D=5$ minimal gauged supergravity, whose bosonic sector is governed by the action
\be
S=\int_\mathcal{M} \star(R+\Lambda) - \frac{1}{2} F \wedge \star F + \frac{1}{3 \sqrt{3}} F \wedge F \wedge A.
\ee
It was demonstrated in \cite{Wu:2009a,Wu:2009b} that the Hamilton-Jacobi and charged Klein-Gordon equation are separable for these black holes, as is a modified Dirac equation. The background admits a Killing tensor, however this cannot be expressed as the square of a conformal Killing-Yano tensor. To fully understand the hidden symmetry of this background, it is necessary to consider a generalised conformal Killing-Yano form. The three form $\star F$ furnishes a natural candidate for a torsion form. In fact, we set $\sqrt{3}T = \star F$. With this identification, a generalised conformal Killing-Yano tensor exists, whose square is the Killing tensor.

Another interesting class of spacetimes admitting generalised conformal Killing-Yano are the Kerr-Sen black holes. These extremise the low energy string theory effective action, which in string frame is given by: 
\begin{eqnarray}
S=-\int d^4x&&\sqrt{-g}e^{-\Phi}\bigl(-R+\frac{1}{12}H_{\lambda\mu\nu}H^{\lambda\mu\nu} \nonumber
\\&&-g^{ab}\partial_a\Phi\partial_b\Phi+\frac{1}{8}F_{\mu\nu}F^{\mu\nu}\bigr)\,.
\end{eqnarray}
Where $F, H$ are totally antisymmetric. 

The Kerr-Sen solutions may be found by a solution generating technique, starting with the well known Kerr solution \cite{Sen:1992, WuCai:2003}.  The string-frame fields are given by 
\begin{eqnarray}
ds^2\!&=&\!e^{\Phi}\Bigl\{-\frac{\Delta}{\rho_b^2}\bigl(dt-a\sin^2\!\theta d\varphi\bigr)^2\nonumber \\ 
&& \hspace{-.6cm} +\frac{\sin^2\!\theta}{\rho_b^2}\Bigl[a dt-(r^2+2br+a^2)d\varphi\Bigr]^2  +\frac{\rho_b^2}{\Delta}dr^2+\rho_b^2 d\theta^2\Bigr\}\,,\nonumber\\
{H}\!&=&\!-\frac{2ba}{\rho_b^4}\,{d}t\wedge{d}\varphi\wedge  \Bigl[\bigl(r^2-a^2\cos^2\!\theta\bigr)\sin^2\!\theta{d}r \nn \\ 
&& \hspace{2.5cm} -r\Delta\sin 2\theta {d}\theta\Bigr]\,,\nonumber\\
{A}\!&=&\!-\frac{Qr}{\rho_b^2}\bigl({d}t-a\sin^2\!\theta {d}\varphi\bigr)\,,\nonumber\\
\Phi\!&=&\! 2\ln \left(\frac{\rho}{\rho_b}\right)\,,
\end{eqnarray}
where
\begin{eqnarray}
&\rho^2=r^2+a^2\cos^2\theta\,,\ \quad\  \rho_b^2=\rho^2+2br&\,,\nonumber\\  &\Delta=r^2-2(M-b)r+a^2\,.&
\end{eqnarray}
 The solution describes a black hole with mass $M$, charge $Q$,
 angular momentum $J=Ma$, and magnetic dipole momentum $\mu=Qa$. When
 the twist parameter $b=Q^2/2M$ is set to zero, the solution reduces to the Kerr solution of vacuum general relativity.

The Kerr-Sen black hole admits a generalised conformal Killing-Yano tensor where the torsion is identified as the naturally occurring three-form $H$, \cite{Houri:2010fr}. As can be easily seen from the expression for the fields, when $b=0$ $H$ vanishes and the generalised conformal Killing-Yano reduces to the principal  conformal Killing-Yano tensor of Kerr.
	
As a result of the existence of the generalised conformal Killing-Yano tensor, the equations of motion for a charged particle moving in the Kerr-Sen background separate. Furthermore the natural linear scalar field equations separate, as does the Dirac equation \cite{Houri:2010fr,WuCai:2003}. The properties generalise to the higher dimensional analogues of the Kerr-Sen black hole found by Chow \cite{Chow:2008}.

\subsection{Killing-Yano tensors and G--structures} 
Some special geometries that are known in the literature present reduced holonomy, and among reduced holonomy spaces a number of these is known to admit conformal Killing-Yano tensors and conformal Killing-Yano tensors with torsion. Such spaces are of interest both from the intrinsic geometrical point of view, and as they naturally arise as solutions of gauged and ungauged supergravity theories, and can have potential application in the AdS/CFT correspondence. In this section we will limit ourselves to Riemannian signature. 
 
We will begin with a brief reminder of the concept of reduced holonomy. Given an $\dm$-dimensional  Riemmanian manifold $\M$ with a metric $g$ and a connection $\nabla$, parallel transport of tensors around closed loops will in general result in the tensors being changed by an $O(\dm)$ transformation, $SO(\dm)$ in case $\M$ is orientable. While in flat space the transformation would be trivially the identity, in a generic curved space this need not be so, and this is accounted for by the curvature of the connection. Such non-trivial transformations form a group, the \textit{holonomy group} of $\M$. It is possible to prove that the holonomy group is a closed Lie subgroup of $O(\dm)$, in particular compact \cite{BorelLichnerowicz1952}. A theorem by Ambrose and Singer relates the Lie algebra of the holonomy group to the transformations generated by the curvature 2--form of $\nabla$ \cite{AmbroseSinger1953}. When the holonomy group is strictly a subgroup of $SO(\dm)$ then it is said that the holonomy of $(\M, g, \nabla)$ is \textit{reduced}. Berger \cite{Berger1955} has given a complete classification of the possible holonomy groups for Riemannian manifolds that are simply connected, irreducible and non-symmetric. What is known as Berger's list includes the groups $U(m)$, $SU(m)$, $Sp(m)$, $Sp(m) \cdot Sp(1)$, $G_2$ and $Spin(7)$, which we will discuss in more detail in this section. On the other hand, Riemannian symmetric spaces are locally isometric to homogeneous spaces of the kind $G/H$, which have holonomy $H$. 
 
A concept related to reduced holonomy is that of a $G$-structure. Given the natural $Gl(\dm)$ frame bundle associated to $\M$ and a group $G$, a $G$-structure is a principal $G$-sub-bundle. In several cases the $G$-structure can be prescribed by a set of tensors that do not vanish at any point and that are invariant under the action of $G$, for example an almost complex structure $J$ determines a $Gl(\dm/2, \mathbb{C})$ structure. These cases are relevant for physics since manifold with special tensors appear naturally for examples in solutions of supergravity theories. In fact in this context the formalism of $G$-structures has evolved into a machinery that has been used to classify supergravity solutions. The literature in this respect is quite vast: for an initial list of examples, by no means complete, see \citep{Tod1983,Gauntlet2003,GauntlettNull2003,MacConamhna2004,CarigliaMacConamhna2005,OrtinEtAl2010} and referring works. 
 
There is a relation between $G$-structures given by a set of tensors and reduced holonomy when for examples such tensors are covariantly constant. As $G$ is a subgroup of $O(\dm)$ we can write in terms of Lie algebras $\mathfrak{so}(\dm) = \mathfrak{g} + \mathfrak{g}^\perp$, and decompose the Levi-Civita connection as 
\be 
\nabla = \nabla^\mathfrak{g} + \nabla^{\mathfrak{g^\perp}} \, , 
\ee 
where $ \nabla^{\mathfrak{g}}$ takes values in $\mathfrak{g}$ and, respectively, $ \nabla^{\mathfrak{g^\perp}}$ in $\mathfrak{g}^\perp$. If the tensors are covariantly constant then it must be $\nabla =  \nabla^\mathfrak{g}$, and the holonomy group is necessarily a subgroup of $G$. For example in the case of the groups in Berger's list each of them can be obtained as holonomy group for a given set of covariantly constant forms.

It is possible to generalise the above construction by considering $G$-invariant tensors that have non-zero covariant derivative. The covariant derivative of the $G$-invariant tensors can be decomposed into irreducible representations of $G$, and thus classified. This has been done for the first time , in the case of almost Hermitian manifolds, in \cite{GrayHervella1980}, then for for $G_2$ structures in \cite{GrayFernandez1982}, for $SU(\dm)$ structures in \cite{ChiossiSalamon2002} and for $Sp(m)$ structures in \cite{CabreraSwann2007}. There are  works that generalise these results:  Brozos-V\'azquez, Garc\'ia-R\'io, Gilkey and Hervella study manifolds with pseudo-Riemannian metric and discuss the pseudo-Hermitian and para-Hermitian cases \cite{GilkeyHervellaBrozosVazquezGarciaRio2010}, while Barberis, Dotti and Santillan study properties of the Killing-Yano equation on Lie groups \cite{BarberisDottiSantillan2012}. Santillan has analysed the Killing-Yano equation for a number of structures outside those of the Berger type analysed by Papadopoulos \cite{Santillan2012}. 
 
In \cite{Papadopoulos2007} it has been shown that the Killing-Yano equation can be solved on manifolds that admit forms associated to a $G$-structure where $G$ is one of the groups in Berger's list, and the candidate conformal Killing-Yano form is built from the $G$-invariant tensors. This implies constraints on the covariant derivatives of the tensors, that can be classified as above. In \cite{Papadopoulos2011} similar results were obtained in the case of connections with totally skew-symmetric torsion. When the resulting manifolds admit Killing spinors, a direct construction of conformal Killing-Yano forms in terms of spinor bilinears has been given in \cite{Semmelmann2003,Marco2003}. 
  
The results of \cite{Papadopoulos2007} are obtained as follows. One starts from the definition of a Killing-Yano form, equations \eqref{eq:symplectic_CKY_definition_0}, \eqref{eq:symplectic_CKY_definition} with $\delta h = 0$: 
\be 
\nabla_\lambda \, h_{\mu_1 \dots \mu_r} =  \nabla_{[\lambda} \,  h_{\mu_1 \dots \mu_r]} \, . 
\ee 
By taking $h$ to be one of the $G$-invariant forms associated to one of the groups in Berger's list, automatically $ \nabla^\mathfrak{g} h = 0$ and the equation above becomes a set of constraints for the components of the connection along $\mathfrak{g}^\perp$. By using a frame that is \textit{adapted} to the $G$-structure, i.e. such that the special tensors have canonical components in the frame, calculations tend to simplify and it is possible to compute the restrictions that the Killing-Yano equation imposes on the connection. We now discuss the results found. 
 
\subsubsection{$U(\dm/2)$ structures} 
An almost complex structure $J$, that is a smooth $(1,1)$ tensor field on $\M$, defines a $Gl(\dm /2)$ structure. If the real metric $g$ is hermitian with respect to the complex structure then this induces a $U(\dm /2)$ structure. When the K\"{a}hler form $\omega$, defined  by $\omega (X, Y) = g(X, J \cdot Y)$, is closed the structure becomes an almost K\"{a}hler structure, and if in addition the complex structure is integrable then one has a K\"{a}hler structure. 
 
One can choose an adapted frame $\{ e^i \} = \{ e^\alpha, e^{\bar{\beta}} \}$, such that $g = \delta_{ij} e^i e^j = 2 \delta_{a \bar{\beta}} e^\alpha e^{\bar{\beta}}$, and $\omega = - i \delta_{a \bar{\beta}} e^\alpha \wedge e^{\bar{\beta}}$, and look for conditions under which $\omega$ or $\omega^{\wedge k}$ are Killing-Yano forms. The Gray-Hervella decomposition of the frame connection under $U(\dm/2)$ irreducible representations yields four such representations, usually indicated with $W_1, \dots , W_4$. These arise as such: the tensor $\nabla\omega$ has components in $T^* \M \otimes \mathfrak{u}(\dm/2)^\perp$ of the type 
\be 
\left(\nabla \omega \right)_{ \alpha \beta \gamma} \, , \hspace{0.4cm} \left(\nabla \omega \right)_{ \alpha \bar{\beta} \bar{\gamma}} \, , \hspace{0.4cm} \left(\nabla \omega \right)_{ \bar{\alpha} \beta \gamma} \, , \hspace{0.4cm} \left(\nabla \omega \right)_{ \bar{\alpha} \bar{\beta} \bar{\gamma}} \, . 
\ee 
Since in the adapted frame the components of $\omega$ are constant, $\nabla\omega$ can be re-written in terms of the spin-connection $\Omega_{i,jk}$. Then the irreducible $U(\dm/2)$ representations are \cite{Gillard2005,Santillan2011}
\ba 
&& (W_1)_{\bar{\alpha}\bar{\beta}\bar{\gamma}} = (\Omega)_{[\bar{\alpha},\bar{\beta}\bar{\gamma}]} \, \nn \\ 
&&  (W_2)_{\bar{\alpha}\bar{\beta}\bar{\gamma}} = (\Omega )_{\bar{\alpha},\bar{\beta}\bar{\gamma}} - (\Omega)_{[\bar{\alpha},\bar{\beta}\bar{\gamma}]} \, , \nn \\ 
&& (W_3)_{\bar{\alpha}\beta\gamma} = (\Omega)_{\bar{\alpha},\beta\gamma} - \frac{2}{\frac{\dm}{2} - 1} (\Omega)_{\bar{\delta} ,}{}^{\bar{\delta}}{} _{[\gamma} g_{\beta] \bar{\alpha}} \, \nn \\ 
&&  (W_4)_{\alpha} =  (\Omega)_{\bar{\delta},}{}^{\bar{\delta}}{}_{\alpha} \, . 
\ea 
The Killing-Yano equation for $\omega$  implies directly that it must always be $W_2 = 0 = W_3 = W_4$. This means that $\M$ is almost K\"{a}hler, however its Nijenhuis tensor does not necessarily vanish. A similar calculation for $\omega^{\wedge k}$, with $1< k < \dm /2$, shows that in this case all the $W_i$ vanish and the manifold is strictly K\"{a}hler.

\subsubsection{$SU(\dm/2)$ structures} 
The analysis is similar to that of the preceding case, and in addition there is an $\mathfrak{su}(\dm/2)$ invariant complex volume form $\Psi$. Now $\mathfrak{su}(\dm/2)^\perp = \mathfrak{u}(\dm/2)^\perp \oplus \mathbb{R}$, and the $\mathbb{R}$ factor corresponds to the trace part of the spin-connection 
\be 
(W_5)_{\bar{\alpha}} = \Omega_{\bar{\alpha}, \beta}{}^{\beta} \, , 
\ee 
which defines a fifth invariance class. 
 
Asking that $\Psi$ is Killing-Yano amounts to the condition $W_4 = W_5 = 0$, which yields a balanced hermitian manifold. Asking that both $\omega$ and $\Psi$ are Killing-Yano gives $W_2 = W_3 = W_4 = W_5 = 0$, which corresponds to a special almost K\"{a}hler manifold, and asking for $\omega^{\wedge k}$, $1< k < \dm/2$, and $\Psi$ implies that all the $W_i$ are zero and the manifold is Calabi-Yau.

\subsubsection{$Sp(\dm/4) \cdot Sp(1)$ structures} 
An almost quaternion-Hermitian structure is given by the presence of three almost complex structures $I, J, K$, satisfying the identities f the imaginary unit quaternions, $I J = K$, and cyclic permutations. There is also a fundamental form $\chi = \omega_I^{\wedge 2} + \omega_J^{\wedge 2} + \omega_K^{\wedge 2}$. 
 
The conditions for a single almost complex structure to be Killing-Yano are those discussed above for $U(m)$ structures. To analyse when $\chi$ is Killing-Yano one can choose a frame adapted to $I$. The components of the spin-connection $\Omega$ in $(\mathfrak{sp}(\dm/4) \otimes \mathfrak{sp}(1))^\perp$ are given by the $\omega_J$ traceless part of $\Omega_{i, \alpha \beta}$, plus the component $\tilde{\Omega}_{i, \alpha \bar{\beta}}$ that satisfies $(\tilde{\Omega} \cdot J)_{i, \alpha \beta} = (\tilde{\Omega} \cdot J)_{i,[ \alpha \beta ]}$, and $\tilde{\Omega}_{i, \alpha} {}^\alpha = 0$. When $\dm = 8$ this yields four irreducible representations and therefore $2^4 = 16$ different classes, and for $\dm = 12, 16, \dots$, this gives six irreducible representations and $2^6 = 64$ classes. These classes can be related to $\nabla \chi$. One finds that for $\chi$ to be Killing-Yano all the components of $\Omega$ in $(\mathfrak{sp}(\dm/4) \otimes \mathfrak{sp}(1))^\perp$ must vanish, making $\chi$ parallel and $\M$ quaternionic K\"{a}hler. 
 
\subsubsection{$Sp(\dm/4)$ structures}  
Similarly to the previous section, there are three almost complex structures $I, J, K$. The components of the spin-connection $\Omega$ in $(\mathfrak{sp}(\dm/4) $ are similar, but in this case are not traceless with respect to $\omega_J$: they are given by $\Omega_{i, \alpha \beta}$, plus the component $\tilde{\Omega}_{i, \alpha \bar{\beta}}$ that satisfies $(\tilde{\Omega} \cdot J)_{i, \alpha \beta} = (\tilde{\Omega} \cdot J)_{i,[ \alpha \beta ]}$. Asking that $\omega_I$ is Killing-Yano yields the same conditions as in the section on $U(\dm / 2)$ structures, that is the only non-zero components of the spin connection are of the type $\Omega_{[\alpha, \beta \gamma ]}$. If now one asks that also $\omega_J$ is Killing-Yano then the only available solution is $\Omega_{[\alpha, \beta \gamma ]}= 0$, and the manifold is hyper-K\"{a}hler.

\subsubsection{$G_2$ structures} 
In an adapted basis a $G_2$ invariant $3$-form for a seven-dimensional manifold is written as 
\be 
\phi = c_{abc} \, e^a \wedge e^b \wedge e^c \, , 
\ee 
where the $c_{abc}$ coefficients are the multiplication constants of the imaginary octonions. Since $G_2 \subset SO(7)$ is the automorphism group of the imaginary octonions, then a $G_2$ rotation of the adapted frame will leave the form invariant. The spin connection form $\Omega$, which has values in the adjoint representation of $SO(7)$, decomposes naturally into a part $\Omega^+$ in the adjoint representation of $G_2$, with 14 generators, and a part $\Omega^-$ in the fundamental representation, with 7 generators. When $\Omega^- = 0$ the holonomy is $G_2$ and $\phi$ is closed and co-closed. In the general case instead $\Omega^-$ is parameterised by the $G_2$ torsion classes \cite{GrayFernandez1982}: 
\ba 
d \phi &=& \tau_0 * \phi + 3 \tau_1 \wedge \phi + * \tau_3 \, , \nn \\ 
d *\phi &=& 4 \tau_1 \wedge *\phi + * \tau_2 \, . 
\ea 
The covariant derivative of $\phi$ can be written as 
\be 
\nabla_\lambda \phi_{\mu\nu\rho} = T_{\lambda \alpha} g^{\alpha \beta} *\phi_{\beta \mu\nu\rho} \, , 
\ee 
where the tensor $T$ is the full torsion tensor \cite{Karigiannis2009}  given by 
\be 
T = \frac{\tau_0}{4} g - \tau_{3} + \tau_1 - \frac{1}{2} \tau_2 \, , 
\ee 
$g$ being the metric. Then the Killing-Yano equation for $\phi$ implies that $\tau_1 = 0 = \tau_2 = \tau_3 = \tau_4$. Structures with only a non-vanishing $\tau_0$ torsion class are nearly parallel $G_2$. A similar calculation shows that $*\phi$ is Killing-Yano only is all the torsion components are zero and the manifold is strictly $G_2$.

\subsubsection{$Spin_7$ structures}  
A $Spin_7$ form $\Psi$ on an eight-dimensional manifold is self-dual. If it is to be a Killing-Yano form, then by definition it is co-closed. Being self-dual it must then also be closed. But then the Killing-Yano equation implies that it is covariantly constant, and the manifold is strictly $Spin(7)$. 
 
A similar analysis for the case of the Killing-Yano equation using a covariant derivative with totally anti-symmetric torsion has been done in \cite{Papadopoulos2011}, finding that the Killing-Yano equation in several cases does not determine uniquely the torsion classes, leaving space for a bigger set of manifolds with special holonomy. The manifolds found are the target spaces of supersymmetric non-relativistic particles with torsion couplings. The full conformal Killing-Yano equation \eqref{eq:symplectic_CKY_definition} has not been analysed yet. As it will be seen in sect.\ref{sec:dirac_equation_linear_symmety_operators}, this is useful in studying symmetry operators of the Dirac equation for example. Kubiz\v{n}\'ak, Warnick, Houri and Yasui  discuss a classification of the local form of metrics admitting a principal Killing-Yano tensor with torsion \cite{DavidClaudeHouriYasui2012}. Houri, Takeuchi and Yasui discuss deformations of Sasakian structures in the presence of totally skew-symmetric torsion, the fact they possess a closed conformal Killing-Yano tensor with torsion, and how they arise as solutions of five-dimensional gauged and ungauged supergravity, and eleven dimensional supergravity \cite{HouriTakeuchiYasui2013}. Houri and Yamamoto analyse five-dimensional metrics with a rank-2 Killing-Yano tensor with torsion, such that it has a Killing vector as eigenvector with zero eigenvalue \cite{HouriYamamoto2013}. They find geometries among which there are solutions of minimal supergravity and abelian heterotic supergravity describing charged, rotating Kaluza-Klein black holes.

\section{QUANTUM SYSTEMS\label{sec:quantum_systems}} 
In this section we will present some examples of quantum  systems that admit hidden symmetries of the dynamics, and then discuss the general theory of separation of variables for the Schr\"odinger and Klein-Gordon equations. We will finish with a discussion of the hidden symmetries of the Dirac equation that are linear in momenta, the only ones for which there exists a well established theory. An important application of separability both for the Schr\"odinger and Klein-Gordon equation, and for the Dirac one, will be that of Kerr-NUT-(A)dS black holes. 
 
It is important noticing that hidden symmetries are not exclusively a feature of classical systems, rather they appear in quantum systems as well. The non-relativistic hydrogen atom and the harmonic oscillator that we are going to discuss are universally known systems and for these it will be instructive to compare the classical and quantum treatments of hidden symmetries. 
 
The theory of separation of variables for the  Schr\"odinger and Klein-Gordon equations is closely related to that of the Hamilton-Jacobi equation discussed in sec.\ref{sec:Hamilton-Jacobi}, quantum hidden symmetries requiring extra conditions relative to classical ones. The reader will be able to follow in detail the differences between the two cases. 
 
Lastly, we will present a relatively detailed discussion of hidden symmetries of the Dirac equation that are linear in the momenta. We will make use of the efficient formalism of differential forms, where the sections of the Clifford bundle are identified with differential forms, and the Clifford algebra maps into a non-commutative algebra of forms. This formalism is superior in efficiency and clarity to that of calculus with explicit use of indices. We will discuss hidden symmetries in a curved geometry, as well as the more general concept of hidden symmetries in the presence of flux fields. We will finish discussing the Dirac equation for the Eisenhart-Duval lift metrics. 
 
\subsection{Hidden symmetries of the quantum isotropic oscillator}  
The quantum harmonic oscillator is a physical system rightly of great importance. Rather than providing here a reminder of all the applications the harmonic oscillator has (in fact, one for each system the linearisation of which is worth studying), we provide a quote attributed to Sidney Coleman: "The career of a young theoretical physicist consists of treating the harmonic oscillator in ever-increasing levels of abstraction". 
 
In this section for simplicity we deal with the isotropic oscillator. We begin describing the classical system from the point of view of its dynamical symmetry group. The Hamiltonian function in $\dm$ dimension is 
\be 
H = \sum_{i=1}^\dm \left(\frac{p_i^2}{2m} + \frac{\omega^2}{2} q_i^2 \right) \, . 
\ee 
Using the rescaled variables $\tilde{q}_i = \left(\omega\sqrt{m}\right)^{\frac{1}{2}} q_i$, $\tilde{p}_i = \left(\omega\sqrt{m}\right)^{-\frac{1}{2}} p_i$, the Hamiltonian can be written as 
\be 
\frac{\sqrt{m}}{\omega} H =  \tilde{H} = \frac{1}{2} \sum_{i=1}^\dm \left(\tilde{p}_i^2 + \tilde{q}_i^2 \right) \, . 
\ee 
We define new variables $z_i = \tilde{q}_i + i \tilde{p}_i$, in terms of which $\tilde{H} = \frac{1}{2} z^\dagger z$ and the equations of motion are 
 
\be 
\frac{dz_i}{d\lambda} = - i z_i \, . 
\ee 
Using these variables it becomes evident that there is an action of the group $U(\dm )$ on phase space that preserves the Hamiltonian, with $U(\dm)$ strictly bigger than the isometry group $O(\dm)$. To build conserved charges we can proceed as follows. Let $\{ T_a, a = 1, \dots, \dm^2-1 \}$ be  $\dm \times \dm$ complex, traceless matrices satisfying $T_a^\dagger = - T_a$, that generate the Lie algebra of $SU(\dm)$. With these we can build the real quantity 
\be \label{eq:complex_conserved_constants}
C_a = \frac{1}{2i} z^\dagger T_a z \, . 
\ee 
The only element of the Lie algebra of $U(\dm)$ we are not describing with these is a multiple of the identity matrix, which would give rise to the Hamiltonian through a formula analogue to  \eqref{eq:complex_conserved_constants}. It is straightforward checking that the $C_a$ quantities are conserved: 
\be 
\frac{d C_a}{d\lambda} = \frac{1}{2i} \left[ - z^\dagger T_a (-i z) + (i z^\dagger) T_a z \right] = 0 \, . 
\ee 
A direct calculation shows that 
\ba 
X_{C_a} &=& - 2i \sum_{i=1}^\dm \left[ \partial_{\overline{z}_i} C_a \partial_{z_i} + 2i \partial_{z_i} C_a \partial_{\overline{z}_i} \right] \nn \\ 
&=& - \sum_{i=1}^\dm \left[ T_{a,ij} z_j \partial_{z_i} + \left(T_{a,ij} z_j\right)^* \partial_{\overline{z}_i} \right] \, . 
\ea 
In general, transformations that are not isometries will alter the shape of the trajectory. To illustrate the principle we examine in detail the case $\dm = 2$. Then there are three $T_a$ matrices: 
\be 
T_1 = \left( \begin{array}{cc} 0 & i \\ i & 0 \end{array} \right) \, , \quad 
T_2 = \left( \begin{array}{cc} 0 & 1 \\ -1 & 0 \end{array} \right) \, , \quad 
T_3 = \left( \begin{array}{cc} i & 0 \\ 0 & -i \end{array} \right) \, . 
\ee 
The matrix $T_2$ is real and therefore generates an isometry: $T_2$ in fact is the generator of the $SO(2)$ Lie algebra acting on configuration space by $\delta \tilde{q}_1 = \epsilon \tilde{q}_2$, $\delta \tilde{q}_2 = - \epsilon \tilde{q}_1$, and the same for the $\tilde{p}_i$. This is a rotation in the $1,2$ plane. The matrix $T_3$ instead generates a forward time evolution in the $(\tilde{q}_1, \tilde{p}_1)$ variables, and a backward time evolution in the $(\tilde{q}_2, \tilde{p}_2)$ variables. Lastly, the finite transformation associated to $T_1$ can be obtained by exponentiation: 
\be 
z^\prime = \exp\left( s T_1 \right) z = \left( \begin{array}{cc} \cos s & i \sin s \\ i \sin s & \cos s \end{array} \right) z  \, , 
\ee 
where $s$ is a continuous, real parameter, or equivalently 
\be \label{eq:harmonic_oscillator_transformation} 
\left\{ \begin{array}{l} \tilde{q}_1^\prime = \cos s \, \tilde{q}_1 - \sin s \, \tilde{p}_2 \, , \quad \tilde{p}_1^\prime = \cos s \,\tilde{p}_1  + \sin s \, \tilde{q}_2 \, ,  \\ 
 \tilde{q}_2^\prime = \cos s \, \tilde{q}_2 - \sin s \, \tilde{p}_1 \, , \quad \tilde{p}_2^\prime = \cos s \, \tilde{p}_2  + \sin s \, \tilde{q}_1  \, . 
\end{array} \right. 
\ee
Consider for example the trajectory $\lambda \mapsto (z_1 (\lambda), z_2 (\lambda))$ 
\be \label{eq:harmonic_initial_trajectory}
\left\{ \begin{array}{l}  \tilde{q}_1 = A \cos \lambda \, , \quad \tilde{p}_1 = - A \sin \lambda \, ,  \\ 
 \tilde{q}_2 = A \sin \lambda \, , \quad \tilde{p}_2 =  A \cos \lambda \, ,  
\end{array} \right. 
\ee 
$A>0$ constant. This is such that $\tilde{q}_1^2 + \tilde{q}_2^2 = A^2$, and similarly for the $\tilde{p}$ variables, describing circles in the $\tilde{q}$ and $\tilde{p}$ planes with maximum amplitude $A$. Applying transformation \eqref{eq:harmonic_oscillator_transformation} with $s = \frac{\pi}{4}$ we get 
\ba 
&& \tilde{q}_1^\prime = 0 \, , \quad \tilde{p}_1 = 0 \, , \nn \\ 
&& \tilde{q}_2^\prime = \sqrt{2} A \sin \lambda \, , \quad \tilde{p}_2^\prime =  \sqrt{2} A  \cos \lambda \, , 
\ea 
which is a harmonic motion along the $2$ direction, with amplitude $\sqrt{2} A$ which is the correct factor in order to maintain the same energy.

In fact, applying the transformation \eqref{eq:harmonic_oscillator_transformation} we can obtain all possible trajectories of a given energy $E$, since the space of trajectories of a given energy $E$ in phase space is $2$-dimensional, and we can act with the independent transformations generated by $T_1$, $T_2$ and $T_3$: the action of $U(2)$ on such space is transitive. The same holds for a generic dimension $\dm > 2$: $U(\dm)$ has $\dm^2$ generators, with $\dm$ of these  of the diagonal type, similar to the $T_3$ matrix of the $\dm = 2$ case. These act on each individual pair of variables $(\tilde{q}_i, \tilde{p}_i)$ as the forward or backward time evolution. Next, there are $\dm (\dm - 1)$ independent non-diagonal transformations. Since the space of trajectories of a given energy $E$ is $2 (\dm -1)$ dimensional, these are enough to provide a transitive action. We can then say that under the action of $U(\dm)$ there is a unique representative in each family of trajectories of energy $E$, $E>0$. 
 
The quantum model follows a similar description. The Hamiltonian is 
\be 
H = \sum_{i=1}^\dm \left(\frac{P_i^2}{2m} + \frac{\omega^2}{2} Q_i^2 \right) \, , 
\ee 
with $[Q_l, P_m] = i \hbar \delta_{lm}$. The analog of the $z$, $z^\dagger$ variables are now ladder operators: first define $\tilde{Q}_i = \sqrt{\frac{m^{1/2} \omega}{2\hbar}} Q_i$, $\tilde{P}_i = \sqrt{\frac{1}{2\hbar m^{1/2} \omega}} P_i$, and 
\be 
a_i = \tilde{Q}_i + i \tilde{P}_i \, . 
\ee 
Then the commutation relations become $[ a, a^\dagger ] = \mathbf{1}$ and the Hamiltonian is written as  
\be \label{eq:harmonic_oscillator_Hamiltonian}
H = \frac{\hbar \omega}{\sqrt{m}} \left( a^\dagger a + \frac{\dm}{2} \right) \, . 
\ee 
For any matrix $T$ that is a linear combination with real coefficients of the $T_a$ defined above, the operator $\hat{U} = \exp \left( a^\dagger T a \right)$ is unitary and acts on $a$ as 
\be 
\hat{U}^\dagger \, a \, \hat{U} = \exp \left( T \right) a \, , 
\ee 
and therefore it induces a unitary transformation on $a$ that leaves the Hamiltonian invariant. Conversely, all unitary transformations on $a$ can be written for some operator $\hat{U}$ in this form. Given the vacuum state $|0\rangle$ such that $a_i |0\rangle = 0$ $\forall i$ and $U |0\rangle = |0\rangle$, a generic state is written as 
\be \label{eq:harmonic_oscillator_state}
|l_1, \dots, l_\dm \rangle = \frac{1}{\sqrt{l_1!}} \dots \frac{1}{\sqrt{l_\dm!}} (a_1^\dagger)^{l_1} \dots (a_\dm^\dagger)^{l_\dm} |0\rangle \, , 
\ee 
and has energy $E = \frac{\hbar \omega}{\sqrt{m}} \left( \sum_i l_i + \frac{\dm}{2} \right) = \frac{\hbar \omega}{\sqrt{m}} \left(  L + \frac{\dm}{2} \right)$. There is a degeneracy in the energy levels that is not explained only by the $SO(\dm)$ symmetry, however one can notice that the state above transforms as a $U(\dm)$ tensor with $L = \sum_i l_i$ indices on $\mathbb{C}^\dm$. Acting with $U(\dm)$ transformations one can recover all possible combinations of indices and therefore $U(\dm)$ acts transitively on the states of a given energy $E$. Lastly, it is worth rephrasing the result as follows. States as per \eqref{eq:harmonic_oscillator_state} are solutions of the Schr\"{o}dinger equation associated to the Hamiltonian \eqref{eq:harmonic_oscillator_Hamiltonian}. The operators $\hat{U}$   are \textit{symmetry operators}, meaning that $[\hat{U}, H] = 0$. Therefore they transform solutions of the Schr\"{o}dinger equations into solutions. This theme of quantum dynamical symmetries being associated to operators that transform solutions of a differential equation into solutions will keep appearing in the rest of this section.

\subsection{Non-relativistic hydrogen atom\label{sec:hydrogen_atom}}  
For a short historical discussion of the inverse square central potential the reader can refer to the previous sec. \ref{sec:examples_Kepler} on the classical Kepler problem. Quantum mechanically several features are either kept unchanged or follow with appropriate changes. The Schr\"{o}dinger equation for the non-relativistic hydrogen atom is 
\be 
\left( - \frac{\hbar^2}{2m} \nabla^2 - \frac{e^2}{r} - E \right) \psi(\vec{r}) = 0 \, . 
\ee 
Since there is a manifest $SO(3)$ isometry one can separate variables using spherical coordinates, such that 
\be 
\nabla^2 = \left( \frac{1}{r} \partial_r r \right)^2 + \frac{1}{r^2} \Delta_{S^2} \, , 
\ee 
where $\Delta_{S^2}$ is the Laplacian on the $2$-sphere 
\be 
\Delta_{S^2} = \frac{1}{\sin\theta} \partial_\theta \left( \sin\theta \partial_\theta \right) + \frac{1}{\sin^2\theta} \partial_\phi^2 \, , 
\ee 
which has the spherical harmonics $Y^l_m (\theta, \phi)$ as eigenfunctions with eigenvalues $-l(l+1)$, where $l= 0,1,2, \dots$ and $-l \le m \le +l$ are integers. Assuming 
\be 
\psi (\vec{r}) = \frac{1}{r} R(r) Y^l_m (\theta, \phi ) \, 
\ee 
one gets an ordinary differential equation for $R$: 
\be 
\left(\frac{d^2}{dr^2} - \frac{l(l+1)}{r^2} +  \frac{2me^2}{\hbar^2 r} + \frac{2mE}{\hbar^2} \right) R(r) = 0 \, . 
\ee 
$L^2$ solutions with appropriate boundary behaviour are indexed by a radial quantum number $n$ and the eigenvalues of the energy for bound states are 
\be 
E_N = - \frac{mc^2 \alpha^2}{2 N^2} \, , 
\ee 
where $\frac{mc^2 \alpha^2}{2} \sim 13.6 eV$, and $N = n + l + 1$, with $0 \le l \le N - 1$ for a given value of $n$. Therefore there is a degeneracy in the states of a given $E_N$: there are $\sum_{l=0}^{N-1} (2l+1) = N^2$ states, many more then one would expect from the $SO(3)$ symmetry alone. 
 
The reason for this is the quantum mechanical version of the Runge-Lenz vector, associated again to an $O(4)$ group of dynamical symmetries. The quantum mechanical version of eq.\eqref{eq:RungeLenz_def} is given by 
\be \label{eq:RungeLenz_def_QM}
\vec{A} = \frac{1}{2} \left( \vec{p} \times \vec{L} - \vec{L} \times \vec{p} \right) - m e^2 \, \frac{\vec{r}}{r} \, , 
\ee
where all quantities in this section are intended in the sense of operators. The reason why \eqref{eq:RungeLenz_def_QM} is used instead of \eqref{eq:RungeLenz_def_QM} is that the former presents the right operator ordering to make the result self-adjoint, as showed initially by Pauli \cite{Pauli1926}. The quantum Runge-Lenz vector is again a conserved quantity, in the sense that $[\vec{A}, H] = 0$. The $SO(4)$ algebra relations hold in the following form 
\ba \label{eq:L_A_algebra_QM}
\left[ L^i , L^j \right] &=& i \hbar \, \sum_{k = 1}^3 \epsilon^{ijk} L^k \, , \nn \\ 
\left[ L^i , A^j \right] &=&  i \hbar \, \sum_{k = 1}^3 \epsilon^{ijk} A^k \, ,  \nn \\ 
\left[ A^i , A^j \right] &=& - 2 i m H  \hbar \, \sum_{k = 1}^3 \epsilon^{ijk} L^k \, , 
\ea
so that exponentiating these operators one find the $SO(4)$ group of dynamical symmetries, $O(4)$ adding parity. Once again, the conditions $\left[\vec{L}, H\right] = 0$, $\left[\vec{A}, H\right] = 0$ mean that $O(4)$ transformations will transform solutions of the Schr\"{o}dinger equation into solutions, while $\left[L^2 , \vec{A} \right] = 2 i \hbar \vec{A}$ means that they will not in general preserve the square of the angular momentum. Therefore the $O(4)$ dynamical group transformations deforms different orbitals associated with the same energy one into the other , and it is possible to show that this action is transitive on the states of a given energy.

\subsection{Schr\"{o}dinger and Klein-Gordon equations\label{sec:Klein_Gordon}} 
 
In this section we study the following equation for a scalar field $\psi$ 
\be \label{eq:Schrodinger_Klein-Gordon}
\left( - \frac{\hbar^2}{2} \Delta + V (q) - E \right) \psi= 0 \, , 
\ee 
where $\Delta = \nabla_\mu \left( g^{\mu\nu} \right) \nabla_\nu$, $g$ is a metric, $\nabla$ is the covariant derivative with respect to the Levi-Civita connection and $E$ a constant. When the metric is Riemannian \eqref{eq:Schrodinger_Klein-Gordon} becomes the time independent Schr\"{o}dinger equation in a potential, while if the metric is pseudo-Riemannian it becomes the relativistic Klein-Gordon equation. With this in mind, in the reminder of this section we will refer to it loosely by Schr\"{o}dinger equation. 
  
\subsubsection{The general theory of separation of variables\label{sec:Schrodinger_general_theory}}  
A key role in the early development of quantum theory and of the Schr\"{o}dinger equation was played by the hydrogen atom, that had a simple analytical solution. As seen in section \ref{sec:hydrogen_atom}, the Schr\"{o}dinger equation separates due to the symmetry of the problem. In fact, systems for which the Schr\"{o}dinger equation separates are extremely important as they provide examples for which we can obtain analytic solutions. In this section we will admit arbitrary signature metrics but we will exclude the technically more involved case of null second class coordinates. The first important point to make, as mentioned in sec.\ref{sec:examples_spinning_particle}, is that in general the operator version of a quantity that is conserved for the analogous classical problem of motion in the metric $g$ with potential $V$ will not provide a good quantum number, due to anomalous contributions to the conservation law. Such anomalies arise in a curved background, and where pointed out first by Carter \cite{Carter1977}. Carter found that for a rank 2 Killing tensor $K$ they are proportional to the term 
\be 
K^{\mu} {}_\lambda R^{\lambda \nu} - R^{\mu} {}_\lambda K^{\lambda \nu} = g^{\nu \rho} \left[ K , R \right]^\mu {}_\rho \, , 
\ee 
where $R$ is the Ricci tensor and by $\left[ K , R \right]$ we mean the matrix commutator of $K$ and $R$ thought of as linear operators on the tangent space. No anomalous terms rise for Killing vectors instead. Therefore the quantum symmetries will be non-anomalous only if $K$ and $R$ commute and admit common eigenvalues. The term above arises from quantum integrability conditions in a curved geometry. For a classically separable system it says that the Killing web has to be compatible with the presence of curvature. The equation 
\be 
KR = RK 
\ee 
is called the \textit{Robertson condition}. It is clearly satisfied if the Ricci tensor is zero or, more generally, if the space is Einstein. There is another notable case: every time the Killing tensor $K$ is obtained by taking the square of a Killing-Yano tensor the Robertson condition will be automatically satisfied. 
 
We can make these statements more precise. We start by defining separability of the Schr\"{o}dinger equation according to Benenti \cite{Benenti2002outline,Benenti2002a,benenti2002b}. There are two natural definitions of separability. We say that the Schr\"{o}dinger equation \eqref{eq:Schrodinger_Klein-Gordon} is \textit{freely separable} if it admits a \textit{complete separated solution} 
\be 
\psi (q, c) = \prod_{\mu = 1}^\dm \psi_\mu (q^\mu, c ) \, , 
\ee 
that depends on $2 \dm$ parameters $c = (C_M) = ( c_1, \dots, c_{2\dm})$ that satisfy the completeness condition 
\be 
\det \left[ \frac{\partial u_\mu}{\partial c_M} \; \, \frac{\partial v_\mu}{\partial c_M} \right] \neq 0 \, , 
\ee 
where $u_\mu = \frac{\psi_\mu^\prime}{\psi_\mu}$, $v_\mu = \frac{\psi_\mu^{\prime\prime}}{\psi_\mu}$. This notion of separability is directly linked to classical separability in orthogonal coordinates, and can occur only for this type of coordinates, the classical separability being a necessary but in general not sufficient condition. In fact one can prove the following: 
\begin{Theorem} 
The Schr\"{o}dinger equation is freely separable if and only if there exists a characteristic
Killing tensor K (with simple eigenvalues and normal eigenvectors) such that $d(KdV ) = 0$, and the Robertson condition holds. 
\end{Theorem} 
As discussed in sec.\ref{sec:Hamilton-Jacobi} this implies that the Hamilton-Jacobi equation for the Hamitonian $H = \frac{1}{2} g^{\mu\nu} p_\mu p_\nu + V$ is separable in orthogonal coordinates. When expressed in orthogonal coordinates, the Robertson condition becomes $R^{\mu\nu} = 0$ for $\mu \neq \nu$. 
 
There exists another notion of separability of the Schr\"{o}dinger equation that relates to classical separability in non-orthogonal coordinates. We say that $\psi$ is a \textit{reduced separated solution}  if it can be written as 
\be \label{eq:reduced_separable_solution}
\psi (q, c) = \prod_{a=1}^m \psi_a (q^a, c) \cdot \prod_{\alpha = m+1}^\dm \exp \left( k_\alpha q^\alpha \right) \, . 
\ee 
This time there are $2m$ parameters $c = (c_A) = (c_1, \dots, c_{2m})$ in the $\psi_a$ part of the solution, and $\dm - m$ parameters $k_\alpha$ in the remaining part. This solution is called reduced because one is reducing the initial possible freedom of having arbitrary separated functions, trading this off for a specific exponential form. In total therefore there are $\dm + m$ free parameters, and the completeness condition is expressed as 
\be 
\det \left[ \frac{\partial u_a}{\partial c_A} \; \, \frac{\partial v_a}{\partial c_A} \right] \neq 0 \, , 
\ee 
where $u_a = \frac{\psi_a^\prime}{\psi_a}$, $v_\mu = \frac{\psi_a^{\prime\prime}}{\psi_a}$, $a = 1, \dots, m$. There is a theorem that is the analogue of the previous one \cite{Benenti2002a}: 
\begin{Theorem} 
The Schr\"{o}dinger equation is reductively separable if and only if there exists a non degenerate Killing web $(\mathcal{S}, D)$, where $D$ is a vector space spanned by Killing vectors as in sec.\ref{sec:Hamilton-Jacobi}, such that: (i) the potential V is D-invariant, DV = 0; (ii) $K$ is a characteristic Killing tensor with pairwise and pointwise distinct eigenvalues, orthogonal to the leaves of $\mathcal{S}$, and such that $d(KdV = 0)$ (iii) the spaces orthogonal to D are invariant under the Ricci tensor R, interpreted as a linear operator, and the restrictions to these spaces of R and K commute. 
\end{Theorem} 
Note that condition $(iii)$ above is equivalent to saying that the essential eigenvectors are eigenvectors of the Ricci tensor R, or Ricci principal directions. Then the coordinates $q^\alpha$ are ignorable and condition $(iii)$ can be written as $R^{ab} = 0$ for $a\neq b$. 

There exists another, more general type of separability, the so called R-separability \cite{Moon1952a,Moon1952b,Moon1971,KalninsMiller1978,KalninsMiller1980remarkable,KalninsMiller1982,KalninsMiller1984,ChanachowiczChanuMcLenaghan2009}.  
A solution $\psi$ of the Schr\"{o}dinger equation is said to be R-separable if it can be written as 
\be \label{eq:R_separability_definition}
\psi (q, c) = \mathcal{R} (q) \prod_{\mu = 1}^\dm \psi_\mu (q^\mu, c ) \, .  
\ee 
Notice how the $\mathcal{R}$ factor does not depend on the constants $c$. Kalnins and Miller gave an intrinsic characterisation of R-separability for the Helmholtz equation (no potential term) using a set of commuting operators obeying appropriate conditions \cite{KalninsMiller1982intrinsic,KalninsMiller1983}, while Chanu and Rastelli considered orthogonal coordinates and described conditions for R-separability for a single fixed value of $E$, which they called \textit{fixed energy R-separation} (FER-separation), and allows for a larger number of classes of separable coordinates \cite{ChanuRastelli2006}. In their definition the function $\psi$ in \eqref{eq:R_separability_definition} depends on $2 \dm - 1$ parameters $c_I = (c_1, \dots, c_{2 \dm -1})$ and satisfies the condition 
\be 
\operatorname{rank} \left[ \frac{\partial u_\mu}{\partial c_I} \; \, \frac{\partial v_\mu}{\partial c_I} \right] = 2 \dm -1  \, , 
\ee  
where $u_\mu = \frac{\psi_\mu^\prime}{\psi_\mu}$, $v_\mu = \frac{\psi_\mu^{\prime\prime}}{\psi_\mu}$. 
Chanu and Rastelli show that the coordinates allowing FER-separation as above are necessarily orthogonal and conformally separable (sec.\ref{sec:Hamilton-Jacobi_null}). In fact, they prove a theorem with necessary and sufficient conditions for FER-separation. We first define the \textit{contracted Christoffel symbols}  
\be 
\Gamma^\mu := g^{\rho\sigma} \Gamma^\mu_{\rho\sigma} \, , \quad \Gamma_\mu := g_{\mu\nu} \Gamma^\nu \, ,  
\ee 
We also define a \textit{conformal St\"{a}ckel metric} as a metric $g$ for which there exists a function $\Lambda$ such that  $\overline{g}^{\mu\mu} = \frac{g^{\mu\mu}}{\Lambda}$ (no sum) is a St\"{a}ckel metric, i.e. it corresponds to a row of the inverse of a St\"{a}ckel matrix $\theta_{\mu}^{(\nu )}$, sec.\ref{sec:Hamilton-Jacobi_riemannian_orthogonal}. Lastly we define a \textit{pseudo-St\"{a}ckel factor} as a function $f$ that can be written as $f = \sum_\mu \tilde{g}^{\mu\mu} \phi_\mu (q^\mu)$.  where $\tilde{g}$ is a conformal St\"{a}ckel metric. 
In terms of these one has 
\begin{Theorem} 
FER-separation for the Schr\"{o}dinger equation \eqref{eq:Schrodinger_Klein-Gordon} holds if and only if: 1) the coordinates are orthogonal; 2) the coordinates are conformally separable; 3) the function $\mathcal{R}$ is a solution of 
\be \label{eq:R_equation} 
 \partial_\mu \ln \mathcal{R} =\frac{1}{2} \Gamma_\mu 
\ee 
modulo separated factors; 4) the function $E - V + \frac{\hbar^2}{2} \frac{\Delta \mathcal{R}}{\mathcal{R}}$ is a pseudo-St\"{a}ckel factor for $g$. 
\end{Theorem}

\noindent From the theorem it follows that the metric 
\be 
\overline{g}^{\mu\nu} = \frac{g^{\mu\nu}}{ E - V + \frac{\hbar^2}{2} \frac{\Delta \mathcal{R}}{\mathcal{R}}}  \, , 
\ee 
is a St\"{a}ckel metric, associated to standard separation of variables. As such, the geodesic Hamilton-Jacobi equation for $\overline{g}$ is separable, and there exist a complete set of independent Killing tensors in involution, which for $g$ become conformal Killing tensors. This therefore is a necessary condition for R-separation. Another result of \cite{ChanuRastelli2006} is that if the conditions of the theorem are realised for two different values of the energy $E_1 \neq E_2$, then the same coordinates allow R-separation for all values of $E$.

\subsubsection{Klein-Gordon equation for higher dimensional rotating black holes} 
In this section we follow the special notation of sec.\ref{sec:higher_dimensional_black_holes}, including the suspension of the Einstein sum convention in relation to indices of type $\mu$ or $i$.  
 
We have seen in section \ref{sec:higher_dimensional_black_holes} that Kerr-NUT-(A)dS black holes and their off-shell generalisations, the canonical metrics with a principal conformal Killing-Yano tensor, admit a tower of mutually commuting and independent rank two Killing tensors and Killing vectors. According to the theory discussed in sec.\ref{sec:Hamilton-Jacobi} this implies that the Hamilton-Jacobi equation is separable. Also, since the Killing tensors are built as the square of Killing-Yano tensors, then the Robertson condition of section \ref{sec:Schrodinger_general_theory} is satisfied and the Schr\"{o}dinger equation is reductively separable. Krtou\v{s} and Sergyeyev proved the reduced separability by explicitly constructing a set of commuting operators for the Schr\"{o}dinger equation and showing that a reduced separated solution of the form \eqref{eq:reduced_separable_solution} is a common eigenfunction for all of them \cite{Pavel2008complete}.  The operators are given by 
\ba 
\mathcal{L}^{(j)} &=& - i \hbar \xi^{(j) a} \nabla_a \, , \nn \\ 
\mathcal{K}^{(j)} &=& - \hbar^2 \nabla_a \left( K^{(j) ab} \nabla_b \right) \, . 
\ea 
These operators are shown to mutually commute by an explicit calculation 
\ba 
\left[ \mathcal{L}^{(i)} , \mathcal{L}^{(j)} \right] &=& 0 \, , \\ 
\left[ \mathcal{L}^{(i)} , \mathcal{K}^{(j)} \right] &=& 0 \, , \\  
\left[ \mathcal{K}^{(i)} , \mathcal{K}^{(j)} \right] &=& 0 \, . 
\ea 
A reduced separated solution is sought of the form 
\be 
\psi = \prod_{\mu = 1}^N \psi_\mu (x_\mu)  \prod_{k=0}^{N + \epsilon - 1} \exp \left( \frac{i}{\hbar} \Psi^k \psi_k \right) \, .  
\ee 
This diagonalises the mutually commuting operators 
\ba 
 \mathcal{L}^{(i)} \psi &=& \Psi^i \psi \, , \nn \\ 
 \mathcal{K}^{(j)} \psi &=& \Xi^j \psi \, , \nn \\  
\ea 
if the functions $\psi_\mu$ satisfy the following set of ordinary differential equations 
\be 
\left( X_\mu \psi_\mu^\prime \right)^\prime + \epsilon \frac{X_\mu}{x_\mu} \psi_\mu^\prime + \frac{1}{\hbar^2} \left( \tilde{\Xi}_\mu (x_\mu) - \frac{\tilde{\Psi}_\mu^2 (x_\mu)}{X_\mu} \right) \psi_\mu = 0 \, , 
\ee 
where the functions $\tilde{\Psi}_\mu$ and $\tilde{\Xi}_\mu$ are defined by 
\ba 
\tilde{\Psi}_\mu &=& \sum_{k=0}^{N + \epsilon - 1} \Psi_k (-x_\mu^2)^{N-1-k} \, , \nn \\ 
\tilde{\Xi}_\mu &=& \sum_{k=0}^{N + \epsilon - 1} \Xi_k (-x_\mu^2)^{N-1-k} \, . 
\ea

\subsection{Dirac equation\label{sec:quantum_dirac_equation}} 
The Dirac equation, has been derived in 1928, and among its successes are the description of the relativistic hydrogen atom and the prediction of the existence of anti-particles. It is possible to study relativistic spin $\frac{1}{2}$ particles on a curved background by writing the Dirac equation with a curved metric. With time, mainly inspired by unification theories such as String/M--Theory and by cosmological models, people have studied extensions of the equation to higher dimensions than four. 
 
Hidden symmetries of the dynamics are found studying the Dirac equation in appropriate backgrounds, as we will see in this section, and in its classical limit, the spinning particle, as seen in section \ref{sec:examples_spinning_particle}. The key geometrical objects associated to hidden symmetries are conformal Killing-Yano tensors. This can be compared to the case of the classical Hamilton-Jacobi equations, the Schr\"{o}dinger equation and the Klein-Gordon equation, where the main objects playing a role are Killing vectors and rank 2 Killing-St\"{a}ckel tensors, as described in sections \ref{sec:Hamilton-Jacobi}, \ref{sec:Eisenhart-Duval}, \ref{sec:Klein_Gordon}.  With the special tensors one can build either conserved quantities in the classical theory, or symmetry operators in the quantum mechanical one. 
For the Dirac equation there is as yet no complete theory of separation of variables. There is a clear picture for what concerns symmetry operators of first order in the derivatives, and several important known examples fit into this case, such as the  Dirac equation in the Kerr metric, the higher dimensional Kerr-NUT-(A)dS metrics or the Eisenhart-Duval lift metrics that we will discuss in this section. Open avenues for future research are the description of symmetry operators of order higher than one in the derivatives for the Dirac operator, and the construction of a theory of separation of variables.

\subsubsection{Gamma matrices and differential forms} 
The typical objects that appear in the theory of symmetry operators of the Dirac equation are commutators and anti-commutators of Clifford bundle valued differential operators. Analysing such objects using explicit index tensor calculus leads to long and complicated formulas. A natural formalism that allows for practical symbolic calculations is instead that which identifies sections of the Clifford bundle with differential forms \cite{BennTucker:book}. Then operations of commutation and anti-commutation can be translated into the language of differential forms. Also, the properties of conformal Killing-Yano tensors, which are originally defined as differential forms, automatically lift to those of appropriate differential operators defined on the Clifford bundle. 

We will use the following conventions. The base spacetime is a pseudo-Riemannian spin manifold $M$ of dimension ${\dm}$ with metric ${g_{\mu\nu}}$ and local coordinates $\{ x^\mu \}$. We use lowercase Greek indices to denote  components of spacetime tensors, associated to general diffeomorphism transformations, and lowercase latin indices to denote components associated to $SO(m, \dm-m)$ transformations, where $m\ge0$. The Clifford bundle has fibers with a Clifford algebra generated by the gamma matrices ${\gamma^\mu}$: these connect the Clifford bundle with the tangent space. The gamma matrices satisfy 
\begin{equation}\label{ggmetric}
    \gamma^\mu\,\gamma^\nu + \gamma^\nu\, \gamma^\mu = 2g^{\mu\nu}\;. 
\end{equation} 
Using this relation any element $\slash\hspace{-5.5pt}\alpha$ of the Clifford algebra can be reduced to a sum of antisymmetric products ${\gamma^{\mu_1\dots \mu_p} :=\gamma^{[\mu_1}}\dots\gamma^{\mu_p]}$: 
\begin{equation}\label{clobrepr}
    \slash\hspace{-5.5pt}\alpha
     = \sum_p \frac1{p!}\, \alpha^{(p)}_{\mu_1\dots \mu_p} \gamma^{\mu_1\dots \mu_p}\;.
\end{equation} 
The coefficients are given by anti-symmetric forms $\alpha^{(p)}_{\mu_1\dots \mu_p} \in \Omega^{(p)}(\mathcal{M})$, the rank-$p$ exterior bundle, giving a unique representation. We therefore have an isomorphism ${\gamma_*}$ of the Clifford bundle with the exterior algebra ${\Omega(\mathcal{M})=\bigoplus_{p=0}^{\dm} \Omega^p(\mathcal{M})}$ of inhomogeneous antisymmetric forms: $\slash\hspace{-5.5pt}\alpha = \gamma_* \alpha$, where $\alpha = \sum_p  \alpha^{(p)}$ is an inhomogeneous form. We will typically write $\alpha$ instead of $\slash\hspace{-5.5pt}\alpha$ to describe an element of the Clifford algebra every time the context is not ambiguous.

The metric musical isomorphism allows raising and lowering : if $\omega$ is a 1--form and $V$ a vector, we denote the corresponding vector and 1--form as ${\omega^\sharp}$ and ${V^\flat}$, respectively, with the natural extension to higher rank tensors.

We remind the reader that we have defined the wedge product between forms in eq.\eqref{eq:wedge_product_definition}, and the hook operation in eq.\eqref{hook}. 
 
We can decompose a product of any two rank $p$ and $q$ gamma matrices $\gamma^{\mu_1\dots \mu_p}$ and $\gamma^{\nu_1\dots \nu_q}$ over single gamma matrices by using the Clifford algebra relation \eqref{ggmetric}. In terms of product of forms, for $\alpha\in\Omega^p(\mathcal{M})$,  $\beta\in\Omega^q(\mathcal{M})$ Clifford bundle forms, with $p\leq q$, the Clifford product is explicitly written as 
\be
\alpha \beta = \sum_{m=0}^p \frac{ (-1)^{m(p-m) + [m/2]}}{m!} \alpha \cwedge{m}  \beta  \, . \label{usefulProduct1}
\ee
$\cwedge{m}$ is a contraction operator that is defined by recursion: 
\ba
\alpha\cwedge{0}\beta &=& \alpha \wedge \beta  \, , \nn \\
\alpha\cwedge{k}\beta &=& (X_a \hook \alpha ) \cwedge{k-1} (X^a \hook \beta) \qquad (k\ge 1) \, , \nn \\
\alpha\cwedge{k}\beta &=& 0 \qquad\qquad\qquad\qquad\qquad\, (k<0) \, .  \label{cwedgedef}
\ea
In the equations above the $X^a$ vectors are an orthonormal basis defined as follows. Given a set of $\dm$--beins $\left\{ e^a_\mu \right\}$, we can build $\dm$ 1--forms $e^a = e^a_\mu dx^\mu$, with $X^a = (e^a)^\sharp$ a dual vector basis. The $e^a$ are mapped under $\gamma_*$ to a set of matrices that satisfy an equation analogue to \eqref{ggmetric} but with the flat metric $\eta^{ab}$ instead of $g^{\mu\nu}$. The coefficients $e^a_\mu$ or their inverse $E_a^\mu$ act as transformation matrices to exchange curved spacetime indices with flat locally freely falling ones. 
 
We lift the covariant derivative on $\Omega(\mathcal{M})$ to one on $\gamma_* \left( \Omega(\mathcal{M}) \right)$ with the following definition for any $\alpha$ in the Clifford bundle 
\begin{equation}\label{covderform}
    \nabla_{\!a}\alpha = \partial_a \alpha - \omega{}_a \cwedge{1} \alpha \, , 
\end{equation}
 where $\partial_a \alpha = X_a  [\alpha] = E_a^\mu \partial_\mu \alpha$, $\omega_a$ is the connection 2-form ${\omega_a= \frac12\omega_{abc} e^b\wedge e^c}$ and $\omega_{abc}$ are the components of the spin connection. 
 
If we treat the forms $\{e^a \}$ as a single object defined on $Cliff(\mathcal{M}) \times T(\mathcal{M})$ with an $SO(n)$ vector index, then its full covariant derivative is zero: 
\begin{equation}\label{covdervielbein}
    \nabla_{\!a} e^b = \partial_a e^b + \omega_a {}^b {}_c e^c  - \omega{}_a \cwedge{1} e^b = 0 \, . 
\end{equation}

Lastly we introduce the parity operator that acts on an inhomogeneous form $\alpha = \sum_{p} \alpha^{(p)}$ as 
\be \label{eq:parity}
\eta \alpha = \sum_{p=0} (-1)^p \, \alpha^{(p)} \, . 
\ee

In this formalism the Dirac operator is written as ${D\equiv e^a\nabla_{\!a}=\nabla_{\!a} e^a}$, the exterior derivative acting on forms  ${d = e^a \wedge \nabla_{\!a} = \nabla_{\!a}\, e^a \wedge}$, and the co-differential  ${\delta = - X^a \hook \nabla_{\!a} = -\nabla_{\!a}\, X^a \hook}$.

\subsubsection{Linear symmetry operators\label{sec:dirac_equation_linear_symmety_operators}} 
Given a local differential operator such as the Dirac operator $D$, an R--commuting operator $S$ is an operator that satisfies 
\be 
DS=RD
\ee
for some operator $R$. These operators transform solutions of the equation $D\psi = 0$ into solutions: given $\psi'=S\psi$, this satisfies $D\psi'=DS\psi=RD\psi=0$, so that $\psi'$ is also a solution. In particular {\em commuting operators} are important, as their eigenvalues yield quantum numbers characterizing the solution, analogues of constants of motion.

First order symmetry operators for the Dirac equations are well known. They have been described in arbitrary dimension and signature \cite{MclenaghanSpindel1979,CarterMcLenaghan1979,KamranMcLenaghan1984,BennCharlton1997,BennKress2004}. 
However, in general first order symmetry operators are not sufficient to describe separability.  Fels and Kamran have shown  that there exist systems with a separable Dirac equation that is explained by  symmetry operators of order higher than one \cite{FelsKamran1990}. There exist cases where second order symmetry operators have been built, see \cite{McLenaghanRastelli2011} and references therein. Nevertheless, there is no general construction that is  known for arbitrary dimension, nor necessary and sufficient conditions for separability are known. 
 
As seen in sec.\ref{sec:symplectic_special_tensors} Killing-Yano tensors are special forms satisfying eq.\ref{eq:symplectic_CKY_definition}. The equations generalises to the case of inhomogeneous forms. R--symmetry operators of the massless Dirac operator that are first-order in derivatives can be written uniquely in terms of conformal Killing-Yano forms, independently of the dimension $\dm$ or the signature, as shown by  
Benn, Charlton, and Kress \cite{BennCharlton1997, BennKress2004}. The generic form of these operators is 
\be
S=S_h+\alpha D\,, 
\ee
where $\alpha$ is an arbitrary inhomogeneous form, representing the freedom to add an operator proportional to $D$, while $S_h$ is given by 
 \be\label{SOProp1}
S_h=X^a\hook h\, \nabla_a+\frac{\pi-1}{2\pi}d h-\frac{n-\pi-1}{2(n-\pi)}\delta h\, ,  
\ee
h being an inhomogeneous conformal Killing-Yano form $h$ obeying \eqref{eq:symplectic_CKY_definition}. $\pi$ is the degree operator defined in sec.\ref{sec:symplectic_special_tensors}. 

In the special case of strict commutation, the operator $S$  splits into the Clifford even and Clifford odd parts \cite{MarcoDavidPavel2011} 
\be\label{Scom}
  S=S_\mathrm{e}+S_\mathrm{o}\,,
\ee
where
\begin{align}
    S_\mathrm{e}&= K_{h_\odd} \equiv X^a\hook h_\odd\nabla_{\!a}
       + \frac{\pi-1}{2\pi}dh_\odd\;,
       \label{Kdef}\\
    S_\mathrm{o}&= M_{h_\even} \equiv e^a\wedge h_\even\nabla_{\!a}
       - \frac{\dm-\pi-1}{2(\dm-\pi)}\delta h_\even\, , \label{Mdef}
\end{align}
where ${h_\odd}$ is an odd Killing-Yano form and $h_\even$ is an even closed conformal Killing-Yano form.

\subsubsection{Separability in Kerr-NUT-(A)dS metrics\label{Dirac:KerrNUTAdS}} 

Especially interesting  is the case when one has a {\em complete set of mutually commuting operators} and their common eigenfunctions can be found by separating variables. The corresponding eigenvalues then completely characterize the separated solution and play the role of separation constants. 
 
One non-trivial example is given by the higher dimensional Kerr-NUT-(A)dS black hole metrics of sec.\ref{sec:higher_dimensional_black_holes}, where the Dirac equation is separable \cite{OotaYasui2008} and the separability is fully accounted for by a complete set of linear symmetry operators that are mutually commuting and admit common separable spinorial eigenfunctions.
 
In this background there exist $N+\eps$ Killing vectors $\xi^{(0)}, \dots, \xi^{(N-1+\eps)}$ and $N$ closed conformal Killing-Yano forms $h^{(j)}$. We associate to them operators $K_{\xi^{(0)}}, \dots, K_{\xi^{(N-1+\eps)}}$ and, respectively, $M_{h^{(1)}}, \dots, M_{h^{(N-1)}}$. Using the results of section \ref{sec:dirac_equation_linear_symmety_operators} we know that each of these commutes with the Dirac operator $D$. Cariglia, Krtou\v{s} and Kubiz\v{n}\'ak have shown that all the operators are mutually commuting \cite{MarcoDavidPavel2011}. Therefore, they can be simultaneously diagonalised and it is possible to construct common spinorial eigenfunctions. 
 The solution to the eigenvalue problem 
\be\label{eigenvaluetilde}
K_{\xi^{(k)}} \chi = i\,\psc{k}\chi\;,\quad M_j \chi = m_j \chi\;,
\ee
can be found in the tensorial R--separated form \cite{MarcoDavidPavel2011_2}
\begin{equation}\label{tens_sep}
    \chi = R \exp\bigl({\textstyle i\sum_k\psc{k}\psi_{k}}\bigr)\,
           \bigotimes_\nu \chi_\nu\;,
\end{equation} 
where $\left\{\chi_\nu \right\}$ is an $N$-tuple of 2-dimensional spinors and ${R}$ is a  Clifford bundle-valued prefactor. $\chi_\nu$ is a function of the variable ${x_\nu}$, $\chi_\nu=\chi_\nu(x_\nu)$, and satisfies the equation 
\begin{equation}\label{chieq}
\begin{split}
&\Biggl[\Bigl(
    \frac{d}{dx_\nu}+\frac{X_\nu'}{4X_\nu}
    +\frac{\tilde \Psi_\nu}{X_\nu}\iota_{\lst{\nu}}
    +\frac{\eps}{2x_\nu}\Bigr)\,\sigma_{\lst{\nu}} \\
&\hspace{70pt}- \,\frac{\bigl(- \iota_{\lst{\nu}} \bigr)^{\!N\!-\!\nu}}{\sqrt{|X_\nu|}}
   \Bigl(\eps \frac{i\sqrt{-c}}{2x_\nu^2} +m_\nu\Bigr)\Biggr]\,\chi_\nu=0\, , 
\end{split}
\end{equation}
where 
\begin{equation}\label{psfdef}
    \psf{\mu} = \sum_k \psc{k} (-x_\mu^2)^{N{-}1{-}k}\; , 
\end{equation}
and 
\be  \label{eq:chf_mu_definition}
m_\nu =  \sum_{j} (-i)^j m_j \left( -\iota_{\lst{\nu}} x_\nu \right)^{N-1-j} \, .
\ee
$\iota_{\lst{\nu}}$ is an operator acting only the $2$-spinor $\chi_\nu$ as a $\sigma_3$ Pauli matrix, and $\sigma_{\lst{\nu}}$ acts similarly as as $\sigma_1$ matrix. 
 
The common eigenspinor \eqref{tens_sep} is the solution given in \cite{OotaYasui2008}, that was found with a direct calculation that proved separability. The eigenvalues $\psc{k}$ and $m_j$ are mapped into the arbitrary integration constants found there. A similar construction holds for weakly charged rotating black holes \cite{CarigliaEtAl2012}.

\subsubsection{Fluxes\label{sec:Dirac_equation_with_flux}} 
In some cases of physical interest the Dirac operator gets modified by the inclusion of flux terms. Typically this happens in backgrounds with scalar potential, including the massive Dirac equation, backgrounds with electromagnetic fields, torsion, or higher order forms in supergravity and superstring related geometries. It is still possible to study operators that R-commute with the modified Dirac operator, \cite{DavidPavelClaude2011}, or that graded commute with it \cite{VercinEtAl2009}. We focus here our attention on the solution generating R-commuting case.  One interesting feature arising is that, differently from the cases discussed so far with no flux, where the existence of conformal Killing-Yano tensors guarantees having a linear R-commuting operators, in the presence of flux anomalies can appear. This means that even if special tensors exist that satisfy a modified conformal Killing-Yano equation with flux, in general these will not generate linear R-commuting operators unless they satisfy an additional set of constraints. 
 
When fluxes are present they can be represented by an inhomogeneous form $B$. Then the Dirac operator with flux is written as 
\be 
\mathcal{D} = D + B \, . 
\ee 
We can define a bracket operator on forms and on the Clifford bundle by setting 
  \be\label{eq:bracket}
\{\alpha,\beta\}\equiv\alpha\beta-(-1)^{q}\beta\alpha\,,
\ee
for a $p$-form $\beta$ and a $q$-form $\alpha$. Then an $R$-commuting operator $S$ satisfies
\be\label{eq:R_comm_flux}
\{\mathcal{D},S\}= R \mathcal{D}\,
\ee
for some operator $R$.
As before, there is the freedom to add a term $\alpha \mD$, which has the result of at most changing the operator $R$. 
We follow the argument of \cite{BennKress2004} to construct the generic symmetry operator of $\mD$: we look for a special operator $S$, 
\be\label{eq:S}
S=2h^a\nabla_a+\Omega\,,
\ee 
where $h^a$ and $\Omega$ are for now unknown inhomogenous forms. We ask that $S$ satisfied \eqref{eq:R_comm_flux}, with $R$ an inhomogeneous form. 
 
First, one finds that $h^a$ can be generated from an inhomogeneous form $h$ via 
\be\label{eq:h_a}
h^a=X^a\hook h\, .  
\ee
where the form $h$ satisfies the generalised conformal Killing-Yano equation 
\be \label{eq:CKY_with_flux} 
\nabla_a h-\frac{1}{\pi+1}X_a\hook dh+ \frac{1}{\dm-\pi+1}e_a\wedge\delta h + \{B,\omega_a\}_\perp=0 \, .  
\ee 
Here $(\cdot)_\perp$ is the following projector operator for inhomogeneous forms: 
\ba\label{eq:projector}
(\alpha_a)_\perp& \equiv &\alpha_a-\frac{1}{\pi+1}X_a\hook\bigl(e^b\wedge \alpha_b\bigr) \nn \\ 
&& -\frac{1}{\dm-\pi+1}e_a\wedge\bigl(X^a\hook\alpha_a\bigr)\, , 
\ea 
which satisfies $e^a\wedge (\alpha_a)_\perp=0=X^a\hook (\alpha_a)_\perp$.  
 
Second, the $\Omega$ term in \eqref{eq:S} is given by 
\ba \label{eq:Omega} 
\Omega &=& \frac{\pi-1}{\pi}dh-\frac{\dm-\pi-1}{\dm-\pi}\delta h-\frac{1}{\dm-\pi}X^a\hook \{B,h_a\} \nn \\ 
&& -\frac{1}{\pi}e^a\wedge\{B,h_a\}+ f +\epsilon\, ,\quad
\ea 
where $f$ and $\epsilon$ are an arbitrary $0$-form and, respectively, $n$-form. $R$ is given by 
\be \label{eq:R_form} 
R = -\frac{2\eta}{\dm-\pi}\delta h +\frac{2\eta}{\dm-\pi}X^a\hook \{B,h_a\}-2\eta\epsilon\, , 
\ee 
$\eta$ being the parity operator defined in \eqref{eq:parity}. 
In general it is not sufficient that $h$ satisfies \eqref{eq:CKY_with_flux} to generate an R-symmetry operator $S$ as given by \eqref{eq:S}. There is one last condition to be satisfied: 
\ba \label{eq:anomaly_flux_case}
&& \delta\left(\frac{1}{\dm-\pi}X^a\hook \{B,h_a\}\right)-d\left(
\frac{1}{\pi}e^a\wedge\{B,h_a\}\right) -\nabla^a\{B,h_a\} \nonumber\\
&& 
+\bigl\{B,\Omega\}-2(\eta h^a)\nabla_aB- R B+df-\delta\epsilon=0\,. 
\ea
This condition does not exist in the case of zero flux, the conformal Killing-Yano equation being sufficient in that case. For this reason \eqref{eq:anomaly_flux_case} is sometimes referred to as an absence of anomalies for a generalised conformal Killing-Yano tensor to produce an R-commuting operator in the presence of flux. 
 
One may choose the forms $f$ and $\epsilon$ to simplify the anomalies. We give now a number of examples. 

The first example is the Dirac equation with potential, setting $B=iV$. One gets the conditions 
\be
\{V,h_a\}=2X_a\hook (V h_e)\,,\quad \{V,h_a\}_\perp=0\,,
\ee
where $h_e$ is the even part of $h$.
Then $h$ obeys a conformal Killing-Yano equation with no flux. Fixing $f=0=\epsilon$ the anomaly condition is 
\ba\label{eq:scalar_potential_condition}
&& -2e^a\nabla_a(Vh_e)+2V\Omega_o-2(\eta h^a)\nabla_aV \nn \\ 
&& +\frac{2\eta}{\dm-\pi}\delta h V=0\,.
\ea
There are two interesting specific cases: i) $h$ is odd, $h=h_o$, and ii) $h$ is even, $h=h_e$.
If $h$ is odd eq. \eqref{eq:scalar_potential_condition} becomes 
\be\label{conV1}
(dV)^\sharp\hook h-\frac{V}{\dm-\pi}\delta h=0\,.
\ee
For example, when $V=m=const$ then it must be $\delta h=0$. If $h$ is even eq. \eqref{eq:scalar_potential_condition} becomes 
\be\label{conV2}
dV\wedge h +\frac{V}{\pi}dh=0\,,
\ee 
and when $V=m$ it must be $dh=0$. It is known that symmetry operators of the massive Dirac equation are given in terms of Killing--Yano tensors of odd rank or in terms of closed conformal Killing--Yano tensors of even rank \cite{BennCharlton:1997}. 

The second example we consider is when $h$ is a $1$-form. In this case eq.\eqref{eq:CKY_with_flux} becomes a conformal Killing vector equation. The anomaly condition \eqref{eq:anomaly_flux_case} with $f = 0 = \epsilon$ becomes 
to
\begin{equation}
\mathcal{L}_{h^\sharp} B = -\frac{\delta h}{\dm}(\pi-1)B.
\end{equation}
This is equivalent to invariance under the conformal symmetry generated by $h^\sharp$ of the action
\begin{equation}
S = \int(\overline{\psi} D \psi + \overline{\psi} B \psi)\, d^\dm q \, , 
\end{equation}
which gives rise to the modified Dirac equation. 
 
The last example is given by $U(1)$ and $3$-form fields. If the Dirac spinor is coupled to a skew-symmetric torsion and a $U(1)$ field then we can take 
\be
B=iA-\frac{1}{4}T\,,
\ee
where $A$ is a 1-form and $T$ a 3-form. As mentioned in sec.\ref{connection_with_torsion}, specifically eq.\eqref{eq:natural_Dirac_torsion}, the factor of $-\frac{1}{4}$ arises naturally if we want to interpret $T$ as the torsion tensor of the covariant derivative $\nabla^T = \nabla + T$, where $\nabla$ is the Levi-Civita connection. 
 
In general the connection with torsion will act on forms as 
\be \label{eq:connection_with_torsion}
\nabla_a^T\alpha =\nabla_a\alpha+\frac{1}{2}(X_a\hook T)\cwedge{1}\alpha\,.
\ee 
One can associate with it the following two operations: 
\ba\label{def2}
\delta^T\alpha &=&-X^a\hook\nabla^T_a\alpha=\delta\alpha-\frac{1}{2}T\cwedge{2}\alpha\,,\nonumber\\
d^T\alpha&=&e^a\wedge \nabla^T_a\alpha=d\alpha-T\cwedge{1}\alpha\,. 
\ea

The generalised conformal Killing-Yano equation \eqref{eq:CKY_with_flux} becomes 
\be\label{eq:Ttwistor}
\nabla_a^Th-\frac{1}{\pi+1}X_a\hook d^Th+\frac{1}{n-\pi+1}e_a\wedge \delta^T h=0\,,
\ee
and each $p$-form component of $h$ has to satisfy the equation separately. This equation was originally introduced in \cite{KubiznakEtal:2009b}. The $\Omega$ and $R$ forms are given by 
\ba
\Omega&=&\frac{\pi-1}{\pi}dh-\frac{n-\pi-1}{n-\pi}\delta^Th+2iA\cwedge{1}h
 \nn \\ 
&& +\frac{2-\pi}{2\pi}T\cwedge{1}h 
-\frac{1}{2}T\cwedge{2}h\!+\!\frac{1}{12}T\cwedge{3}h\!+\! f \!+\!\epsilon\,,\nonumber\\
R &=& -\frac{2\eta}{n-\pi}\delta^Th-2\eta\epsilon\,, 
\ea
and  $S$ by 
\ba\label{eq:realsymetry} 
S&=&2X^a\hook h \nabla_a+\frac{\pi-1}{\pi}dh+2iA\cwedge{1}h \nn \\ 
&& +\frac{2-\pi}{2\pi}T\cwedge{1}h
-\frac{1}{2}T\cwedge{2}h+\frac{1}{12}T\cwedge{3}h+f\,,
\ea
which is the symmetry operator derived in \cite{HouriEtal:2010a}, up to the arbitrary 0-form $f$ and in the case when $A=0$.

The anomaly condition \eqref{eq:anomaly_flux_case} can be re-expressed in the form 
\be
2i(dA)\cwedge{1}h+A^{(cl)}+A^{(q)}-df+\delta\epsilon=0\,, \label{maxtoran}
\ee
where 
\ba
A^{(cl)}&=&\frac{1}{\pi-1}d(d^Th)-\frac{1}{2}dT\cwedge{1}h \nn \\ 
&& -\frac{1}{n-\pi+3}T\wedge\delta^Th\,,\\
A^{(q)}&=&\frac{1}{n-\pi-1}\delta(\delta^Th)-\frac{1}{6(\pi+3)}T\cwedge{3}d^Th \nn \\ 
&& +\frac{1}{12}dT\cwedge{3}h\,,
\ea
where one can recognise the the `quantum' and `classical' anomalies discussed in \cite{HouriEtal:2010a}. It is worth noticing that the $A$ and torsion anomalies are not coupled. If 
$h$ is a homogeneous form of rank $p$ then the first three terms in
\eqref{maxtoran} are of rank $p$, $p+2$ and, respectively, $p-2$: so each must be zero independently. Some of these terms might be further simplified using the freedom of the $f$ and $\epsilon$ terms. When $T=0$ this equation has been discussed in \cite{VercinEtAl2009}, and when $A= 0$ in \cite{HouriEtal:2010a}. 
 
There are two main examples in the literature of generalised closed conformal Killing-Yano tensors that satisfy the anomaly equation. One is the generalized closed conformal Killing--Yano tensor, with torsion given by the 3-form flux $H$ \cite{Houri:2010fr}, in Kerr--Sen black hole spacetimes in generic dimension: it satisfies the anomaly equation with $f=0=\epsilon$ \cite{Sen:1992, CveticYoum:1996bps, Chow:2008}. The other is minimal five-dimensional supergravity, with the torsion identified with the dual of Maxwell field $T=*F/\sqrt{3}$, for the  most general black hole solution \cite{ChongEtal:2005b}: in this case one has to choose $\epsilon=0$ and $f=-\frac{1}{12}T\wedge\!\!\!\!_{{\!}_3}\,h$ \cite{KubiznakEtal:2009b}. One can also recover the symmetry operator
\be
S=2X^a\hook h \nabla_a+\frac{3}{4}dh+2iA\cwedge{1}h\,
\ee
 for the massive minimally coupled Dirac equation with torsion obtained by Wu \cite{Wu:2009b} noticing that $T\wedge\!\!\!\!_{{}_1}\,h=0=T\wedge\!\!\!\!_{{\!}_2}\,h$.

More examples with a $5$- and $7$-form are discussed in \cite{DavidClaudePavel2010}.

\subsubsection{Lift and reduction for Eisenhart-Duval metrics\label{sec:Dirac_EisenhatDuval}}
In this section we discuss the main features of the lift and reduction procedure for solutions of the Dirac equation in Eisenhart-Duval  metrics. We use the notation introduced in sec.\ref{sec:Eisenhart-Duval}. 
 
The Eisenhart-Duval lift allows relating the massive Schr\"{o}dinger equation in Riemannian $d$--dimensional spacetime $\M$ to the massless free Klein-Gordon equation in $\dm+2$--dimensional Lorentzian Eisenhart-Duval spacetime $\h{\M}$ by projecting on the base space, see for example \cite{HorvathyZhang2009}. Similarly, dimensional reduction of the massless Dirac equation on $\h{\M}$ can be put in correspondence with the L\'evy-Leblond equation \cite{LevyLeblond1967}, its non-relativistic counterpart on $\M$. Duval, Horv\'athy and Palla derived the L\'evy-Leblond equation from a lightlike reduction from 4 and 5 dimensions in \cite{DuvalHorvathyPalla1996}. 
 
In order to set up the relationship, we examine how to relate spinorial quantities on $\M$ and $\h{\M}$. 
Spinors on $\M$ have dimension $2^{\left[ \frac{\dm}{2} \right]}$, while spinors on $\h{\M}$ have dimension $2^{\left[ \frac{\dm +2}{2} \right]}= 2 \cdot 2^{\left[ \frac{\dm}{2} \right]}$. We introduce the Pauli matrices 
\begin{equation}\label{eq:Pauli_matrices}
\begin{gathered}
\sigma_1 \equiv
    \left(\begin{array}{cc}
        0 & 1 \\
        1 & 0 \\
      \end{array}\right)
    \;,\quad
    \sigma_2 \equiv
    \left(\begin{array}{cc}
        0 & -i \\
        i & 0 \\
      \end{array}\right)
    \; , \quad     
\sigma_3 \equiv
    \left(\begin{array}{cc}
        1 & 0 \\
        0 & -1 \\
      \end{array}\right)
    \; ,  
\end{gathered}
\end{equation}
and define $\sigma^\pm = \frac{\sigma_1 \pm i \sigma_2}{2}$. These satisfy $(\sigma^\pm)^2 = 0$, $\left\{ \sigma^+ , \sigma^- \right\} = \mathbb{I}$. 
 
Now we can relate the gamma matrices for $\M$, $\gamma^a$, with those on $\h{\M}$, $\h{\gamma}^A$, using the following representation: 
\be \label{eq:gamma_matrices_representation} 
\left\{ \begin{array}{rcl} 
\h{\gamma}^+ &=& \sigma^+ \otimes \mathbb{I} \, ,  \\ 
\h{\gamma}^- &=& \sigma^- \otimes \mathbb{I} \, , \\ 
\h{\gamma}^a &=& \sigma_3 \otimes \gamma^a \, . 
\end{array} \right. 
\ee 
Covariant derivatives of a spinor $\h{\psi}$ on $\h{\M}$ can be written in terms of quantities defined on $\M$ using the explicit form of the spin connection  \eqref{eq:EisenhartDuval_spin_connection}: 
\be 
\left\{ \begin{array}{rcl} 
\h{\nabla}_- &=& \partial_v \, ,  \\ 
\h{\nabla}_+ &=& \left(\frac{V}{m} \partial_v + \partial_t \right) - \frac{1}{2m}\h{\gamma}^+ dV - \frac{e}{4m} F \, , \\ 
\h{\nabla}_a &=& \nabla_a - \frac{e}{m} A_a \partial_v + \frac{e}{4m} \h{\gamma}^+ \h{\gamma}^b F_{ba}  \, . 
\end{array} \right. 
\ee
This allows expressing the Dirac operator on $\h{\M}$ as 
\ba 
\h{D} = \h{\gamma}^A \h{\nabla}_A &=&  \h{\gamma}^- \partial_v + \h{\gamma}^+ \left( \frac{V}{m}\partial_v + \partial_t + \frac{e}{4m} F \right) \nn \\ 
&& + \h{\gamma}^a \left( \nabla_a - \frac{e}{m} A_a \partial_v \right) \, . 
\ea 
The symmetry operators of eq.\eqref{SOProp1} for a Killing vector $\h{K}$ are given by 
\be 
\h{S}_{\h{K}} = \h{\nabla}_{\h{K}} + \frac{1}{4} \h{d} \h{K} \, . 
\ee 
In the specific case of the two Killing vectors $\h{X}^+$ and $\h{X}^- - \frac{V}{m} \h{X}^+$ they take the form 
\be 
K_{\h{X}^+} = \partial_v \, , 
\ee 
and 
\be 
K_{(\h{X}^- - \frac{V}{m} \h{X}^+)} = \partial_t \, . 
\ee
Since they both commute with the Dirac operator $\h{D}$ on $\h{\M}$, then one can ask that solutions $\h{\psi}$ on $\h{\M}$ of $\h{D} \h{\psi} = 0$ are eigenspinors of the two operators. However, for the purpose of recovering the L\'evy-Leblond equation it is sufficient to consider the less restrictive condition of $\h{\psi}$ being eigenspinor only of $K_{\h{X}^+}$: 
\be \label{eq:Eisenhart_Duval_reduction_partial_v_psi} 
\partial_v \h{\psi} = im \h{\psi} \, . 
\ee
Choosing the $im$ eigenvalue reduces the action of the Dirac operator to 
\be 
\h{D} = i m  \h{\gamma}^- + \h{\gamma}^+ \left[ i V + \partial_t + \frac{e}{4m} F \right] + \h{\gamma}^a \mathcal{D}_a \, ,  
\ee
where  $\mathcal{D}_a = \nabla_a - i e A_a$ is the $U(1)$ covariant spinor derivative on $\M$, and $\mathcal{D} = \Gamma^a \mathcal{D}_a$ represents the Dirac operator on $\M$ with A flux. 
 
One can decompose a spinor $\h{\psi}$ on $\h{\M}$ following the gamma matrices representation \eqref{eq:gamma_matrices_representation}: 
\be \label{eq:oM_spinor_split}
\h{\psi} = \left( \begin{array}{c} \chi_1 \\ \chi_2 \end{array} \right) \, , 
\ee 
where  $\chi_1$ and $\chi_2$ are spinors on $\M$. Finally, the massless Dirac equation on $\h{\M}$, $\h{D} \h{\psi} = 0$, decomposes into the two equations 
\be \label{eq:Levy-Lebond} \left\{ \begin{array}{lcl} 
\partial_t \chi_2 + \mathcal{O} \chi_2 + \mathcal{D} \chi_1  &=& 0 \, ,  \\  i m \chi_1 - \mathcal{D} \chi_2 &=&0  \, , \end{array} \right.  
\ee 
where the operator $\mathcal{O}$ is given by $\mathcal{O} = i V  + \frac{e}{4m} F$. In this way one obtains the non-relativistic L\'evy-Leblond equation for a particle of mass $\tilde{m} = \frac{m}{2}$, in curved space, with scalar and vector potential  and with an additional term $\frac{e}{4m} F$. This extra term induces an anomalous gyromagnetic factor $g=3/2$ and was not present in the original work by L\'evy-Leblond. Therefore the Eisenhart-Duval metric allows a geometrical derivation of the L\'evy-Leblond equation. This is ultimately possible since it is possible to embed the Bargmann group -- the central extension of the Galilei group that leaves invariant the Schr\"{o}dinger equation and the L\'evy-Leblond equation -- in the de Sitter  group $O(1,\dm +1)$ \cite{DuvalBurdetKunzlePerrin1985}. 

While it is always possible to dimensionally reduce the Dirac equation on $\h{\M}$ to the L\'evy-Leblond  equation on $\M$, sometimes it is possible through reduction to obtain a fully relativistic Dirac equation on $\M$. To do this it is necessary to ask that $\partial_t \h{\psi} = 0$ in addition to \eqref{eq:Eisenhart_Duval_reduction_partial_v_psi}, and the non-trivial projection 
\be \label{eq:spinor_projection}
i V \chi_1 = \mathcal{O} \chi_2 \, .  
\ee 
This projection will be satisfied only for specific combinations of $V$ and $F$. For what concerns hidden symmetry operators, from the lift point of view there exist symmetry operators on $\M$ generated by conformal Killing-Yano tensors that cannot be lifted on $\h{\M}$. From the point of view of reduction instead there are Killing-Yano and closed conformal Killing-Yano forms on $\h{\M}$ that arise as lifts of Killing-Yano and closed conformal Killing-Yano forms on $\M$, but such that their corresponding symmetry operators cannot be dimensionally reduced. These Killing-Yano and closed conformal Killing-Yano forms are exactly those that in lower dimension would generate anomalous symmetry operators. In the remaining cases it is possible to establish a correspondence between symmetry operators on $\M$ and $\h{\M}$.  More details and explicit formulas can be found in \cite{Marco2012}.

\section{GEODESICS ON LIE GROUPS\label{sec:geodesics_lie_groups}} 
In this section we describe a relatively new framework where hidden symmetries of the dynamics have been applied, that of Lie groups. Lie groups naturally have a rich geometrical structure and therefore dynamical systems associated to them provide good candidates for the study of hidden symmetries. Nevertheless, to our knowledge the study of such systems has started only recently: Barberis, Dotti and Santillan studied the Killing-Yano equation on a number of Lie groups with a left-invariant metric \cite{BarberisDottiSantillan2012}, while Cariglia and Gibbons analysed the Toda chain which can be described as geodesic motion on a Lie group, and for which a generalised Eisenhart lift metric, as described in sec.\ref{sec:n_plus_one_Eisenhart_lift}, naturally arises \cite{GaryMarco2013}. 
 
In this section we describe the Toda chain result. A brief historical digression can help understanding where this analysis fits. It is a known result by Perelemov and Olshanetsky that there exist a number of physical systems describing $\dm$ interacting particles on a line that are integrable both classically and quantum mechanically \cite{PerelomovOlshanetsky1981,PerelomovOlshanetsky1983}. Some of these systems involve pairwise interaction via a potential $V(q)$ of the following five types: $V_1 = \frac{1}{q^2}$, $V_2 = \frac{1}{\sinh^2 q}$, $V_3 = \frac{1}{\sin^2 q}$, $V_4 = \wp (q)$, $V_5 = \frac{1}{q^2} + \omega^2 q^2$, where $\wp (q)$ is the Weierstrass elliptic  function, and $q$ the relative distance. The first three potentials therefore are special cases of the fourth, as similarly the fifth potential includes the first. Separately, the Toda chain is considered which is made by particles interacting only with the nearest neighbour via an exponential potential $V = \exp (q)$. For all the potentials apart from $V_4$ it is shown how the motion can be obtained as an appropriate projection of geodesic motion on a higher dimensional Lie group. Thus one may ask the following question: for the systems above, expect $V_4$, we know that there exist two different higher dimensional spaces such that the original motion is the projection of geodesic motion in higher dimension. The first such space is the appropriate Lie group described in \cite{PerelomovOlshanetsky1981}, while the second one is the Eisenhart lift of section \ref{sec:Eisenhart-Duval}. Is there a relationship between these two spaces? 
 
The question has been first asked in the case of Toda systems in \cite{GaryMarco2013}, and the answer in that case is that the Lie group is endowed with a generalised Eisenhart lift metric, that can be appropriately reduced to the standard Eisenhart lift metric. The answer for the other systems listed in \cite{PerelomovOlshanetsky1981} is unknown as yet. 
 
\subsection{Geodesic motion for the Toda chain} 
  
\subsubsection{The system} 
The Toda system describes a chain of particles on a 1-dimensional line, interacting via an exponential, nearest neighbour potential. It was first presented by Toda in \cite{Toda1967}, in particular showing that among solutions there are solitonic "travelling waves".  Henon \cite{Henon1974} and Flaschka \cite{Flaschka1974} constructued constants of motion, and showed that the Toda chain is a finite dimensional analog of the KdV equation. 
  
The Hamiltonian of the non-periodic Toda system is given by  
\be \label{eq:non_periodic}
H(p,q) =  \sum_{i=i}^\dm \frac{p_i^2}{2} + V(q) = \sum_{i=i}^\dm \frac{p_i^2}{2} + \sum_{i=1}^{\dm -1}g_i^2 e^{2 (q_i - q_{i+1})} \, . 
\ee 
The particles interact with their nearest neighbour via an exponential potential. The equations of motion are 
\ba 
\dot{q}_i &=& p_i \, , \nn \\ 
\dot{p}_1 &=& - 2 g_1^2 e^{2(q_1 - q_2)} \, , \nn \\ 
\dot{p}_i &=& - 2 g_i^2 e^{2(q_i - q_{i+1})} + g_{i-1}^2 e^{2(q_{i-1} - q_i)} \, , \;\; i >1 \, . 
\ea 
Perelomov and Olshanetsky displayed a Lax pair for the system, given by the matrices  
\ba 
L_{ij} &=& \delta_{ij} p_j + g_{i-1} \delta_{i,j+1} + g_i \, e^{2(q_i - q_{i+1})} \delta_{i, j-1} \, , \label{eq:Lax_L} \\ 
M_{ij} &=& 2 g_i \, e^{2(q_i - q_{i+1})} \delta_{i, j-1}  \label{eq:Lax_M} \, . 
\ea 
This means that the equations of motion can be rewritten in the form $\dot{L} = [L,M]$. 
Agrotis, Damianou and Sophocleous showed the Toda system is superintegrable \cite{Damianou2006}. It is important for our purposes noticing that the properties of integrability and existence of the Lax pair are valid for all choices of the coupling constants. 
 
We can lift the Lax pair to the Eisenhart lift space by setting 
\ba 
\mathcal{L}_{ij} &=& \delta_{ij} p_j + p_y g_{i-1} \delta_{i,j+1} + p_y g_i \, e^{2(q_i - q_{i+1})} \delta_{i, j-1} \, , \nn \\ 
\mathcal{M}_{ij} &=& 2 p_y g_i \, e^{2(q_i - q_{i+1})} \delta_{i, j-1} \nn \, . 
\ea 
Since the original Lax pair was defined for all values of the coupling constants, and since $p_y$ is constant, then it will still be the case that $\dot{\mathcal{L}} = [\mathcal{L},\mathcal{M}]$, and that we can build invariants according to $\mathcal{I}_i = \frac{1}{2^i} Tr \mathcal{L}^i$, $i= 1, \dots, \dm$. However, this time the $\mathcal{I}_i$ are polynomials in the momenta $p_\mu = \{p_1, \dots, p_\dm, p_y\}$ of degree $i$, and therefore they must correspond to rank $i$ Killing tensors $K_{(i)}^{M_1 \dots M_i}$, $M= 1, \dots, \dm+1$, according to the formula 
\be 
\mathcal{I}_i = \frac{1}{i!} \, K_{(i)}^{M_1 \dots M_i} p_{M_1} \dots p_{M_i} \, . 
\ee 
So using this technique we can build new examples of non-trivial higher rank Killing tensors. We can apply the same reasoning to the generalised Eisenhart lift of eq.\eqref{eq:Eisenhart_2nd_type_generalised_Hamiltonian}, thus obtaining a Lax pair and Killing tensors for the generalised lift metric. This technique will work for the lift of any system that admits a Lax pair where the entries of the Lax matrix $L$ are homogeneous in the mixed variables $(p,g)$.

\subsubsection{The lift on a Lie group} 
Olshanetsky and Perelomov \cite{PerelomovOlshanetsky1981} showed that the Toda chain can be obtained by projecting to lower dimension the geodesic equations on the Lie group $X^- = SO(\dm) \backslash SL(\dm, \mathbb{R})$. The elements of $X^-$ are symmetric positive-definite $\dm\times \dm$ matrices $x$ with real components and unit determinant, $\det x = 1$. In the rest of this section we mostly follow the notation used in the original work. 
 
Real symmetric positive definite matrices $x \in X^-$ can be written according to a Cholesky $UDU$ decomposition 
\be \label{eq:n_x_2}
x = Z h^2 Z^T  \, , 
\ee 
where $h^2$ is diagonal and with positive elements, and $Z \in \mathcal{Z} \subset SL(\dm, \mathbb{R})$, the subgroup of upper triangular matrices with units on the diagonal. We parameterise $h^2$ as 
\be 
h^2 = diag [ e^{2q_1}, \dots, e^{2q_\dm} ] \, , 
\ee 
where the $q_i$ fields, of the dilaton type, will be identified with the positions of the particles in the Toda chain \eqref{eq:non_periodic}. The condition $\det h^2 = 1$ implies the restriction 
\be \label{eq:det1}
\sum_{a=1}^\dm q_a = 0 \, , 
\ee 
which amounts to a choice of the position of the system's centre of mass. 
 
The equation of motion on $X^-$ is given by 
\be \label{eq:geodesic}
\frac{d}{dt} \left( \dot x (t) x^{-1} (t) \right) = 0 \, .  
\ee 
In \cite{PerelomovOlshanetsky1981} it is shown that if one chooses specific geodesics for which 
\be \label{eq:review_specific_geodesic} 
Z^{-1} \dot{Z} = M \, , 
\ee 
with $M$ given by \eqref{eq:Lax_M}, then \eqref{eq:geodesic} implies the Lax equations $\dot{L} = [L,M]$. 

It is worth noticing that the condition \eqref{eq:review_specific_geodesic}, which selects specific geodesics, is equivalent to say that $Z^{-1}$ satisfies the evolution equation \eqref{eq:Lax_evolution_matrix}. Flaschka in 1974 displayed a different Lax pair, where the evolution matrix is orthogonal \cite{Flaschka1974}.

\subsubsection{The generalised Eisenhart metric} 
In this section we will endow $X^-$ with a right invariant metric whose geodesics correspond to trajectories in the $\dm$-dimensional non-periodic Toda system when appropriately projected. We will see that the higher dimensional metric is a generalised Eisenhart lift metric. 
 
In order to present the results we need a small number of results on upper unitriangular matrics. We will indicate a generic such matrix as $Z$. They form a subgroup  of $SL(\dm, \mathbb{R})$, with Lie algebra $\mathcal{Z}$ generated by the strictly upper triangular matrices $M_{ab}$, $a,b = 1, \dots, \dm$, $a < b$, with components 
\be 
\left( M_{ab}\right)_{ij} = \delta_{ia} \delta_{jb} \, . 
\ee 
From the definition it follows that $M_{ab}^2 = 0$. 
$M$ matrices satisfy the product rule 
\be \label{eq:Mproduct}
M_{ab} M_{cd} = \delta_{bc} M_{ad} \, . 
\ee 
We will also use diagonal matrices $D$, with elements $(D)_{ij} = d_i \delta_{ij}$ (no sum). They obey mutual product rules 
\be \label{eq:MD}
M_{ab} D = d_b M_{ab} \, , \quad D M_{ab} = d_a M_{ab} \, . 
\ee 
In particular, we will use a basis of diagonal, zero trace matrices given by $M_a$, $a=1, \dots, \dm -1$: 
\be 
M_a = diag \left[ 0, \dots, 1, \dots, 0, -1 \right] \, , 
\ee 
with the $1$ term in the $a$-th position. It is convenient to also define $M_\dm = 0$. 
 
Analogous properties are satisfied by lower diagonal matrices, which we will indicate by $\Mb_{ab}$, $a,b = 1, \dots, \dm$, $a > b$, with components 
\be 
\left( \Mb_{ab}\right)_{ij} = \delta_{ia} \delta_{jb} \, , 
\ee 
as they satisfy 
\be \label{eq:MD_2}
\Mb_{ab} D = d_b \Mb_{ab} \, , \quad D \Mb_{ab} = d_a \Mb_{ab} \, .  
\ee 
The mixed $M \Mb$ product rule is 
\be 
M_{ab} \, \Mb_{cd} = \delta_{bc} \left( M_{ad} + \Mb_{ad} + \delta_{ad} \mathbb{I}_a \right) \, , 
\ee 
where $\mathbb{I}_a$ is the diagonal matrix with a $1$ in the $a$-th element and zero otherwise. We use these matrices to parameterise the Lie algebra of $SL(\dm , \mathbb{R})$ using the set $\{ M_{ab}, \Mb_{ab}, M_a \}$.

We use the $M$ matrices to parameterise a generic matrix $Z \in \mathcal{Z}$ as 
\be 
Z = \exp \left(\sum_{a<b} \omega_{ab} M_{ab}\right) \, . 
\ee 
We have in mind the specific case $\dot{Z} Z^{-1} = M$, where $M$ is the second matrix in the Lax pair. This implies  $\dot{\omega}_{a,a+1} = 2 g_a e^{2(q_a - q_{a+1})}$, and  $\dot{\omega}_{ab} = 0$ otherwise, so we can simplify our initial assumption and set to zero all $\omega_{ab}$  unless $b = a+1$.  Doing this is equivalent to considering a $2(\dm -1)$-dimensional submanifold $\tilde{X}$ of $X^-$, with coordinates $q_a$ and $\omega_{a,a+1}$. We will write $\omega_a$ instead of $\omega_{a,a+1}$. The parameterisation becomes 
\be 
Z = \exp \left(\sum_{a=1}^{\dm -1} \omega_a M_{a, a+1}\right) \, . 
\ee 
$\tilde{X}$ is a totally geodesic submanifold of $X^-$ since by construction geodesics on $\tilde{X}$ can be obtained from generic geodesics on $X^{-}$ by imposing $\dot{\omega}_{ab} = 0$ for $b>a+1$. It is convenient defining $\omega_{0} = 0 = \omega_n$. 
 
Using repeatedly \eqref{eq:Mproduct} one finds that for $a<b$ 
\be 
Z_{ab} = \frac{1}{(b-a)!} \omega_a \omega_{a+1} \dots \omega_{b-1} \, 
\ee 
and 
\be 
Z^{-1}_{ab} = (-1)^{b-a} Z_{ab} \, . 
\ee 

In order to display the right invariant metric we first build right invariant forms on $\tilde{X}$, invariant under the full group $SL(\dm, \mathbb{R})$, calculating the form $dx x^{-1} = \rho^A M_A$, $A = 1, \dots, \dm^2 -1$. Following \cite{PerelomovOlshanetsky1981} we write 
\ba \label{eq:Toda_intermediate} 
 \dot x (t) x^{-1} (t) &=& 2Z \left[ \frac{1}{2} Z^{-1} \dot{Z} + 
diag \left[ \dot{q}_1 \dots  \dot{q}_\dm \right]  \right. \nn \\ 
&& \hspace{0cm} \left. + \frac{1}{2} h^2 \left( Z^{-1} \dot{Z} \right)^T h^{-2} \right] Z^{-1}  . 
\ea
We can analyse separately the three terms that arise from the terms in square brackets. The first one is 
\be 
 Z^{-1} dZ = \sum_{a=1}^{\dm-1} d\omega_a M_{a,a+1} \, , 
\ee 
These are left invariant forms of the subgroup $\mathcal{Z}$. The second term is calculated in \cite{GaryMarco2013} and is seen to be given by 
\ba 
&& 2 \sum_{a=1}^{\dm-1} dq_a M_a + 2 \sum_{b<c} \Big[ \sum_{a=1}^{\dm-1} d q_a \left( \delta_{ac} Z_{ba} + (-1)^{c-b} \delta_{ab} Z_{ac} \right. \nn \\ 
&& \hspace{2.5cm} \left. + (-1)^{c-a} Z_{ba} Z_{ac} - \delta_{cn} Z_{bc} \right) \Big] M_{bc} \, , 
\ea 
while the third term by 
\ba 
&&  \sum_{a=1}^{\dm-1} e^{-2(q_a - q_{a+1})} d\omega_a \Mb_{a+1,a} + e^{-2(q_1-q_2)} \omega_1 d \omega_1 M_1 \nn \\ 
&&  + \sum_{a=2}^{\dm-1} \left( e^{-2(q_a - q_{a+1})} \omega_a d\omega_a - e^{-2(q_{a-1} - q_a)} \omega_{a-1} d\omega_{a-1} \right) M_a \nn \\ 
&& + \sum_{b<c} \left(\sum_{a=1}^{\dm -1} e^{-2(q_a-q_{a+1})} g_{abc} \, d\omega_a \right) M_{bc} \, . 
\ea 
On the other hand we have the expansion 
\be 
dx x^{-1} = \sum_{a=1}^{\dm-1} \rho_a M_a + \sum_{a<b} \rho_{ab} M_{ab} + \sum_{a>b} \rhob_{ab} \Mb_{ab} \, , 
\ee 
in terms of right invariant forms. Comparing the two expansions we find 
\be 
\rhob_{a+1,a} = e^{-2(q_a-q_{a+1})} d\omega_a \, , 
\ee 
the other $\rhob$ forms being zero on $\tilde{X}$. The $\rhob$ forms correspond to the conserved quantities 
\be \label{eq:general_condition_Z_dot_Z_inverse}
\overline{c}_{a+1,a} = e^{-2(q_a-q_{a+1})} \dot{\omega}_a  \, , 
\ee 
such that setting $\overline{c}_{a+1,1} = 2 g_a$ we recover the condition $\dot{Z} Z^{-1} = M$, with $M$ defined in \eqref{eq:Lax_M}. Next, we find the forms 
\ba \label{eq:rho_a_forms}
\rho_1 &=& 2 dq_1 + e^{-2(q_1 - q_2)} \omega_1 d\omega_1 \, , \nn \\ 
\rho_a &=& 2 dq_a + \left( e^{-2(q_a - q_{a+1})} \omega_a d\omega_a \right. \nn \\ 
&& \left. - e^{-2(q_{a-1} - q_{a})} \omega_{a-1} d\omega_{a-1} \right) \, , \quad a>1 \, . 
\ea 
These give rise to the conserved quantities 
\ba 
\lambda_1 &=& \dot{q}_1 + g_1 \omega_1  \, , \nn \\ 
\lambda_a &=& 2 \dot{q}_a + g_a \omega_a - g_{a-1} \omega_{a-1}  \, , \quad a>1 \, , 
\ea
which imply the equations of motion arising from \eqref{eq:non_periodic}. The forms $\rho_a$ are defined for $a=1, \dots, \dm -1$. It is  useful defining an $\dm$-th form $\rho_\dm$, linearly dependent on the other ones, by setting $a=\dm$ in eq.\eqref{eq:rho_a_forms} above: 
\ba 
\hspace{-0.5cm} \rho_\dm =  2 dq_\dm  - e^{-2(q_{\dm-1} - q_{\dm})} \omega_{\dm-1} d\omega_{\dm-1} = - \sum_{a=1}^{\dm-1} \rho_a \, . 
\ea

Lastly, there are the forms of type $\rho_{ab}$. In fact we only explicitly need the following ones: 
\ba \label{eq:forms_rho_a_a_plus_one}
&& \rho_{a,a+1} = 2 \omega_a (dq_{a+1} - dq_a ) + d\omega_a \nn \\ 
&& + \frac{\omega_a}{2} \left[ \left( \omega_{a+1}  e^{-2(q_{a+1} - q_{a+2})} d\omega_{a+1} - \omega_a  e^{-2(q_a - q_{a+1})} d\omega_a \right) \right. \nn \\ 
&& \left. \hspace{-0.5cm} - \left( \omega_a  e^{-2(q_a - q_{a+1})} d\omega_{a-1} - \omega_{a-1}  e^{-2(q_{a-1} - q_a)} d\omega_{a-1} \right) \right]  . 
\ea 
 
Any right invariant metric that generate the conserved quantities given above. In particular, we can make the following choice: 
\ba \label{eq:metric_generic_case} 
\hat{g} &=& \sum_{a=1}^{\dm} \left(\frac{\rho_a}{2}\right)^2 + \sum_{a=1}^{\dm-1} \frac{\rhob_{a+1,a} \rho_{a,a+1}}{2} \nn \\ 
&=& \sum_{a=1}^{\dm-1 } \left(\frac{\rho_a}{2}\right)^2 + \left( \sum_{a=1}^{\dm-1} \frac{\rho_a}{2} \right)^2 + \sum_{a=1}^{\dm-1} \frac{\rhob_{a+1,a} \rho_{a,a+1}}{2} \nn \\ 
&=& \sum_{a=1}^\dm dq_a^2 + \frac{1}{2} \sum_{a=1}^{\dm-1} e^{-2(q_a - q_{a+1})} d\omega_a^2 \, . 
\ea 
We can immediately recognise that this is the generalised Eisenhart lift metric in the same format as given by \eqref{eq:Eisenhart_2nd_type_generalised_metric} (the coupling constants are at this stage hidden in the definition of the $\omega$ variables, eq.\eqref{eq:general_condition_Z_dot_Z_inverse}). 
  
Performing a dimensional reduction to the standard Eisenhart lift is straightforward. We can define the variable 
\be 
y= \sum_{a=1}^{\dm-1} g_a \omega_a \, . 
\ee 
On allowed trajectories it satisfies 
\be 
\dot{y} = \sum_{a=1}^{\dm-1} 2 g_a^2 e^{2(q_a-q_{a+1})} = 2 V(q) \, , 
\ee 
where the potential $V$ is given in \eqref{eq:non_periodic}. On the other hand, on trajectories we also have 
\be 
\frac{1}{2} \sum_{a=1}^{\dm-1} e^{-2(q_a - q_{a+1})} \dot{\omega_a}^2 = 2 V(q) \, , 
\ee 
so we have the equality between forms 
\be 
\frac{1}{2} \sum_{a=1}^{\dm-1} e^{-2(q_a - q_{a+1})} d\omega_a^2 = \frac{dy^2}{2V(q)}  
\ee 
on all allowed trajectories. Then this identity holds on the whole span of the trajectories and we can re-write the metric as 
\be 
\hat{g} = \sum_{a=1}^\dm dq_a^2 + \frac{dy^2}{2V(q)}  \, , 
\ee 
which is the standard Eisenhart metric. 
 
We remark that the isometries of the metric $\hat{g}$, given by the Killing vectors associated to the right invariant forms of $SL(\dm, \mathbb{R})$, become transformations that leave  unchanged the higher dimensional Hamiltonian 
\be 
\mathcal{H} = \sum_{a=1}^\dm  \frac{p_{q_a}^2}{2} + \sum_{a=1}^{\dm-1} p_{w_a}^2 e^{2(q_a - a_{a+1})} \, , 
\ee 
and hence are transformations that mix $q$ and $\omega$ coordinates while leaving the energy of the original Toda chain unchanged. Consider as an example the conserved quantities calculated above. The first ones are of the kind 
\be 
 \overline{c}_{a+1,a} = 2 p_{\omega_a}   \, , 
\ee 
which generate trivial translations of the $\omega$ variables.  
 
Next,  the quantities
\ba 
&& \lambda_1 = p_{q_1} + p_{\omega_1} \omega_1 \, , \nn \\ 
&& \lambda_a = p_{q_a} + p_{\omega_a} \omega_a - p_{\omega_{a-1}} \omega_{a-1} \, , a > 1 \, , 
\ea 
generate the transformations 
\ba 
&& \delta q_1 = \epsilon \, , \nn \\ 
&& \delta \omega_1 =  \epsilon \omega_1 \, , \\ 
&& \delta_{p_{\omega_1}} = - \epsilon p_{{\omega_1}} \, , 
\ea 
and 
\ba 
&& \delta q_a = \epsilon \, , \nn \\ 
&& \delta \omega_a =  \epsilon \omega_a \, , \delta \omega_{a-1} =  - \epsilon \omega_{a-1}  \\ 
&& \delta_{p_{\omega_a}} = - \epsilon p_{\omega_a} \, , \delta_{p_{\omega_{a-1}}} =  \epsilon p_{{\omega_{a-1}}} \, , 
\ea 
where $\epsilon$ is an infinitesimal parameter. Their meaning is that it is possible to make a constant shift of one of the variables $q_a$ while keeping the energy unchanged by at the same time making a constant rescaling of the $g_a$ and $g_{a-1}$ couplings. In the case of the conserved quantities associated to the conserved forms of equation \eqref{eq:forms_rho_a_a_plus_one}, these give rise to more complicated transformations where $\delta p_{\omega_a}$ is not constant. 
 
Summarising, the symmetries of the metric \eqref{eq:metric_generic_case} give  the dynamical transformations that leave the energy unchanged, where varying $p_{\omega_a}$ are allowed. In particular, the transformations that are of lowest order in the momenta are those given by the Killing vectors, which form an $SL(\dm, \mathbb{R})$ algebra.

\section{CONCLUDING REMARKS\label{sec:final}} 
Dynamical symmetries are the full symmetries of the equations of motion of classical and quantum systems, extending the traditionally considered isometries and providing the complete set of energy preserving transformations in phase space. Such symmetries are known since a long time and have attracted the attention of researchers from different angles and at different stages, however a comprehensive, unified and modern treatment is lacking. It is having in mind these requirements that the present review was written with the intention of providing an initial tool for graduate students and young researchers interested in an introduction to the topic, and for specialists to find a place where several of the more modern results are gathered. We have provided a number of examples in sec.\ref{sec:applications_and_examples} inspired from different areas of physics, relativistic as well as not, some including gravitation, some in flat space, some older and somewhat classical examples as well as more modern ones, like the spinning particle and the quantum dot. We have given an introduction to the theory of spin-zero separation of variables, both in the classical setting, the Hamilton-Jacobi equation discussed in sec.\ref{sec:Hamilton-Jacobi}, and the quantum mechanical version, the Schr\"{o}dinger and Klein-Gordon equations discussed in sec.\ref{sec:Klein_Gordon}. The spin zero case is the simplest possible, and admits a fairly mature theory of the intrinsic characterisation  of separability, but even there we can see at least two areas for further improvement. First, to our knowledge a theory of the intrinsic R-separation for generic non-orthogonal variables and the presence of a scalar and vector potential has not been finalised as yet, although several of the necessary ingredients are already in place. Also, a more directly geometrical interpretation of the R-factor would be desirable: we believe that in this respect embedding the dynamical system in higher dimensions through the Eisenhart lift procedure has the potential to provide a geometrical interpretation for it. Second, we know very little about more complicated transformations in phase space. While the theory of separation of variables mainly deals with canonical transformations associated to phase space functions that are second order polynomials in the momenta, transformations that are of higher order or non-polynomial have not been systematically studied. We know specific examples but there is not yet a complete theory. Even deeper relations among Hamiltonian systems, such as for example the coupling constant metamorphosis discussed in \cite{Hietarinta1984coupling} and following works, are not related to canonical transformations and represent an open window on a more complete understanding. 
 
Outside the realm of the zero spin case, what we know is mostly concrete results and partial theories. The simplest cases with non-zero spin are given by the semi-classical spinning particle and the quantum mechanical Dirac equation, discussed in section \ref{sec:examples_spinning_particle} and, respectively, \ref{sec:quantum_dirac_equation}. A theory of symmetry operators of the Dirac equation in generic dimension exists only for linear operators. Conserved quantities of second order in the momenta for the spinning particle have been discussed recently for the rotating black holes of the Kerr-NUT-(A)dS metrics, and they seem to work in their known form specifically for these metrics. Some partial results for integrability and separability of higher spin equations are also known, but even less is known in this case of a general theory. The possibility of choosing both a set of coordinates and a locally freely falling frame to describe the spin variables represents a much bigger freedom  compared to that of the spin zero case. 
 
We have discussed the Eisenhart lift as a powerful geometerisation of dynamics, that allows to easily find the full group of symmetries of the dynamics, and to extend the theory of separability of variables to the case with scalar and vector potential. In general, it seems the appropriate tool to describe the dynamics of Hamiltonian systems. 
 
Much of the recent interest in hidden symmetries of the dynamics in the General Relativity area has stem from the discovery that the Kerr-NUT-(A)dS metrics admit a principal conformal Killing-Yano tensor and a whole tower of Killing-Yano and Killing tensors. We have given an account of how the construction works and how it relates with the other topics discussed in the review. It would certainly be important to obtain a classification of Lorentzian geometries that admit a principal conformal Killing-Yano tensor: this potentially can provide new metrics that support integrable dynamics. 
 
By studying the generalisation of the conformal Killing-Yano equation to some specific cases when fluxes are present we have shown one possible way to look for hidden symmetries of the dynamics when the field content of the theory is richer. It is therefore a perfectly reasonable expectation that various supergravity backgrounds and other String Theory inspired backgrounds might be compatible with more elaborated dynamical symmetry transformations. Here, once more, our knowledge is partial and a theory exists mainly for the case discussed of spacetimes with specific types of fluxes.  
 
One last subject we touched on is that of geodesics on Lie groups. This subject has been studied relatively little from the point of view of hidden symmetries. In particular, there are two outstanding questions. The first is what is the relation between the known geodesic description in higher dimensions of some integrable systems of particles on a line, and the standard Eisenhart lift. We have seen that the answer is particularly simple when the integrable system is given by the $\dm$-particles Toda chain, however the answer for the other cases is not known yet. Second, it is still an open conjecture whether all such integrable systems admit a similar geodesic description in higher dimensions in terms of symmetric spaces. We wonder whether an answer to the latter question might be related to the former and to the study of dynamical symmetries. 
 
Undoubtedly the study of dynamical symmetries is here to stay: it is both very important for our understanding of physical systems, and at the same time our current knowledge of the subject has plenty of room for expansion and refinement.

\section*{Acknowledgments} 
The author is grateful to G. E. A. Matsas for suggesting to write a review on the subject of hidden symmetries. The author also 
thanks C. M. Warnick who has collaborated to the review at an initial stage, and G. W. Gibbons and D. Kubiz\v{n}\'ak for useful discussions.

\newpage
\bibliographystyle{apsrmp}
\bibliography{Bibliography_Review_11Mar2014}

\end{document}